\documentclass[american,superscriptaddress,aps,prd,preprint,floatfix,letterpaper]{revtex4}
\usepackage[latin1]{inputenc}
\usepackage{float}
\usepackage{amsmath}
\usepackage{graphicx}
\usepackage{amssymb}
\providecommand{\tabularnewline}{\\}

\usepackage[bookmarks=true]{hyperref}
\usepackage{slashed}
\usepackage{lscape}

\usepackage{babel}
\makeatother
\begin{document}

\title{Non-perturbative Renormalization of Quark Bilinear Operators and $B_K$
using Domain Wall Fermions}

\newcommand\riken{RIKEN-BNL Research Center, Brookhaven National Laboratory,
                  Upton, NY 11973, USA}
\newcommand\bnlaf{Brookhaven National Laboratory, Upton, NY 11973, USA}
\newcommand\edinb{SUPA, School of Physics, The University of Edinburgh,
                  Edinburgh EH9 3JZ, UK}
\newcommand\cuaff{Physics Department, Columbia University, New York, NY 10027, USA}
\newcommand\glasg{SUPA, Department of Physics \& Astronomy, University of
                  Glasgow, Glasgow G12 8QQ, UK}
\newcommand\kek{Institute of Particle and Nuclear Studies, KEK, Tsukuba, Ibaraki 305-0801, Japan}
\newcommand\kanaz{Institute for Theoretical Physics,  Kanazawa University,
                  Kakuma, Kanazawa, 920-1192, Japan}
\newcommand\soham{School of Physics and Astronomy, University of Southampton,
                  Southampton SO17 1BJ, UK}

\author{Y.~Aoki}        \affiliation{\riken}
\author{P.A.~Boyle}     \affiliation{\edinb}
\author{N.H.~Christ}    \affiliation{\cuaff}
\author{C.~Dawson\footnote{Current affiliation: Department of Physics, 
                        University of Virginia, Charlottesville, VA}}
                        \affiliation{\riken}
\author{M.A.~Donnellan} \affiliation{\soham}
\author{T.~Izubuchi}    \affiliation{\riken}
                        \affiliation{\kanaz}
\author{A.~J\"uttner}   \affiliation{\soham}
\author{S.~Li}          \affiliation{\cuaff}
\author{R.D.~Mawhinney} \affiliation{\cuaff}
\author{J.~Noaki}       \affiliation{\kek}
\author{C.T.~Sachrajda} \affiliation{\soham}
\author{A.~Soni}        \affiliation{\bnlaf}
\author{R.J.~Tweedie}   \affiliation{\edinb}
\author{A.~Yamaguchi}   \affiliation{\glasg}
\collaboration{RBC and UKQCD Collaborations}
%
%
\noaffiliation{BNL-HET-07/11, CU-TP-1180, Edinburgh 2007/12, KANAZAWA-07-10, RBRC-681,  SHEP-07-20}


\date{December 6, 2007}

\begin{abstract}
We present a calculation of the renormalization coefficients of
the quark bilinear operators and the $K-\overline{K}$ mixing parameter
$B_K$.  The coefficients relating the bare lattice operators to
those in the RI/MOM scheme are computed non-perturbatively and
then matched perturbatively to the $\overline{\textrm{MS}}$ scheme.
The coefficients are calculated on the RBC/UKQCD 2+1 flavor
dynamical lattice configurations.  Specifically we use a
$16^3 \times 32$ lattice volume, the Iwasaki gauge action at
$\beta=2.13$ and domain wall fermions with $L_s=16$.
\end{abstract}

\maketitle

\section{Introduction}

The RBC and UKQCD collaborations have recently performed the first
simulations with 2+1 flavor domain wall fermions \cite{Antonio:2006px,
Allton:2007hx,Boyle:2007fn}.  Much interesting phenomenology requires the
conversion of bare lattice quantities to a less arbitrary and
more perturbatively amenable continuum scheme. In particular, this
is true for the determination of weak matrix elements such as $B_{K}$
and for the Standard Model parameters such as quark masses. Of
course, physical quantities are independent of the choice of
renormalization procedure, nevertheless theoretical predictions are
often given in terms of the parameters of the theory ($\alpha_{s}$ and
quark masses) which require renormalization.  In addition, for many
processes (e.g. $K-\bar K$ mixing) the amplitudes are factorized
into products of perturbative Wilson coefficient functions and
operator matrix elements which contain the long-distance effects.
The Wilson coefficients and operator matrix elements need to be
combined with both evaluated in the same renormalization scheme. The
purpose of this paper is to determine the factors by which matrix
elements computed in our numerical simulations should be multiplied
in order to obtain those in the $\overline{\rm{MS}}$ scheme which is
conventionally used for the evaluation of the coefficient
functions.

In principle, for a sufficiently small lattice spacing $a$ and a
sufficiently large renormalization scale $\mu$, it is possible to
perform the renormalization of the bare lattice operators using
perturbation theory. However, in practice the coefficients of
lattice perturbation theory are frequently large leading to a poor
convergence of the series and even with attempts such as
\textit{tadpole improvement} to resum some of the large
contributions, it appears that the typical $n$-loop correction is
numerically of $O(\alpha_s^n)$, in contrast to continuum
perturbation theory where the corresponding contributions are of
$O((\alpha_s/4\pi)^n)$. A related difficulty is the choice of the
best expansion parameter ($\alpha_s$), for example between some
tadpole improved lattice coupling or the $\overline{\rm{MS}}$
coupling. In practice, at one-loop order, different reasonable
choices can lead to significantly different results. For the
quark bilinear operators and $B_K$ considered in this paper, we
present the perturbative results and illustrate these points in
Section~\ref{sec:Pert_Theory}.

The main purpose of this paper is to avoid the uncertainties present
when using lattice perturbation theory by implementing the
Rome-Southampton RI/MOM non-perturbative renormalization
technique~\cite{Martinelli:1994ty}. The key idea of this technique
is to define a sufficiently simple renormalization condition such
that it can be easily imposed on correlation functions in any
lattice formulation of QCD, or indeed in any regularization
- that is, the condition is regularization invariant (RI).
We therefore introduce counter-terms for any regularization such that
a Landau gauge renormalized n-point correlation function with standard
MOM kinematics at some scale $\mu^2$ has its tree level value. This
condition is simple to impose whenever the renormalized correlation
function is known in any regularization. It applies equally well to
both perturbative expansions to any order and to non-perturbative
schemes such as the lattice, and thus RI/MOM is a very useful
interface for changing schemes.  In particular, only continuum
perturbation theory and the lattice regularization are required to
obtain physical results from a lattice calculation.

Our choice of lattice action is important for the efficacy of the
RI/MOM technique. With domain-wall fermions, $O(a)$
errors and chiral symmetry violation can be made arbitrarily small
at fixed lattice spacing by increasing the size of the fifth
dimension. This allows us to avoid the mixing between operators
which transform under different representations of the
chiral-symmetry group; this is a very significant simplification
compared to some other formulations of lattice QCD. The action
and operators are also automatically $O(a)$ improved.

Another important property of DWF is the existence of (non-local)
conserved vector and axial currents. This will be discussed in
detail below.

In this paper we study the renormalization of the quark bilinear
operators $\bar\psi\Gamma\psi$, where $\Gamma$ is one of the 16
Dirac matrices, and of the $\Delta S=2$ four-quark operator $O_{LL}$.
Table~\ref{tab:z_npr_summary} contains a summary of our results, relating bare operators in the lattice theory
with Domain Wall Fermions and the Iwasaki gauge action at $\beta=2.13$ ($a^{-1}=1.729(28)\,$GeV, see Section~\ref{sec:Simulation-Details} for further details) to those in two
continuum renormalization schemes.
Columns three through five give the three independent Z factors which,
when multiplying the appropriate bilinear lattice operator, convert
that operator into one normalized according to either the RI/MOM or
$\overline{\textrm{MS}}$(NDR) schemes.  The final column contains
the combination of factors needed to convert a lattice result for
the parameter $B_K$ into the corresponding RI/MOM or
$\overline{\textrm{MS}}$(NDR) value.

The plan of the remainder of the paper is as follows. In the
following section (Section~\ref{sec:Pert_Theory}) we start by
reviewing the perturbative evaluation of the renormalization
constants; the results can later be compared with those obtained
using the non-perturbative procedures. In Section~\ref{sec:Simulation-Details},
we begin the description of the non-perturbative computations
with a brief introduction to the details of our simulation and to
the computation of the quark propagators which are the basic
building blocks for all our subsequent calculations. In
Section~\ref{sec:Renormalization-for-Quark-Bilinears} we give a
short introduction to the regularization independent (RI/MOM)
scheme. In this section we also discuss the renormalization of
flavor non-singlet bilinear operators, including the check of
the Ward-Takahashi identities.  The discussion of the
renormalization of the four-quark operators and the results for
the renormalization constant for $B_K$ are presented in
Section~\ref{sec:Renormalization-Coefficients-for-Bk}.
Section~\ref{sec:Conclusions} contains a brief summary and our
conclusions.

\section{Perturbation Theory\label{sec:Pert Theory}}
\label{sec:Pert_Theory}

Before proceeding to describe our non-perturbative evaluation of
the renormalization constants we briefly review the corresponding
(mean field improved) perturbative calculations. Specifically, we
present perturbative estimates for the renormalization constants
of the quark bilinears and $B_{K}$. These can then be compared to
those obtained non-perturbatively below. The ingredients for the
perturbative calculations and a detailed description of the
procedure can be found in refs.~\cite{Aoki:2002iq,Aoki:2003uf}.

Writing the domain wall height as $M=1-\omega_{0}$, the bare value
of $\omega_{0}$ in our simulation is $\omega_{0}=-0.8$. The mean
field improved value of $\omega_{0}$ is then given by
\begin{equation}
\omega_{0}^{\textrm{MF}}=\omega_{0}+4(1-u)\simeq-0.303\,,\end{equation}
 where the link variable is defined by $u={\cal P}^{1/4}$ and ${\cal P}=0.588130692$
is the value of the plaquette in the chiral limit.

We define the renormalization constant, $Z_{O_{i}}$, which relates
the bare lattice operator, $O_{i}^{\textrm{Latt}}(a^{-1})$, to the
corresponding renormalized one in the $\overline{\textrm{MS}}$
scheme at a renormalization scale of $\mu=a^{-1}$ by:
\begin{equation} O_{i}^{\overline{\textrm{MS}}}(a^{-1})=Z_{i}\,
O_{i}^{\textrm{Latt}}(a^{-1}).\end{equation}
 Here $i=S,P,V,A,T$ for the scalar and pseudoscalar densities, vector
and axial-vector currents and tensor bilinear and $i=B_{K}$ for
the $\Delta S=2$ operator which enters into the
$K^{0}$\,-\,$\bar{K}^{0}$ mixing amplitude (or more precisely for
the ratio of the $\Delta S=2$ operator and the square of the local
axial current, which is the relevant combination for the
determination of $B_{K}$). The one-loop, mean field improved
estimates for the $Z_{i}$ are: \begin{eqnarray}
Z_{S,P} & = & \frac{u}{1-(\omega_{0}^{\textrm{MF}})^{2}}\,\frac{1}{Z_{\omega}^{\textrm{MF}}}\left(1-\frac{\alpha_{s}\, C_{F}}{4\pi}\,5.455\,\right)\label{eq:zsp}\\
Z_{V,A} & = & \frac{u}{1-(\omega_{0}^{\textrm{MF}})^{2}}\,\frac{1}{Z_{\omega}^{\textrm{MF}}}\left(1-\frac{\alpha_{s}\, C_{F}}{4\pi}\,4.660\,\right)\\
Z_{T} & = & \frac{u}{1-(\omega_{0}^{\textrm{MF}})^{2}}\,\frac{1}{Z_{\omega}^{\textrm{MF}}}\left(1-\frac{\alpha_{s}\, C_{F}}{4\pi}\,3.062\,\right)\\
Z_{B_{K}} & = & 1-\frac{\alpha_{s}}{4\pi}\,1.470\,,\end{eqnarray}
where $C_F$ is the second Casimir invariant $C_F=(N^2-1)/2N$ for
the gauge group $SU(N)$.  Here $\sqrt{Z_{w}}$ is the quantum
correction to the normalization factor $\sqrt{1-\omega_{0}^{2}}$
of the physical quark fields (the factors depending on this
normalization cancel in the evaluation of $Z_{B_{K}}$).
 At one-loop order in perturbation theory \begin{equation}
Z_{w}=1+\frac{\alpha_{s}\,
C_{F}}{4\pi}\,5.251\,.\label{eq:zw}\end{equation}
 In obtaining the coefficients in Eqs.\,(\ref{eq:zsp})\,--\,(\ref{eq:zw})
we have interpolated linearly between the entries for $M=1.30$ and
$M=1.40$ in tables III and IV of ref.\,\cite{Aoki:2002iq} to the
mean-field value of $M=1.303$. Since the mean-field value of $M$
is so close to the quoted values at $M=1.30$, we prefer this
procedure to using the general interpolation formula quoted in
\cite{Aoki:2002iq}. The difference between the two procedure is
negligible compared to the remaining systematic uncertainties.

In order to estimate the numerical values of the renormalization
constants we have to make a choice for the expansion parameter,
i.e. the coupling constant $\alpha_{s}$. Here we consider two of
the possible choices, the mean-field value as defined in eq.(62)
of ref.~\cite{Aoki:2003uf} and the $\overline{\textrm{MS}}$
coupling, both defined at $\mu=a^{-1}$. The mean field improved
coupling constant is given by \begin{equation}
\frac{1}{g_{\textrm{MF}}^{2}(a^{-1})}=\frac{{\cal
P}}{g_{0}^{2}}+d_{g}+c_{p}+N_{f}d_{f}\,,\end{equation}
 where $g_{0}$ is the bare lattice coupling constant ($g_{0}^{2}=6/\beta$),
and the remaining parameters are defined in
ref.\,\cite{Aoki:2003uf} and take the numerical values
$d_{g}=0.1053$, $c_{p}=0.1401$ and for
$\omega_{0}^{\textrm{MF}}=-0.303$, $d_{f}=-0.00148$. We therefore
obtain \begin{equation} \alpha_{\textrm{MF}}(1.729\,\textrm{
GeV})=0.1769\,.\end{equation} Such a value of the coupling is
significantly lower than that in the $\overline{\textrm{MS}}$
scheme at the same scale, for which we take,
$\alpha^{\overline{\textrm{MS}}}(1.729\,\textrm{ GeV})=0.3138$.

The difference in the two values of the coupling constant leads to
a significant uncertainty in the estimates of the renormalization
constants at this order, as can be seen in Table~\ref{tab:alpha
diff Z diff}. The need to eliminate this large uncertainty is the
principle motivation for the use of non-perturbative
renormalization. The entries in Table \ref{tab:alpha diff Z diff}
are the factors by which the matrix elements of the bare lattice
operators should be multiplied in order to obtain those in the
$\overline{\textrm{MS}}$(NDR) scheme at the renormalization scale
$\mu=1.729$\,GeV.

Finally we perform the renormalization group running from
$\mu=1.729$\,GeV to obtain the normalization constants at other
scales, and in particular at the conventional reference scale of
$\mu=2$\,GeV (see Table~\ref{tab:latt pert Z}). In each case we
use the highest order available for the anomalous dimension; two
loops for $B_{K}$, three loops for the tensor operator and four
loop for the scalar/pseudoscalar densities. This is the same
procedure which we use for the non-perturbatively renormalized
normalization constants below and the details and references to
the anomalous dimensions are presented in
sections~\ref{sub:Zm_running}, \ref{sub:Zq_running} and
\ref{sub:Zbk_running} below. The numbers in Table~\ref{tab:latt
pert Z} are the factors by which the matrix elements of the bare
lattice operators should be multiplied in order to obtain those in
the $\overline{\textrm{MS}}$(NDR) scheme at $\mu=2$\,GeV. The
entries in the first column indicate which coupling was used in
matching between the bare lattice operators and the
$\overline{\textrm{MS}}$(NDR) scheme at $\mu=1.729$\,GeV, i.e.
before the running to other scales.

\section{Simulation Details\label{sec:Simulation-Details}}

The calculations described below were performed on the 2+1 flavor dynamical lattice configurations generated
by the RBC and UKQCD collaborations \cite{Allton:2007hx}. The lattices
were generated with the Iwasaki gauge action at $\beta=2.13$ and
the domain-wall fermion action with $L_{s}=16$. The size of the lattices
used in this work is $16^{3}\times32$. The lattice spacing is $a^{-1}=1.729(28)$GeV
and the residual mass $m_{\mathrm{res}}=0.00315(2)$ in lattice units \cite{lin:2007pt}.
We have 3 independent ensembles with light sea quark mass 0.01, 0.02 and
0.03 respectively.  The strange sea quark mass is fixed at 0.04. For each
ensemble, we have used 75 configurations, starting from trajectory number
1000 and with trajectory separation 40.

Following the Rome-Southampton RI/MOM non-perturbative renormalization
procedure\cite{Martinelli:1994ty,Blum:2001sr}, the lattices are first
fixed in Landau gauge. Then, on each gauge-fixed configuration,
we measure the point-point quark propagators $S\left(x,x_{0}\right)$
with periodic boundary conditions in space and time, where $x_{0}$
is the source position and $x$ is the sink. We have chosen four different
sources to generate the propagators,\begin{equation}
x_{0}\in\left\{ \left(0,0,0,0\right),\left(4,4,4,8\right),\left(7,7,7,15\right),\left(12,12,12,24\right)\right\} .\label{eq:source-pos}\end{equation}
Next, a discrete Fourier transform is performed on the propagators,\begin{equation}
S\left(p,x_{0}\right)=\sum_{x}S\left(x,x_{0}\right)\exp\left[-ip\cdot\left(x-x_{0}\right)\right]\,,\label{eq:fourier-prop}\end{equation}
where \begin{equation}
p_{\mu}=\frac{2\pi}{L_{\mu}}n_{\mu}\label{eq:p_mu},\end{equation}
$n_\mu$ is a four-vector of integers and\begin{align}
L_{x}=L_{y}=L_{z} =16 \quad L_{t} =32.\label{eq:L_mu}\end{align}
For the $n_\mu$ we take values in the ranges
\begin{equation}
n_{x},n_{y},n_{z} \in\left\{ -2,-1,0,1,2\right\}\qquad\textrm{and}\qquad
n_{t}  \in\left\{ -4,-3,-2,-1,0,1,2,3,4\right\}\label{eq:n_mu}
\end{equation}
and require that the squared amplitude of the lattice momenta is in the range
$0\leq p^{2}\lesssim2.5$.  In this paper, for simplicity of notation we
frequently use lattice units for dimensionful quantities such as $p$ and
$m$.  When we particularly wish to emphasize the nature of the
discretization errors we explicitly reinstate the lattice spacing,
writing for example, $(ap)^2$ or $(am)^2$.

\section{Renormalization of Quark Bilinears
\label{sec:Renormalization-for-Quark-Bilinears}}

We now discuss how Green functions computed on the lattice can be  
used to obtain the non-perturbative renormalization
constants relating bilinear operators defined on the lattice to
those normalized first according to the RI/MOM and then the
$\overline{\rm{MS}}$ scheme.  In the first two subsections below,
Sections~\ref{sec:mass_wf_npr} and~\ref{sub:cal-lambda} we briefly
introduce the definitions and notation that we use in the rest of
this paper. Some of these are extracted from the earlier RBC paper
on quenched lattices \cite{Blum:2001sr} and are included here for
completeness.

The definition of renormalization factors $Z_q$ and $Z_m$ for the
quark wave function and mass in the RI/MOM scheme are given in
Section~\ref{sec:mass_wf_npr}.  The basic amputated quark-bilinear
vertex functions are defined in Section~\ref{sub:cal-lambda} and
the conditions defining the RI/MOM scheme are written down.  Since our
calculations are necessarily performed at finite momenta, the
effects of chiral symmetry breaking coming from both the non-zero
quark masses and spontaneous chiral symmetry breaking are visible.
We discuss these effects in detail in Section~\ref{sec:ch_sym_breaking}
for the important case of the vector and axial vector vertex functions.

Next, as a consistency check for our methods, we discuss the accuracy
with which our off-shell vertex amplitudes satisfy the axial and vector
Ward-Takahashi identities in Sections ~\ref{sec:axial-ward-id} and
\ref{sec:vector-ward-id} respectively.  In the later section, the
determination of $Z_S/Z_q$ is also discussed.  In Section~\ref{sub:Zq_running}
we compare the observed scale dependence of $Z_m$ with that predicted
by perturbation theory and interpret the differences as coming from
$(a\mu)^2$ errors.  These are removed to determine first $Z_m^{RI/MOM}$
and then $Z_m^{\overline{MS}}$.  A similar determination of $Z_q^{RI/MOM}$ and
then $Z_q^{\overline{MS}}$ is presented in Section~\ref{sub:Zq_running}.
Finally, in Section~\ref{sub:Zt_running}, results for the tensor
vertex renormalization factor $Z_T$ are obtained.

\subsection{Quark mass and wavefunction renormalization}
\label{sec:mass_wf_npr}

First, we define the renormalization
coefficients for the quark field and the quark mass as the ratio between
the renormalized quantities and their bare counterparts,\begin{align}
q_{\mathrm{ren}}\left(x\right) & =Z_{q}^{\frac{1}{2}}q_{0}\left(x\right)\label{eq:qren}\\
m_{\mathrm{ren}} & =Z_{m}m_{0}.\label{eq:mren}\end{align}
where $q_{\mathrm{ren}}$ and $q_{0}$ are the renormalized and the
bare quark wavefunction, and $m_{\mathrm{ren}}$ and $m_{0}$ are
the renormalized and the bare quark mass. With domain-wall fermions,
\begin{equation}
m_{0}=m_{f}+m_{\mathrm{res}}\label{eq:m0}\end{equation}
where $m_{f}$ is the input quark mass and $m_{\mathrm{res}}$ is
the residual mass. The renormalized propagator (in momentum space) is
\begin{equation}
S_{\mathrm{ren}}\left(p,m_{\mathrm{ren}}\right)
  =\left.Z_{q}S_{0}\left(p,m_{0}\right)\right|_{m_{0}=m_{\mathrm{ren}}/Z_{m}}
\label{eq:Sren}
\end{equation}
where $p$ is the momentum of the quark propagator.

Since domain wall quarks enter the calculations described here in
three different ways we must be careful to clearly distinguish their
three distinct masses.  As described above, our calculations are
performed with 2+1 flavors of dynamical quarks.  We will use the
variable $m_l$ to label the input mass $m_f$ for the light dynamical
quarks and $m_s$ for that of the dynamical strange quark.  Since we
often evaluate products of propagators which depend on a third quark
mass, that mass is labeled $m_\mathrm{val}$.  In some cases the limit
$m_\mathrm{val} \rightarrow 0$ may be an adequate definition of the
chiral limit.  However, in order to deal with simple results from
which a weak quark mass dependence has been removed we will often
consider the ``unitary'' case $m_\mathrm{val}= m_l$ and take the
limit $m_\mathrm{val}= m_l \rightarrow 0$.  Of course, in this case
$m_s$ remains non-zero but since its value is never changed
this causes no immediate confusion.  Underlying the validity of
the Rome-Southampton renormalization scheme is the use of
infrared-regular renormalization kinematics.  Therefore, as our
renormalization scale $\mu$ becomes larger and future calculations
more accurate, even this weak quark mass dependence will completely
disappear.

As discussed in detail in \cite{Blum:2001sr}, the renormalization
technique requires the existence of a window of momenta such that
\begin{equation}
\Lambda_{\mathrm{QCD}}\ll\left|p\right|\ll a^{-1}.
\label{eq:mom-window}
\end{equation}
In practice, however, violation of these restrictions, especially at
the higher boundary, has to be considered.  Due to the spontaneous
breaking of the chiral symmetry as illustrated by the non-trivial
difference between $Z_{q}/Z_{A}$ and $Z_{q}/Z_{V}$ at low momenta
(see section IV C) we have to rely on the calculation in the
relatively high momentum region, where $\left(ap\right)^{2} \gtrsim 1$.
Fortunately, the effects from breaking the restriction imposed by the
finite lattice spacing $a$ are small and predictable.  They introduce an
error of $\mathcal{O}\left(\left(ap\right)^{2}\right)$ to the
renormalization coefficients which can be removed by quadratic
fitting to the momentum dependence.  A more detailed investigation of
this issue is presented in \cite{Blum:2001sr}.

It is possible in principle to relax the constraint
$|p|\gg\Lambda_{\textrm{QCD}}$ in eq.(19) by performing
\textit{step scaling}, i.e. by matching the renormalization
conditions successively to finer (and also smaller in physical
units) lattices. This is beyond the scope of this paper.

The regularization independent (RI/MOM) scheme is defined such that by
adjusting the renormalization coefficients $Z_{q}$ and $Z_{m}$ at
the renormalization scale $\mu$, and restricting $p$ in a suitable
window, we have:
\begin{align}
\lim_{m_{\mathrm{ren}}\to0}-\frac{i}{12}\mathrm{Tr}\left(
 \frac{\partial S_{\mathrm{ren}}^{-1}}{\partial\slashed{p}}\left(p\right)
     \right)_{p^{2}=\mu^{2}} & =1
\label{eq:RIdef_Zq}\\
\lim_{m_{\mathrm{ren}}\to0}\frac{1}{12m_{\mathrm{ren}}}\mathrm{Tr}\left(
   S_{\mathrm{ren}}^{-1}\left(p\right)\right)_{p^{2}=\mu^{2}} & =1.
\label{eq:RIdef_Zm}
\end{align}
Imposing these conditions on the lattice and taking into account the
dynamical breaking of chiral symmetry at low energies, the additive
mass renormalization $m_{\mathrm{res}}$ and $\mathcal{O}\left(a^{2}\right)$
lattice artifacts, we have the following asymptotic behavior
\cite{Blum:2001sr} for relatively large $p^{2}$:
\begin{equation}
\frac{1}{12}\mathrm{Tr}\left(S_{\mathrm{latt}}^{-1}\left(p\right)\right)
  =\frac{a^{3}\left\langle \bar{q}q\right\rangle }{\left(ap\right)^{2}}C_{1}Z_{q}
   +Z_{m}Z_{q}\left\{ am_\mathrm{val}+am_{\mathrm{res}}\right\}
  +\mathcal{O}\left(\left(ap\right)^{2}\right).\label{eq:TrSinv}
\end{equation}

The left-hand-side of Eq.\eqref{eq:TrSinv} at each of the unitary points
($m_l=m_\mathrm{val}$) is calculated as the inverse of the average over
all propagators, where the average is performed over all sources and
configurations:
\begin{equation}
S\left(p\right)^{-1}=\left\{\frac{1}{N}\sum_{i=1}^{N}
 \left[\frac{1}{n_{source}}\sum_{x_{0}}S_{i}\left(p,x_{0}\right)\right]\right\}^{-1}
\label{eq:S avg}
\end{equation}
where $n_{source}=4$ and $i\in\left\{ 1,2,\cdots,N\right\} $ labels
each configuration.  (For brevity, in this equation and in the following we
will suppress the subscript {}``latt''on all the lattice propagators.
In the following text, unless otherwise specified, all the propagators
$S\left(p\right)$ without a subscript denote the lattice propagators.)
Because of possible correlations between propagators with different sources
calculated on the same configuration, each group of 4 propagators from the
same configuration is considered as one jackknife bin in the
single-elimination procedure.

\begin{figure}[H]
\begin{centering}
\includegraphics[clip]{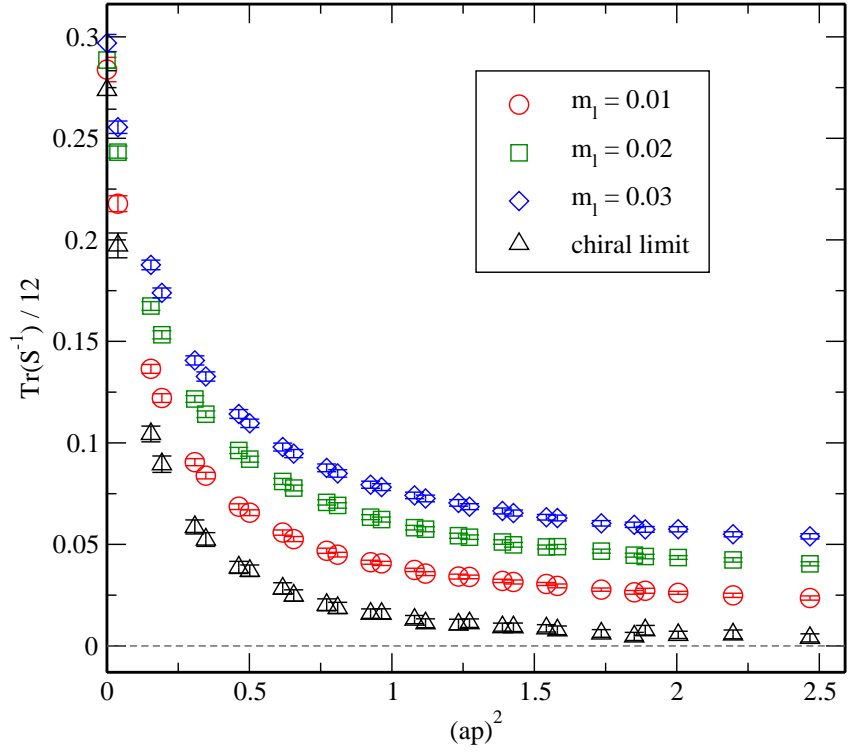}
\par\end{centering}

\caption{The quantity $\frac{1}{12}\mathrm{Tr}\left(S_{\mathrm{latt}}^{-1}\right)$
plotted versus $\left(ap\right)^{2}$ for the unitary mass points $m_l = 0.01$, 0.02
and 0.03 and at the linearly extrapolated, chiral limit $m_l=-m_\mathrm{res}$.
\label{fig:TrSinv_dyn}}
\end{figure}

In Figure~\ref{fig:TrSinv_dyn} we plot the results for
$\frac{1}{12} \mathrm{Tr}[S^{-1}(p)]$ as a function of the momentum and tabulate
the corresponding numerical values in Table~\ref{tab:TrSinv_dyn}. In
addition to our results for non-zero quark mass, we also plot and
tabulate the results extrapolated to the chiral limit where
$m_l=m_\mathrm{val}=-m_\mathrm{res}$ for each momentum.  Two quantities
of interest can be deduced from the mass dependence shown in
Figure~\ref{fig:TrSinv_dyn}.  First, the chiral limit gives a measure
of the spontaneous and explicit chiral symmetry breaking and is given
in the left-most column of Table~\ref{tab:TrSinv_dyn}.  Second determining
the slope with respect to $m_\mathrm{val}$ provides one method to calculate
$Z_{m}Z_{q}$.  This is used later in Section~\ref{sec:vector-ward-id}
to test the vector Ward Identity which relates this to a second method
of computing $Z_m Z_q$.

\subsection{Renormalization of flavor non-singlet fermion bilinears}
\label{sub:cal-lambda}

We now consider the renormalization of quark bilinear
operators of the form $\bar{u}\Gamma d$, where $\Gamma$ is one of the
16 Dirac matrices.  The corresponding renormalization constant
$Z_\Gamma$ is the factor relating the renormalized and bare bilinear
operators:
\begin{equation}
[\bar{u}\Gamma d]_{\textrm{ren}}(\mu)= Z_\Gamma(\mu a)[\bar{u}\Gamma d]_{0}\,,
\label{eq:Z_bi}
\end{equation}
where $\mu$ is the renormalization scale and we will treat only local
operators where the lattice fields $\bar u$ and $d$ in the bilinear
operator $[\bar{u}\Gamma d]_{0}$ are evaluated at the same space-time
point.

Following the Rome-Southampton prescription\cite{Martinelli:1994ty}
for renormalization in the RI/MOM scheme, we define the bare Green
functions between off-shell quark lines, and evaluate their
momentum-space counterparts $G_{\Gamma,0}\left(p\right)$ on the
lattice, averaged over all sources and gauge configurations,
\begin{equation}
G_{\Gamma,0}\left(p\right)=\frac{1}{N}\sum_{i=1}^{N}\left\{ \frac{1}{n_{source}}\sum_{x_{0}}\left[S_{i}\left(p,x_{0}\right)\Gamma\left(\gamma_{5}S_{i}\left(p,x_{0}\right)^{\dagger}\gamma_{5}\right)\right]\right\}. \label{eq:GreenBi_p}
\end{equation}
We then amputate this Green function using the averaged propagators,
\begin{equation}
\Pi_{\Gamma,0}\left(p\right)=S^{-1}\left(p\right)G_{\Gamma,0}\left(p\right) \left(\gamma_{5}\left[S^{-1}\left(p\right)\right]^{\dagger}\gamma_{5}\right)\,,\label{eq:AmpGreenBi_unren}
\end{equation}
where $S^{-1}\left(p\right)$ is calculated according to Eq.~\eqref{eq:S avg}. The bare vertex amplitudes are obtained from the amputated Green functions as follows~\cite{Martinelli:1994ty,Blum:2001sr}: \begin{align}
\Lambda_{S}\left(p\right) & =\frac{1}{12}\mathrm{Tr}\left[\Pi_1\left(p\right)1\right]\label{eq:LmdS}\\
\Lambda_{P}\left(p\right) & =\frac{1}{12}\mathrm{Tr}\left[\Pi_{\gamma_5}\left(p\right)\gamma_{5}\right]\label{eq:LmdP}\\
\Lambda_{V}\left(p\right) & =\frac{1}{48}\mathrm{Tr}\left[\sum_{\mu}\Pi_{\gamma_\mu}\left(p\right)\gamma_{\mu}\right]\label{eq:LmdV}\\
\Lambda_{A}\left(p\right) & =\frac{1}{48}\mathrm{Tr}\left[\sum_{\mu}\Pi_{\gamma_\mu \gamma_5}\left(p\right)\gamma_{5}\gamma_{\mu}\right]\label{eq:LmdA}\\
\Lambda_{T}\left(p\right) & =\frac{1}{72}\mathrm{Tr}\left[\sum_{\mu,\nu}\Pi_{\sigma_{\mu\nu}}\left(p\right)\sigma_{\nu\mu}\right].\label{eq:LmdT}\end{align}
The values of all the five bare vertex amplitudes at the unitary mass points $m_l=m_\mathrm{val}$ are presented in 
Table~\ref{tab:Lmd-all-0.01} through Table~\ref{tab:Lmd-all-0.03}.
Finally, by requiring that the renormalized vertex amplitudes satisfy
\begin{equation}
\Lambda_{i,\mathrm{ren}}=\frac{Z_{i}}{Z_{q}}\Lambda_{i}
  =1\,,\qquad i\in\left\{ S,P,V,A,T\right\} ,
\label{eq:Lmd_ren}
\end{equation}
we can calculate the relevant renormalization constants.

Equations~\eqref{eq:Z_bi} through Eq.~\eqref{eq:Lmd_ren} describe the
schematic procedure used to calculate the renormalization coefficients
of quark bilinears. In practice however, with finite quark masses and
a limited range of momenta, we have to consider lattice artefacts
and other systematic uncertainties. We explain the details in the
following sections.

\subsection{Chiral symmetry breaking and $Z_A - Z_V$}
\label{sec:ch_sym_breaking}

In this section we examine the effects of both the low-energy spontaneous
chiral symmetry breaking present in QCD and our non-zero quark masses
on the large-momentum, off-shell propagators which we are using to
impose non-perturbative renormalization conditions.  A good quantity to
study in order to understand these effects is the difference of the off-shell
vector and axial vector vertex functions.

\subsubsection{Numerical results for $Z_A - Z_V$}

In the limit of a small mass and a large momentum, we expect
\begin{equation}
Z_{A}=Z_{V},\label{eq:Za=Zv}
\end{equation}
or equivalently,
\begin{equation}
\Lambda_{A}(p^2)=\Lambda_{V}(p^2)\label{eq:LmdA=LmdV}
\end{equation}
for $p^2 \gg \Lambda_{\rm QCD}^2,\;m^2$.

However, with finite quark masses and at relatively low momenta
$\Lambda_{V}$ and $\Lambda_{A}$ may receive different contributions
of the form
\begin{equation}
\frac{m_\mathrm{val}^{2}}{p^{2}}
\label{eq:m2term}
\end{equation}
and
\begin{equation}
\frac{m_\mathrm{val}\left\langle \overline{q}q\right\rangle }{p^{4}}.
\label{eq:m-qbq-term}
\end{equation}
Here we are exploiting the $SU_L(3) \times SU_R(3)$ chiral
symmetry of large $L_s$ domain wall fermions which implies that
a difference between $\Lambda_{V}$ and $\Lambda_{A}$ requires the
mixing of $(8,1)$ and $(1,8)$ representations and hence involves
a product of two quantities which transform as $(3,\overline{3})$
and $(\overline{3},3)$.  This requires the two powers of
$m_\mathrm{val}$ in Eq.~\eqref{eq:m2term} and the product
$m_\mathrm{val}\left\langle \overline{q}q\right\rangle$ in
Eq.~\eqref{eq:m-qbq-term}.  The extra factors of $1/p^2$ and $1/p^4$
come from naive dimensional analysis.

\begin{figure}[H]
\begin{centering}
\includegraphics[clip]{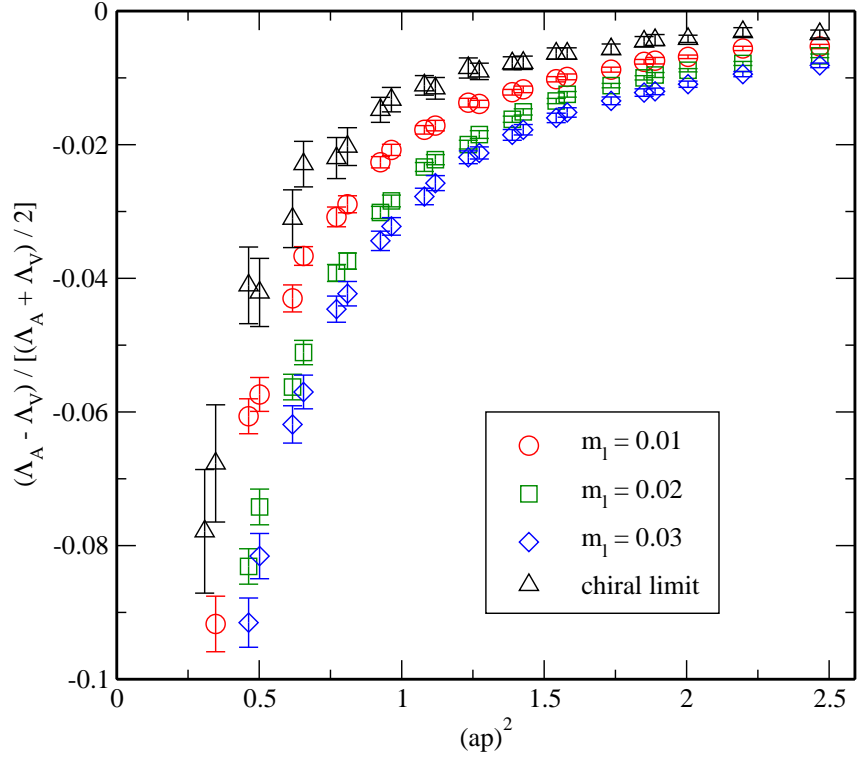}
\par\end{centering}

\caption{The ratio $\frac{\Lambda_{A}-\Lambda_{V}}{\left(\Lambda_{A}+\Lambda_{V}\right)/2}$
plotted as a function of momentum at the unitary mass points $m_\mathrm{val}=m_l$
and in the chiral limit evaluated by linear extrapolation in $m_l$.  The 5-10\%
difference at low momentum decreases rapidly as the momentum increases.  At the scale
$\mu\simeq 2\mbox{ GeV}$, or $\left(ap\right)^{2}\simeq 1.4$, the difference is about
1\%, which contributes to the systematic error in $Z_{B_{K}}$. \label{fig:AVdiff}}
\end{figure}

To determine how much chiral symmetry breaking is present in our
calculation, we examine the relative difference between
$\Lambda_{A}$ and $\Lambda_{V}$.  In Figure~\ref{fig:AVdiff} we
plot the quantity $\frac{\Lambda_{A}-\Lambda_{V}}
{\left(\Lambda_{A}+\Lambda_{V}\right)/2}$ as a function of momentum.
At relative low momenta, $0.5 \le (pa)^2 \le 1$, we observe that this
quantity is quite large $(\sim 5\%)$.  Furthermore, even when we
extrapolate $\frac{\Lambda_{A}-\Lambda_{V}}
{\left(\Lambda_{A}+\Lambda_{V}\right)/2}$ to the chiral
limit, where the terms in Eqs.~\eqref{eq:m2term} and \eqref{eq:m-qbq-term}
both vanish, the difference between $\Lambda_A$ and $\Lambda_V$ does
not vanish. Here to obtain the chiral limit shown in Figure~\ref{fig:AVdiff}
we perform a linear extrapolation
$m_\mathrm{val}+ m_{\rm res} \rightarrow 0$.   While a quadratic
extrapolation gives a similar result, this linear choice is motivated
by the analysis presented in Section~\ref{sec:lin_extrap}.

\begin{figure}[H]
\begin{centering}
\includegraphics[clip,scale=0.5]{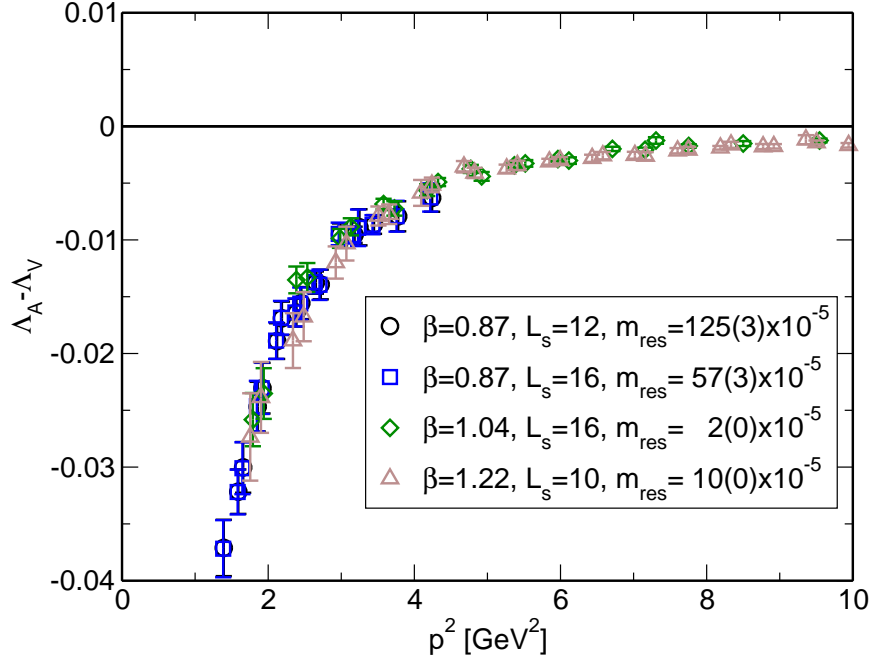}
\par\end{centering}

\caption{The difference $\Lambda_{A}-\Lambda_{V}$ computed using four
different quenched DBW2 lattice ensembles.  These ensembles have quite
different lattice scales.  In addition the values of $L_s$, the extent
in the 5th dimension used in computing the DWF propagators, also varies
significantly.  This provides compelling evidence that the observed
chiral symmetry breaking is not an explicit breaking from finite
$L_{s}$, but rather represents the high energy tail of QCD dynamical
chiral symmetry breaking which would vanish if we were able to perform
the NPR calculation at high enough energy.  The data shown come from
Refs.~\cite{Dawson:2002nr,Aoki:2005ga,Aoki:2006ib}
\label{fig:AVdiff-comp-DBW2} }
\end{figure}

Since the explicit chiral symmetry breaking effects needed to
split $\Lambda_{A}$ and $\Lambda_{V}$ can be argued~\cite{Christ:2005xh} to
be $O(m_{\rm res}^2)$, we would not expect this difference to reflect
explicit, finite-$L_s$, domain wall chiral symmetry breaking.  In fact,
similar deviations between \emph{$\Lambda_{V}$} and \emph{$\Lambda_{A}$}
are seen on lattice ensembles without fermion loops where explicit domain
wall chiral symmetry breaking is expected to be smaller.  This is shown
in Figure~\ref{fig:AVdiff-comp-DBW2} where we plot the same quantities
from a quenched simulation using the DBW2 gauge action.  Thus, it appears
that this difference represents the high energy tail of QCD dynamical
chiral symmetry breaking rather than the explicit chiral symmetry
breaking coming from the finite value of $L_{s}$.

\begin{figure}[H]
\begin{centering}
\includegraphics[clip]{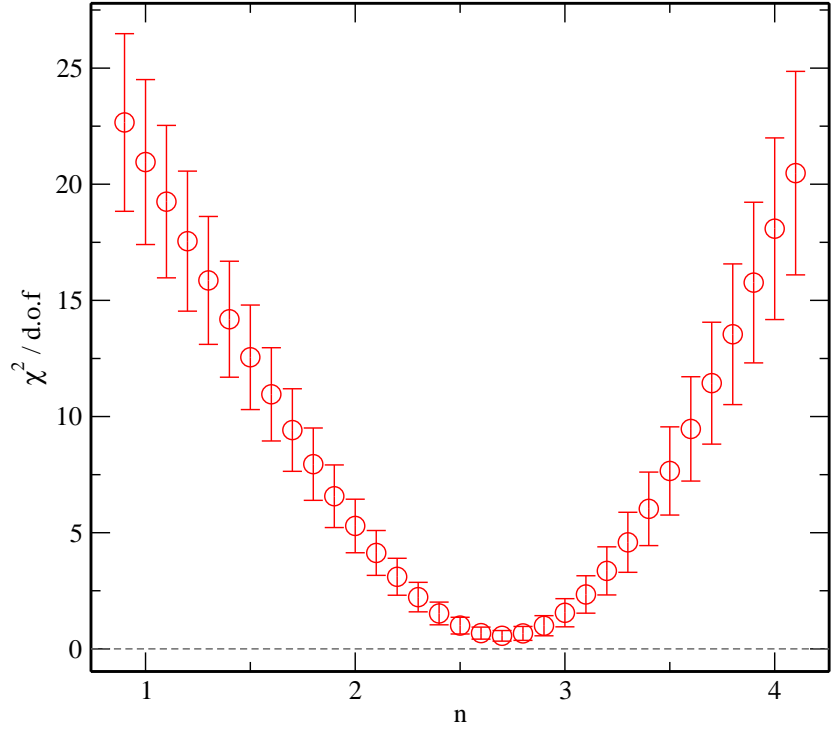}
\par\end{centering}

\caption{The $\chi^{2}/d.o.f$ which results from fitting the momentum
dependence of the quantity
$\frac{\Lambda_{A}-\Lambda_{V}}{\left(\Lambda_{A}+\Lambda_{V}\right)/2}$
(extrapolated to the chiral limit) to the form $p^{-n}$.   We conclude
that the best choice for $n$ lies between 2 and 3 and that it is unlikely
that the term $\left\langle \bar{q}q\right\rangle ^{2}/p^{6}$ gives the
dominant contribution to this chiral symmetry breaking.
\label{fig:AVdiff-pfit-x2-chlim-quad}}
\end{figure}

While the effects of spontaneous chiral symmetry breaking will not vanish
in the limit $m_\mathrm{val} + m_\mathrm{res} \rightarrow 0$, it is
unlikely that the substantial difference found for $\Lambda_A - \Lambda_V$
in the chiral limit can be explained by a dimension-6 condensate such as
\begin{equation}
\frac{\left\langle \bar{q}q\right\rangle ^{2}}{p^{6}}
\label{eq:qbq2-term}
\end{equation}
since it is suppressed by six powers of momentum and appears to be
too small for the size of the breaking we have observed.  We have also
fit the quantity $\frac{\Lambda_{A}-\Lambda_{V}}
{\left(\Lambda_{A}+\Lambda_{V}\right)/2}$ to different powers of $p$,
as is shown in Figure~\ref{fig:AVdiff-pfit-x2-chlim-quad}, and it is
clear that the momentum dependence of the chiral symmetry breaking term
is dominated by $p^{-2}$ or $p^{-3}$, very different from $p^{-6}$ that
naive dimensional analysis suggests should appear in the
$\left\langle \bar{q}q\right\rangle ^{2}$ term above.

\subsubsection{Effects of exceptional momenta}

In fact, we believe that the origin of the difference between $\Lambda_A$
and $\Lambda_V$ is different. Our choice of kinematics corresponds to
so called ``exceptional momenta'', i.e. a momentum transfer is zero. This
invalidates the naive power counting estimates used above and permits the
low-energy, spontaneous chiral symmetry breaking to split
\emph{$\Lambda_{V}$} and \emph{$\Lambda_{A}$} with only a $1/p^2$
suppression for large $p$, as we now explain.  Begin by considering a
general, amputated Feynman graph $\Gamma$ with $F$ external fermion lines
and $B$ external boson lines.  Recall that for connected graphs the
degree of divergence $d$ of $\Gamma$ is defined as $d=4-3F/2-B$.  If the
graph $\Gamma$ is disconnected then its degree of divergence is the sum
of those of its connected components.  Now imagine that each external line
of $\Gamma$ carries an incoming momentum $\lambda\, p_i$ for $1 \le i \le F+B$,
where $\lambda$ is an over-all scale factor.  The asymptotic behavior
for large $\lambda$ of the amplitude corresponding to such a graph will
be $\lambda^{d^\prime}$ where $d^\prime$ is the degree of divergence of
a subgraph $\Gamma^\prime \subseteq \Gamma$.  This subgraph
$\Gamma^\prime$ must be chosen so that i) there exists a routing of
the internal momenta within $\Gamma$ such that all lines carrying
momenta proportional to $\lambda$ lie within $\Gamma^\prime$ and
ii) $\Gamma^\prime$ possesses the least negative degree of divergence
$d^\prime$ of all those subgraphs satisfying i) \cite{Weinberg:1959nj,
Itzykson:1980rh}.  Note, that $\Gamma^\prime$ may equal the original
graph $\Gamma$ and may itself be disconnected.

The most familiar situation is the case of non-exceptional momenta,
defined as a momentum configuration in which no proper partial sum
of the external momenta $p_i$ vanishes.  Under these circumstances
all subgraphs $\Gamma^\prime$ obeying i) must be connected.  (Otherwise
there would be zero momentum transfer between the groups of momenta
entering each of the disconnected components.)  This implies that the
subgraph $\Gamma^\prime$ with the least negative degree of divergence
is one with no additional external lines beyond those already appearing
in $\Gamma$ which in turn implies that this subgraph $\Gamma^\prime$
is the entire graph $\Gamma$.  For the case of the vertex graph of
interest, we deduce a constant behavior (up to logarithms) since
$d=4-1-2\cdot 3/2=0$.  (Here it is convenient to view this vertex graph
as resulting from a normal Feynman graph in which an external vector
boson is coupled to the vertex so the rules discussed above directly
apply.)

This analysis not only gives the leading asymptotic behavior but
also insures that extracting a few extra factors of the mass $m$ or
the chiral condensate $\left\langle \bar{q}q\right\rangle$ will make the
degree of divergence of that graph more negative and hence make its
asymptotic fall-off more rapid, in the fashion suggested by naive power
counting.  For the case of interest, we would like to restrict a subset
of the internal fermion lines of our graph $\Gamma$ to carry only low
momenta so that they will reflect the low-energy, spontaneous chiral
symmetry breaking of QCD.  By definition, these low momentum lines cannot
enter the subgraph $\Gamma^\prime$ discussed above whose degree of divergence
determines the asymptotic behavior of amplitude being studied.  In order
to split \emph{$\Lambda_{V}$} and \emph{$\Lambda_{A}$}, chiral symmetry
breaking transforming as an (8,8) under $SU(3)\times SU(3)$ is required.
This in turn requires that this low energy, excluded subgraph must be joined
to the remainder of the graph by at least four fermion lines.

\begin{figure}[H]
\begin{centering}
\includegraphics[scale = 0.75]{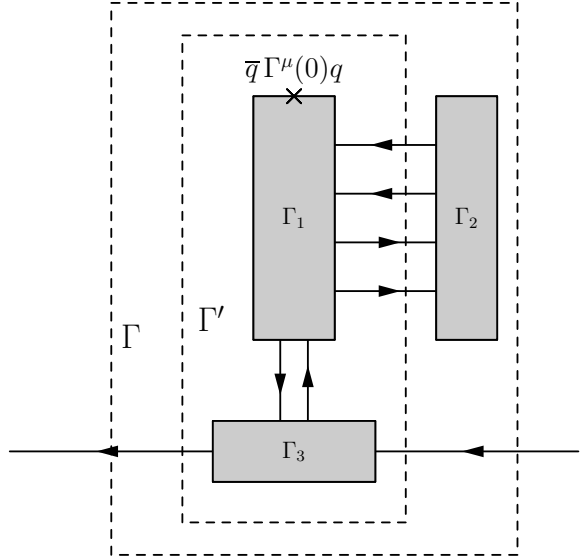}
\par\end{centering}

\caption{The division of a general vertex graph into subgraphs.
If the four-legged, internal subgraph $\Gamma_2$ carries momenta
$p \sim \Lambda_\mathrm{QCD}$ it can introduce low energy, $(8,8)$
chiral symmetry breaking into such an amplitude even in the limit
that the momenta external to the entire diagram $\Gamma$, included
in the outer dashed box, grow large.  As discussed in the text, such
a limit will be suppressed by $1/p^6$ if the external momenta are
non-exceptional but by only $1/p^2$ for the exceptional case.
\label{fig:gen_except}}
\end{figure}

Such a circumstance is illustrated by the general vertex graph $\Gamma$ in
Figure~\ref{fig:gen_except}, contained in the outer dashed box.   Here we have
identified a subgraph $\Gamma_2$ which carries only low momenta and can
therefore transform as (8,8) even in the limit of vanishing quark mass,
$m_\mathrm{val}+m_\mathrm{res}=0$.  For the case of non-exceptional momenta,
we must apply Weinberg's theorem to the subgraph $\Gamma^\prime$, enclosed
in the inner dashed box, through which, by assumption, all of the large
momenta entering the vertex and the two external fermion lines must be routed.
Because of its connections to the subgraph $\Gamma_2$, the subgraph
$\Gamma^\prime$ has six external fermion lines and one external boson line
(connected to the vertex).  The resulting degree of divergence is
$d^\prime = 4 - 1 - 6\cdot 3/2 = -6$, justifying the naive $1/p^6$ behavior
in Eq.~\ref{eq:qbq2-term}.

However, in our case \emph{$\Lambda_{V}$} and \emph{$\Lambda_{A}$}
are being evaluated with zero momentum entering the current vertex
and with a vanishing sum of the two incoming fermion momenta---a
configuration of exceptional momentum.    For such a choice of external
momenta we can divide the subgraph $\Gamma^\prime$ identified above
into two pieces $\Gamma_1$ and $\Gamma_3$.  Because the momenta are
exceptional with no large momenta entering the vertex, we can route
all of the large momenta through $\Gamma_3$.  Since $\Gamma_3$ has
only four external fermion lines, its degree of divergence is
$d_3=4-4\cdot 3/2 = -2$ and the $1/p^6$ behavior above has been replaced
by the much less suppressed $1/p^2$.  If we think of the subgraph
$\Gamma_2$ as a generalized chiral condensate
$\langle 0|\overline{q}q\overline{q}q|0\rangle$ we are seeing the
asymptotic behavior
\begin{equation}
\frac{\langle 0|\overline{q}q\overline{q}q|0\rangle}{p^{2}},
\label{eq:qbq2-term-except}
\end{equation}
very consistent with our numerical results.  Note the discrepancy
in dimensions between Eqs.~\eqref{eq:qbq2-term} and
\eqref{eq:qbq2-term-except} will be made up by four powers of
$\Lambda_\mathrm{QCD}$, the momentum scale to which the subgraph
$\Gamma_2$ is restricted.

\begin{figure}[H]
\begin{centering}
\includegraphics{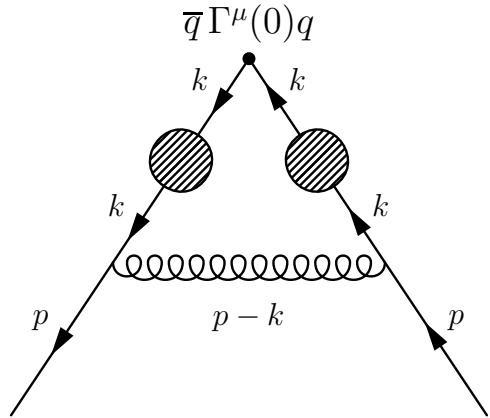}
\par\end{centering}

\caption{Sample diagram in which two low-momentum ($k\simeq\Lambda_{\mathrm{QCD}}$)
fermion propagators appear in a graph which is suppressed at high
momentum only by a single factor of $1/p^{2}$.\label{fig:vertex}}
\end{figure}

A simple class of graph allowing this behavior can be seen in
Figure~\ref{fig:vertex}.  Here the large momentum carried by the
two external fermion lines can be routed through the gluon propagator
that is shown explicitly so that the upper part of the diagram carries
only low momenta.  The large momentum behavior of the gluon propagator
gives the expected $1/p^2$ behavior.  The two general fermion
propagators shown with the shaded ``blobs'' carry small momenta and,
as suggested by Eq.~\eqref{eq:TrSinv}, can show $(3, \overline{3})$ or
$(\overline{3},3)$ chiral symmetry violation even when
$m_\mathrm{val}+m_\mathrm{res}=0$.

\begin{figure}[H]
\begin{centering}
\includegraphics{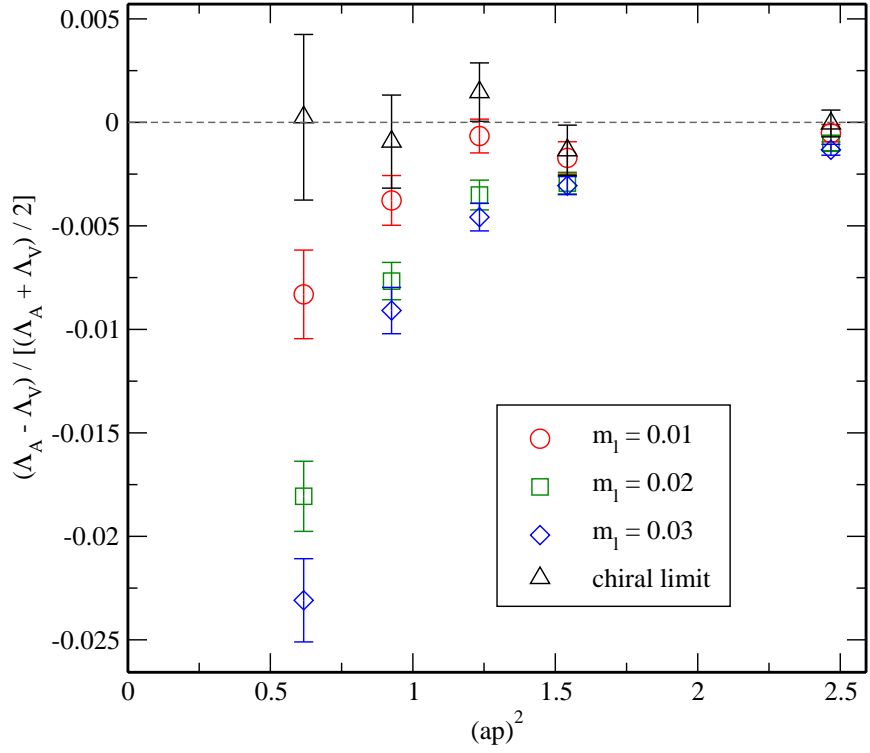}
\par\end{centering}

\caption{The value of $\frac{\Lambda_{A}-\Lambda_{V}}{\left(\Lambda_{A}+\Lambda_{V}\right)/2}$
calculated with non-exceptional kinematics, which requires the sum
of any subset of external momenta be non-zero. With this condition
the chiral symmetry breaking is highly suppressed (as compared to
Fig.~\ref{fig:AVdiff}) and vanishes almost completely over the available
momentum region. \label{fig:AVdiff-nonexp-mom}}
\end{figure}

To confirm this analysis, we have also calculated the difference
between $\Lambda_{A}$ and $\Lambda_{V}$ with non-exceptional
momenta.  We have chosen 5 different momentum scales, each
corresponding to a set of momenta which satisfy the condition
$p_{1}^{2}=p_{2}^{2}=\left(p_{1}-p_{2}\right)^{2}=p^2$ for five
values of $p^2$, as listed in Table~\ref{tab:nonexp-mom1} and Table
\ref{tab:nonexp-mom2}.  We then calculated $\Lambda_{A}$ and $\Lambda_{V}$
with the two external fermions carrying respectively $p_{1}$ and
$p_{2}$. The result is plotted in Figure~\ref{fig:AVdiff-nonexp-mom},
which shows that the chiral symmetry breaking vanishes almost completely
with non-exceptional kinematics at medium to large momenta.

While it would be more satisfactory to perform the calculations
presented in this paper using non-exceptional momenta, the resulting
RI/MOM normalization conditions would not correspond to those for which
perturbative matching calculations have been carried out.  Thus, we
would not be able to relate the quantities which we calculated to
those defined in the $\overline{{\rm MS}}$ scheme.  (Of course, this
difficulty will be removed when the necessary perturbative
calculations have been extended to non-exceptional kinematics.)
A second, less significant advantage of the exceptional momenta
which we use is that the exceptional momentum conditions are
satisfied by a much larger set of discrete lattice momenta
permitting the RI/MOM condition to be satisfied for more fine-grained
sequence of energy scales.

\begin{figure}[H]
\begin{centering}
\includegraphics[clip]{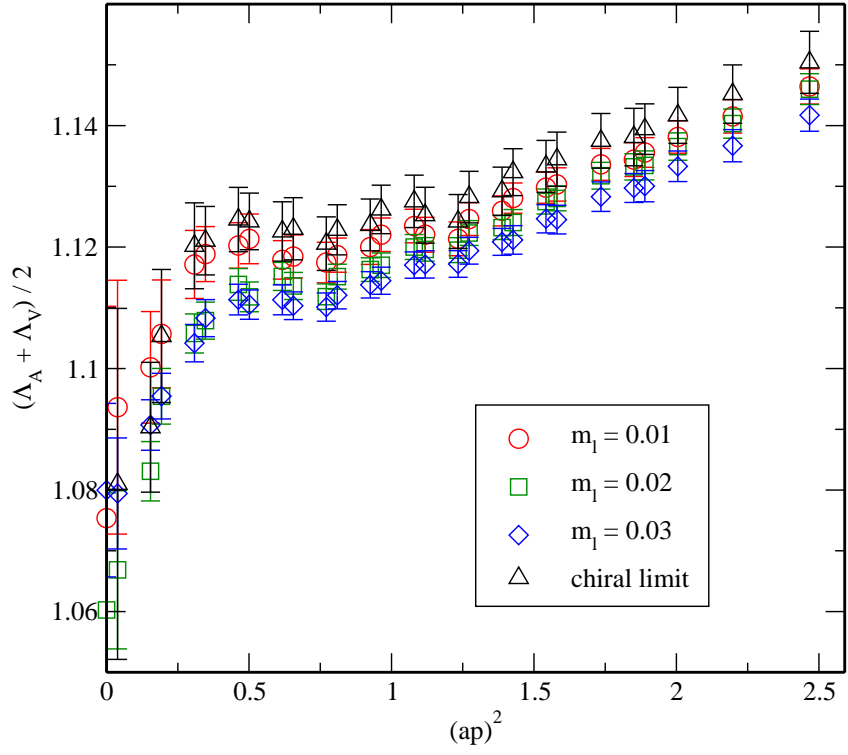}
\par\end{centering}

\caption{The average $\frac{1}{2}\left(\Lambda_{A}+\Lambda_{V}\right)$
plotted as a function of momentum and evaluated for a unitary choice of
masses and in the chiral limit.  The chiral limit is taken using a linear
fit. \label{fig:Lmd_AV}}
\end{figure}

We now return to the calculation with exceptional momenta ($p_{1}=p_{2}$),
at the scale which we are most interested in, that is $\mu\simeq2\mbox{ GeV}$
or $\left(ap\right)^{2}\simeq1.3$, where $\Lambda_{A}$ and $\Lambda_{V}$
have a difference of about 1\%. Since we have no means to determine
which of these two quantities has less contamination from low energy
chiral symmetry breaking we have decided to take the average
$\frac{1}{2}\left(\Lambda_{A}+\Lambda_{V}\right)$ as the central value
for both $Z_{q}/Z_{A}$ and $Z_{q}/Z_{V}$. The difference between
$\Lambda_{A}$ or $\Lambda_{V}$ and $\frac{1}{2}\left(\Lambda_{A}+\Lambda_{V}\right)$
then provides an estimate for one systematic error in our final
results.  The value of $\frac{1}{2}\left(\Lambda_{A}+\Lambda_{V}\right)$
is plotted in Figure~\ref{fig:Lmd_AV}.

\subsubsection{Chiral extrapolation to vanishing quark mass}
\label{sec:lin_extrap}

As discussed above, our use of exceptional momenta implies a $1/p^2$
suppression for both terms behaving as $m_\mathrm{val}$ and
$m_\mathrm{val}^2$.  The added dimension of a $m_\mathrm{val}^2$
term can be provided by a factor of $1/\Lambda_\mathrm{QCD}$
without the need to introduce additional inverse powers of $p$.
For our largest value of $m_\mathrm{val}a = 0.03$, we might
estimate $m_\mathrm{val}/\Lambda_\mathrm{QCD} \approx 0.2$.
This suggests that we should expect a linear rather than quadratic
behavior in $m_\mathrm{val}$ to dominate the small quark mass limit.

\begin{figure}[H]
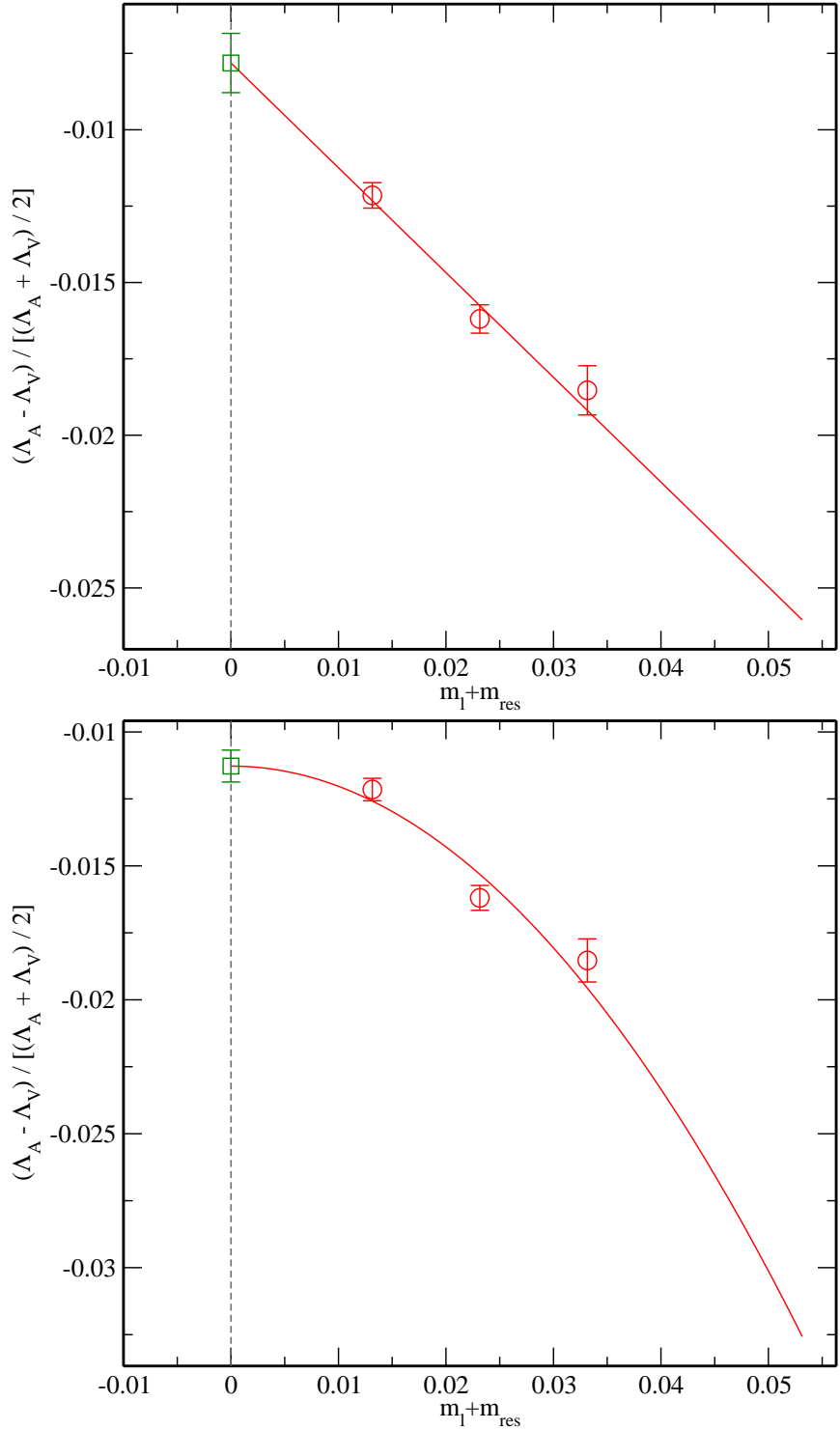

\begin{centering}
\includegraphics[clip]{graphs/AVdiff_dyn_ap2_144.eps}
\includegraphics[clip]{graphs/AVdiff_dyn_m2_ap2_144.eps}
\par\end{centering}

\caption{Comparison of linear (eq.\,(\ref{eq:chiral_fit_linear}) -- top panel) and 
quadratic (eq.\,(\ref{eq:chiral_fit_quad}) -- bottom panel)
fits to the dependence of the chiral symmetry breaking difference
$\frac{\left(\Lambda_{A}-\Lambda_{V}\right)}{\left(\Lambda_{A}+\Lambda_{V}\right)/2}$
on the quark mass $m_\mathrm{val}=m_l$ at the scale $\mu=2.04\text{ GeV}$.
These plots suggest that a
linear description is more accurate.  This conclusion is borne out by
the properties of the actual fits shown in Table~\ref{tab:LA-LV_fits}.
\label{fig:LA-LV_fits}}
\end{figure}

\begin{figure}[H]
\begin{centering}
\includegraphics[clip]{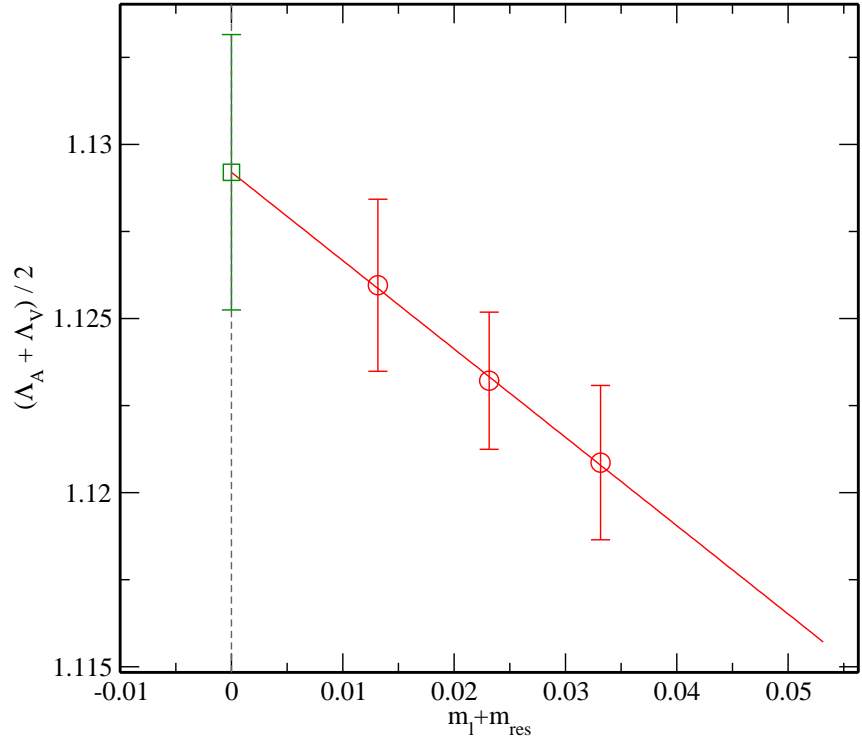}
\par\end{centering}

\caption{A plot showing the linear extrapolation of
$\frac{1}{2}\left(\Lambda_{A}+\Lambda_{V}\right)$ (evaluated at the scale
$\mu=2.04\mbox{ GeV}$) to the chiral limit.  The three data points are evaluated at
the unitary points $m_\mathrm{val}=m_l$. \label{fig:fit_dyn_Zq_Za}}
\end{figure}

The difference $\Lambda_A - \Lambda_V$ discussed above provides a 
good place to study this effect.  This difference reflects the chiral 
symmetry breaking of interest and may make these effects stand out 
with possibly reduced errors because of the statistical correlations 
between the two quantities being subtracted.  Figure~\ref{fig:LA-LV_fits} 
compares linear and quadratic fits to the dependence on the quark 
mass evaluated at unitary points with $m_\mathrm{val}=m_l+m_\mathrm{res}$ 
for $p = 2.04$ GeV.  In Table~\ref{tab:LA-LV_fits} we present the
results of these two fits:
\begin{eqnarray}
\frac{\Lambda_A-\Lambda_V}{(\Lambda_A+\Lambda_V)/2}
                           &=& c_0 + c_1 \frac{m \Lambda_{QCD}}{p^2}
\label{eq:chiral_fit_linear} \\
\frac{\Lambda_A-\Lambda_V}{(\Lambda_A+\Lambda_V)/2}
                           &=& c_0 + c_2 \frac{m^2}{p^2},
\label{eq:chiral_fit_quad}
\end{eqnarray}
for $\Lambda_\mathrm{QCD} = 319.5$ MeV.  As can be seen in the
Table the linear fits are favored.  The linear fits have the smaller
$\chi^2$ and the coefficient $c_1$ is significantly closer to an
expected value of 1 than is the coefficient $c_2$.  Thus, based on
both the theoretical expectation and this empirical evidence, we
will adopt this linear description in the remainder of this paper
and extrapolate our exceptional momentum amplitudes to the chiral
limit using a linear ansatz.  For the case at hand,
Figure~\ref{fig:fit_dyn_Zq_Za} shows this linear extrapolation for
the average  $\frac{1}{2}\left(\Lambda_{A}+\Lambda_{V}\right)$ to
the chiral limit for the momentum $p = 2$ GeV.
Figure~\ref{fig:Lmd_AV} shows the results in the chiral limit as a
function of momentum.  The results in the chiral limit are also presented in
Table~\ref{tab:Lambda_fit}.

\subsection{Axial Ward-Takahashi identity}
\label{sec:axial-ward-id}

Performing an axial rotation on the propagator leads to a relation between
$\Lambda_{P}$ and $\mathrm{Tr}\left(S^{-1}\right)$, the axial Ward-Takahashi
identity\cite{Blum:2001sr}:
\begin{equation}
\Lambda_{P}\left(p\right)=\frac{1}{12}\frac{\mathrm{Tr}\left[S^{-1}\left(p\right)\right]}
                                           {\left(m_{\mathrm{val}}+m_{\mathrm{res}}\right)}.
\label{eq:axial-ward-id}
\end{equation}
In a truly chiral theory or the present DWF calculation in the limit
$L_s \rightarrow \infty$ (when chiral symmetry becomes exact and
$m_\mathrm{res}=0$), this identity will be obeyed configuration
by configuration.  However, for finite $L_s$ and $m_\mathrm{res}\not= 0$,
this relation will hold only after a gauge field average, ({\it e.g.}
$m_{\mathrm{res}}$ is only defined after such an average).  Thus, we should
check Eq.~\eqref{eq:axial-ward-id} on gauge-averaged amplitudes.

\begin{figure}[H]
\begin{centering}
\includegraphics[clip]{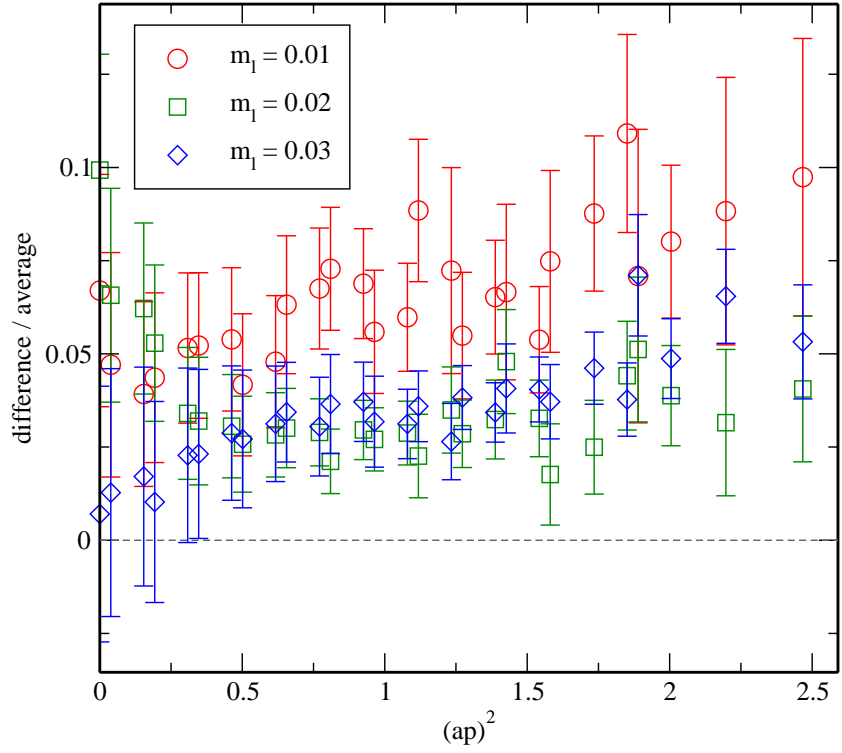}
\par\end{centering}

\caption{The difference between the quantities $\Lambda_{P}$ and
$\frac{1}{12}\frac{\mathrm{Tr}\left(S_{\mathrm{latt}}^{-1}\right)}{m_l+m_{\mathrm{res}}}$,
divided by their average is plotted versus momentum for unitary quark masses.
This provides a test of the axial Ward-Takahashi identity.
\label{fig:check-axial-ward-id}}
\end{figure}

Figure~\ref{fig:check-axial-ward-id} shows the difference between
the l.h.s and r.h.s of Eq.~\eqref{eq:axial-ward-id}, divided by their
average. The case of $m_l=0.01$ shows relatively larger breaking
of $\leq 8$\%, while the other two masses result in a smaller
breaking.  Since for $m=0.01$, the $m_\mathrm{res}$ term, with a value
of 0.00308, represents a 30\% effect, this suggests that the use of
$m_\mathrm{res}$ in the context of Eq.~\eqref{eq:axial-ward-id} may
be accurate at the 25\% level for this lattice spacing.  Note, we
expect violations coming from a dimension-five, anomalous chromo-magnetic
term to be suppressed by a factor of $(pa)^2$ relative to those from
$m_\mathrm{res}$, making this $\leq 8$\% estimate comfortably smaller
than the naive estimate of $(m_\mathrm{res}/0.01)\cdot(p a)^2 \approx 30\%$.
However, the suggested growth in the size of these violations with
increasing $(pa)^2$ may be visible in Figure~\ref{fig:check-axial-ward-id}.

\subsection{Vector Ward-Takahashi identity and the chiral limit of $\Lambda_S$}
\label{sec:vector-ward-id}

Similar to the case with $\Lambda_{P}$, from the continuum vector
Ward-Takahashi identity, we have the relation between $\Lambda_{S}$ and
$\mathrm{Tr}\left(S^{-1}\right)$ \cite{Blum:2001sr}:
\begin{equation}
\Lambda_{S} =
     \frac{1}{12}\frac{\partial\mathrm{Tr}\left[S^{-1}\left(p\right)\right]}
        {\partial m_{\mathrm{val}}}.
\label{eq:vector-ward-id}
\end{equation}
We are able to check our data against this identity using the three
sources, (0,0,0,0), (4,4,4,8) and (12,12,12,24) since it is only
for these three sources that multiple valence mass data are available
for each sea quark mass.

\begin{figure}[H]
\begin{centering}
\includegraphics[clip]{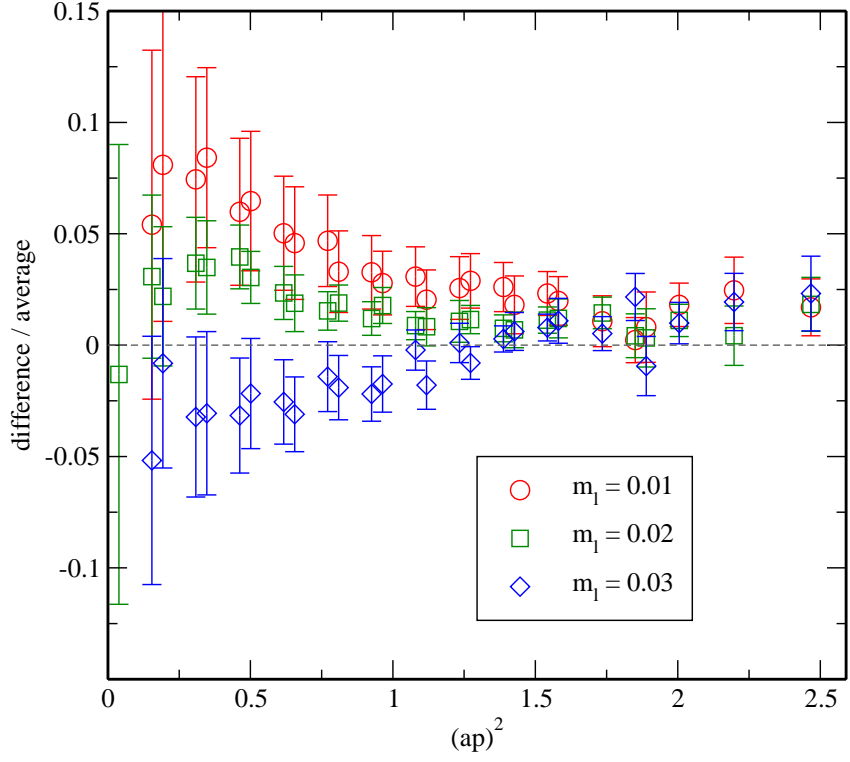}
\par\end{centering}

\caption{The difference between the quantities $\Lambda_{S}$ and
$\frac{1}{12}\frac{\partial\mathrm{Tr}\left[S_{\mathrm{latt}}^{-1}\left(p\right)\right]}
{\partial m_{\mathrm{val}}}$, divided by their average for each sea
quark mass.  The difference appears to zero within errors.  This is
a test of the vector Ward-Takahashi identity.  The plot uses
propagators from three sources.
\label{fig:check-vector-ward-id}}
\end{figure}

Figure~\ref{fig:check-vector-ward-id} shows the difference between
the two sides of Eq.~\eqref{eq:vector-ward-id} divided by their average.
For all three sea quark masses the data agrees well with the vector
Ward-Takahashi Identity (Eq.~\eqref{eq:vector-ward-id}) for medium to large momenta.
Equation~\eqref{eq:vector-ward-id} implies the relation
\begin{equation}
Z_{m}=\frac{1}{Z_{S}},
\label{eq:ZmZs}
\end{equation}
and will use this equation and a calculation of $\Lambda_S$ to
determine the mass renormalization factor in the following
sections.

To extrapolate $\Lambda_{S}$ to the chiral limit, we will improve
upon the discussion in \cite{Blum:2001sr} in two regards.  First,
as explained above, we will exploit the asymptotic properties of
Feynman amplitudes evaluated at exceptional momenta and assume
that the leading mass dependence in the chiral limit will be linear
in $m$.  This is different from the $m^2$ dependence assumed in
Ref.~\cite{Blum:2001sr} where dimensional arguments, appropriate
to the non-exceptional case and leading to the $m^2$ behavior
in Eq.~\ref{eq:m2term} were adopted.

\begin{figure}[H]
\begin{centering}
\includegraphics[clip]{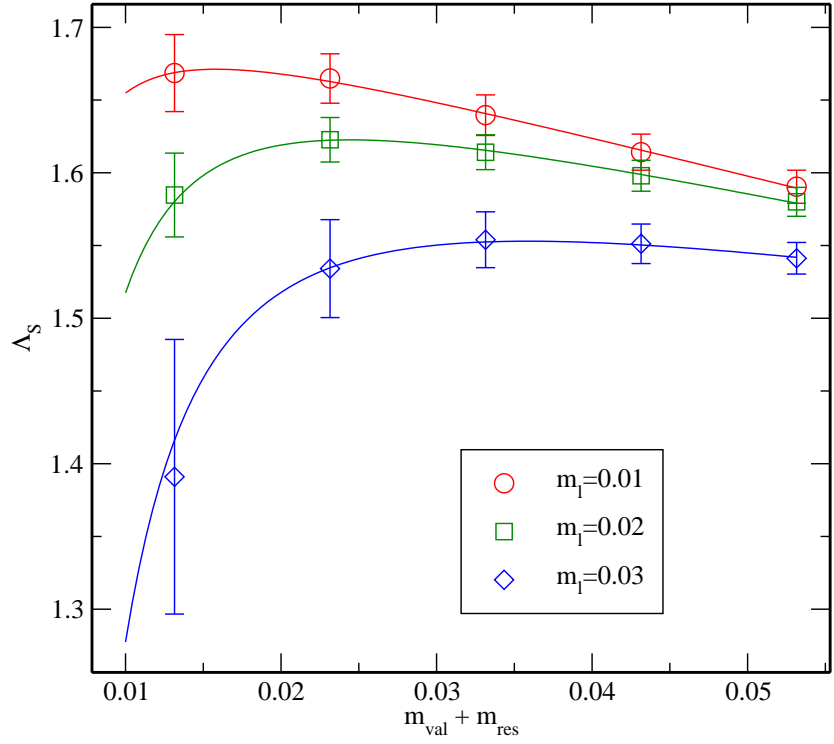}
\par\end{centering}

\caption{The double-pole fit for $\Lambda_{S}$ at $\mu=2.04$ GeV.
The expected decrease in the pronounced $m_\mathrm{val}$ dependence
as the dynamical light quark mass $m_l$ decreases is easily seen.
\label{fig:double_pole}}
\end{figure}

Second, in contrast to that earlier quenched calculation we can
examine the behavior of $Z_S$ as a function of both the valence
and light dynamical quark masses, $m_\mathrm{val}$ and $m_l$
respectively.  In Figure~\ref{fig:double_pole} we plot $\Lambda_S$
as a function of both $m_\mathrm{val}$ and $m_l$.  The three curves
are each a linear plus double pole fit to the valence quark mass
dependence of the form:
\begin{equation}
\Lambda_S(m_\mathrm{val},m_l) = c_0(m_l) + c_1(m_l) m_\mathrm{val}
                           + c_{dp}\frac{m_l^2}{m_\mathrm{val}^2},
\label{eq:double_pole}
\end{equation}
where we have allowed the coefficients $c_0$ and $c_1$ of the constant
and linear terms to vary with the dynamical light quark mass. However,
we have used a common double-pole term with the $m_l^2$ behavior
expected for a theory with two light flavors.  Recall that this double
pole term arises from topological near-zero modes~\cite{Blum:2001sr}
which for two light flavors will be suppressed by two powers of the
light quark mass.  The data in Figure~\ref{fig:double_pole} shows
just this behavior with the sharp turn-over at the smallest value of
$m_\mathrm{val}$ increasing as the light dynamical mass $m_l$ increases.

\begin{figure}[H]
\begin{centering}
\includegraphics[clip]{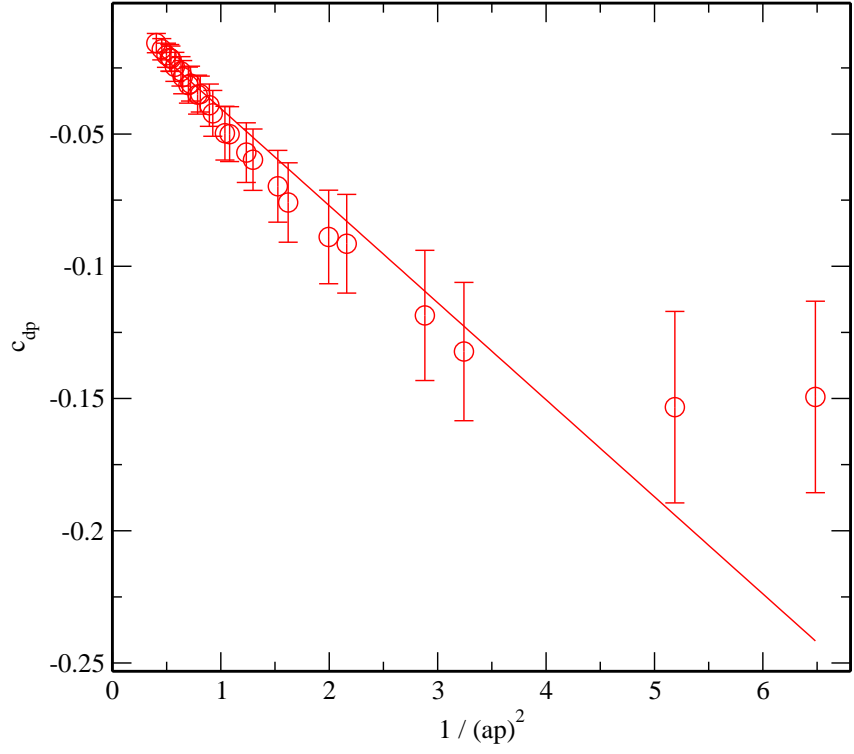}
\par\end{centering}

\caption{Momentum dependence of the double pole coefficient, $c_{dp}$, fit to
the expected $p^{-2}$ behavior.  Good agreement is seen. \label{fig:dp_vs_momentum}}
\end{figure}

This double-pole can be deduced from Eq.~\ref{eq:vector-ward-id}.  As
discussed in Ref.~\cite{Blum:2001sr}, the NLO, $1/p^2$ term derived from
an operator product expansion of the quark propagator on the right-hand
side of this equation is proportional to the chiral condensate
$\langle \overline{q} q\rangle$ \cite{Politzer:1976tv,Becirevic:1999kb}.
Isolated, topological, near-zero modes of the sort that arise from
a gauge field background with non-zero Pontryagin index will contribute
a term to the chiral condensate which behaves as $1/m_\mathrm{val}$.
This implies that the derivative in Eq.~\ref{eq:vector-ward-id} will
yield the double pole, $1/m_\mathrm{val}^2$ hypothesized in
Eq.~\ref{eq:double_pole}.  Such a near-zero mode will also introduce
a factor of $m_l$ into the fermion determinant of the QCD measure
for each light flavor in the theory, hence the expected factor of
$m_l^2$ in the numerator of Eq.~\ref{eq:double_pole}.  In
Figure~\ref{fig:dp_vs_momentum} we show the variation of the double
pole coefficient with the momentum at which the coefficient of the
double pole was extracted.  Also shown in this figure is a fit to the
expected $1/p^2$ behavior which describes the results very well.

This understanding of the double pole terms suggests that a
good strategy for extracting the chiral limit of $\Lambda_S$
first takes the limit of vanishing $m_l$ to remove this NLO
double pole term and then extrapolates to $m_\mathrm{val}=0$.
In the present case, we perform the simpler linear fit using
the unitary points to obtain $\Lambda_S = Z_q/Z_S$ since we do
not have the complete partially quenched results for each of our
four sources.

\subsection{Mass renormalization and renormalization group running}
\label{sub:Zm_running}

To calculate the mass renormalization constant $Z_{m}$, as defined
in Eq. \eqref{eq:mren}, we can either directly take the derivative
of $\mathrm{Tr}\left[S_{\mathrm{latt}}^{-1}\left(p\right)\right]$,
by following Eq. \eqref{eq:TrSinv},
\begin{equation}
Z_{m}Z_{q} = \frac{1}{12}\frac{\partial\mathrm{Tr}\left[S_{\mathrm{latt}}^{-1}\right]}
         {\partial m_{\mathrm{val}}}
\label{eq:Zm-1}
\end{equation}
or we can use the Ward-Takahashi identity,
\begin{equation}
Z_{m}=\frac{1}{Z_{S}}.
\label{eq:Zm-2}
\end{equation}
With the analysis described in the above sections, we find that the
method with the smallest statistical uncertainty is to use $1/Z_{S}$
as the value of $Z_{m}$. To remove the factor $Z_{q}$ from $Z_{q}/Z_{S}$
(which is equal to $\Lambda_{S}$ and can be calculated as described in
Section~\ref{sub:cal-lambda}), we use the ratio $Z_{q}/Z_{A}$
calculated in Section~\ref{sec:ch_sym_breaking}, as well as the value
$Z_A=0.7161(1)$ obtained in Ref.~\cite{lin:2007pt} using hadronic
matrix elements. We therefore determine $Z_m$ in the RI/MOM scheme
by computing separately the three factors on the right-hand side of
\begin{equation}
Z_{m}^{\mathrm{RI/MOM}}\left(p\right) =
        \left[\frac{Z_{q}}{Z_{S}}\left(p\right)\right]
        \left[\frac{Z_{A}}{Z_{q}}\left(p\right)\right]
        \left(\frac{1}{Z_{A}}\right).
\label{eq:Zm}
\end{equation}
Table~\ref{tab:Zm-comb} contains the values of
$Z_{m}^{\mathrm{RI/MOM}}\left(p\right)$ for a variety of momentum scales.

After obtaining the lattice value of $Z_{m}^{\mathrm{RI/MOM}}$ at different
momenta, we divide it by the predicted renormalization group running
factor to calculate the scale invariant quantity $Z_{m}^{\mathrm{SI}}$.
The four-loop running formula we use is \cite{Chetyrkin:1999pq}:
\begin{equation}
Z_{m}^{\mathrm{SI}} =
  \frac{c\left(\alpha_{s}\left(\mu_{0}\right)/\pi\right)}
        {c\left(\alpha_{s}\left(\mu\right)/\pi\right)}
            Z_{m}^{\mathrm{RI/MOM}}\left(\mu\right)
\label{eq:Zm-SI}
\end{equation}
where $\mu_{0}$ is chosen such that $\left(a\mu_{0}\right)^{2}=2$,
a value that lies within the fitting range used below.   For
completeness we present in the appendices the detailed procedure
for running $\alpha_{s}$ at four-loops (Appendix~\ref{sec:alpha_s})
and the form of running factors (Appendix~\ref{sec:Zm run factor}).

\begin{figure}[H]
\begin{centering}
\includegraphics[clip]{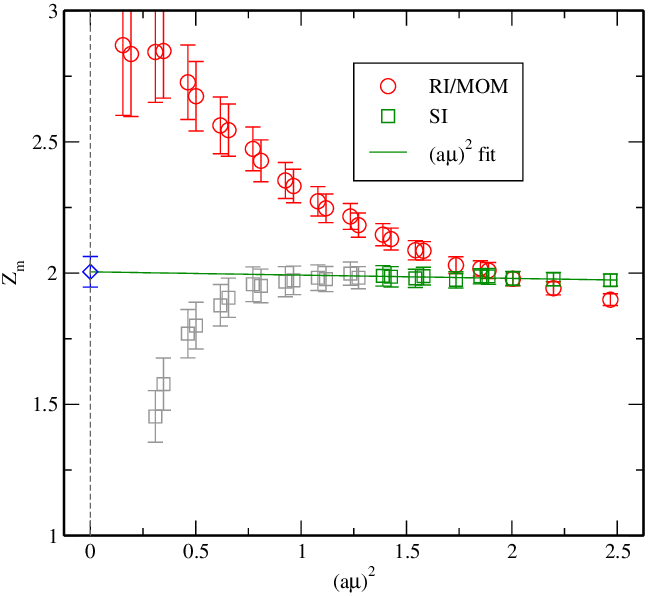}
\par\end{centering}

\caption{The quantities $Z_{m}^{\mathrm{RI/MOM}}\left(\mu\right)$ and
$Z_{m}^{\mathrm{SI}}\left(\mu\right)$ plotted versus the square of the
scale $a \mu$.  Here $Z_{m}^{\mathrm{SI}}\left(\mu\right)$ is obtained
by dividing $Z_{m}^{\mathrm{RI/MOM}}\left(\mu\right)$ by the predicted
perturbative running factor.  Shown also is the linear extrapolation of
$Z_{m}^{\mathrm{SI}}\left(\mu\right)=Z_{m}^{\mathrm{SI}}+ c \left(a\mu\right)^{2}$
using the momentum region $1.3<\left(a\mu\right)^{2}<2.5$ to remove
lattice artifacts.
\label{fig:Zm-RGfit}}
\end{figure}

\begin{figure}[H]
\begin{centering}
\includegraphics[clip]{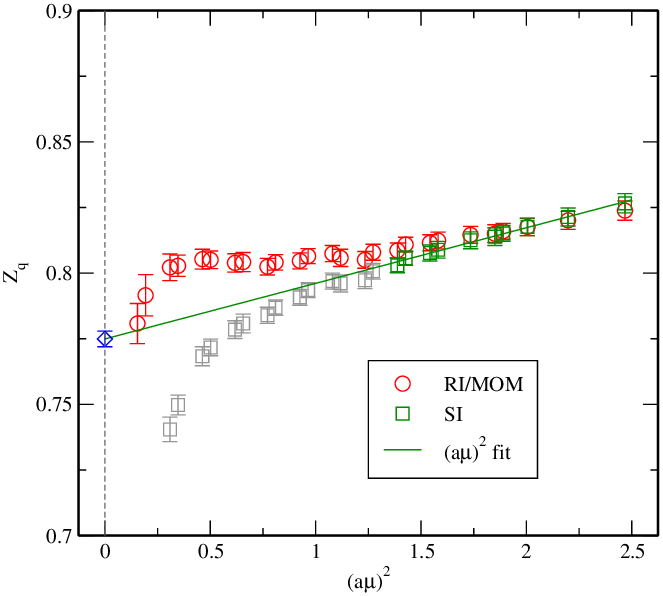}
\par\end{centering}

\caption{The quantities $Z_{q}^{\mathrm{RI/MOM}}\left(\mu\right)$ and
$Z_{q}^{\mathrm{SI}}\left(\mu\right)$ plotted versus the square of the
scale $a \mu$.  Here $Z_{q}^{\mathrm{SI}}\left(\mu\right)$ is obtained
by dividing $Z_{q}^{\mathrm{RI/MOM}}\left(\mu\right)$ by the predicted
perturbative running factor.  Shown also is the linear extrapolation of
$Z_{q}^{\mathrm{SI}}\left(\mu\right)=Z_{q}^{\mathrm{SI}}+ c \left(a\mu\right)^{2}$
using the momentum region $1.3<\left(a\mu\right)^{2}<2.5$ to remove
lattice artifacts.
\label{fig:Zq-RGfit}}
\end{figure}

As Figure~\ref{fig:Zm-RGfit} shows, the quantity $Z_{m}^{\mathrm{SI}}$ 
is remarkably independent of the scale $\mu$.  However, in spite of 
the name, for other cases, the scale-invariant $Z$ factors do show 
noticeable scale dependence and an additional correction is 
warranted.  (See, for example, Figure~\ref{fig:Zq-RGfit}.) We believe 
that the primary reason for this lack of scale invariance is the 
presence of lattice artifacts, namely the finite lattice spacing which 
introduces a small error of $\mathcal{O}\left(\left(a\mu\right)^{2}\right)$.  
Such an error can be reduced by removing the $\mu^{2}$ dependence in
$Z_{m}^{\mathrm{SI}}$.  To do this we fit this momentum dependent
$Z_{m}^{\mathrm{SI}}$ to the form $A+B\left(a\mu\right)^{2}$ over
the momentum range $1.3<\left(a\mu\right)^{2}<2.5$ and then take
the $\left(a\mu\right)^{2}\to 0$ limit of that fit to remove the
$\mu^{2}$ momentum dependence.  We interpret the outcome as the
\emph{true} $Z_{m}^{\mathrm{SI}}$.  Note, we are ignoring
possible $\mu$ dependence arising from the absence of higher order
terms in the matching factor.  Such scale dependence can only be
removed by even higher order computation of the perturbative
matching factor and such a correction is expected to be very small.
While this procedure represents a negligible correction for this 
case of $Z_m^{\mathrm{SI}}$, it will have a more significant effect
in the cases considered below.

\begin{figure}[H]
\begin{centering}
\includegraphics[clip]{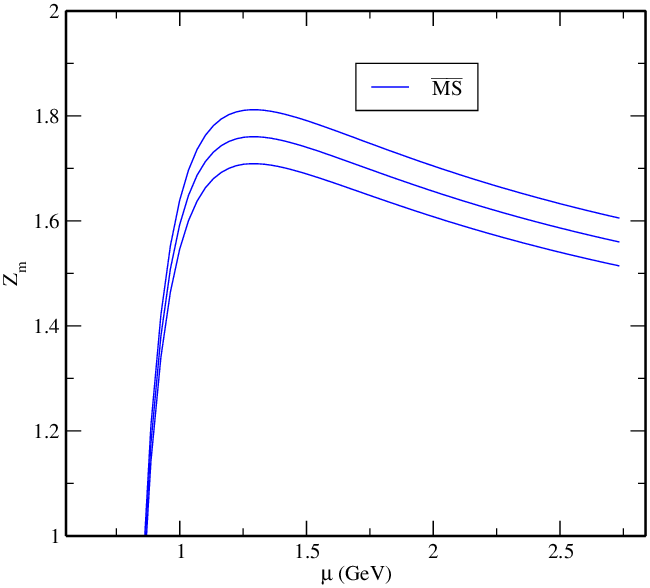}
\par\end{centering}

\caption{The mass renormalization factor $Z_{m}$ expressed in the
$\overline{\mathrm{MS}}$ scheme.  These results are obtained by
applying the perturbative running factor to $Z_{m}^{\mathrm{SI}}$.
The value we are interested in is
$Z_{m}^{\overline{\mathrm{MS}}}\left(\mu=2\mbox{ GeV}\right)$.
The upper and lower curves show the statistical errors.
\label{fig:Zm-RGimp}}
\end{figure}

Our ultimate goal is to determine $Z_{m}^{\overline{\mathrm{MS}}}$
which connects the bare lattice quark mass to its continuum counterpart
defined according to the $\overline{\mathrm{MS}}$ scheme, at the
renormalization scale $\mu=2$ GeV, because the corresponding
continuum renormalization is conventionally done in this scheme.
So we again use Eq. \eqref{eq:Zm-SI} to calculate
$Z_{m}^{\mathrm{RI/MOM}}$(2 GeV) from the scale-independent
value of $Z_{m}^{\mathrm{SI}}$.  Then we multiply it with the
three-loop matching factor, which will also be explained in
Appendix~\ref{sec:Zm run factor}, to match the
$Z_{m}^{\mathrm{RI/MOM}}$(2 GeV) to the $\overline{\mathrm{MS}}$ scheme.
The final step is shown in Figure~\ref{fig:Zm-RGimp} and the results
are given in Table~\ref{tab:Zm-comb}.  The renormalization constant
at the desired scale is
\begin{equation}
Z_{m}^{\overline{\mathrm{MS}}}\left(2\mbox{ GeV}\right)
   = 1.656\pm0.048\left(\mbox{stat}\right)\pm0.150\left(\mbox{sys}\right).
\label{eq:Zm-result}
\end{equation}
The systematic error is determined by adding in quadrature
our estimates of three different types of systematic error which
we will now discuss.

The first is the effect on $Z_m$ of the difference between determining
$Z_{q}/Z_{A}$ from $\frac{1}{2}\left(\Lambda_{A}+\Lambda_{V}\right)$ or
from $\Lambda_{A}$.  This contributes an error of $\pm 0.011$ to
$Z_{m}^{\overline{\mathrm{MS}}}\left(2\mbox{ GeV}\right)$.  Next we
must assign an error to our use of three-loop matching factor, given in
Eq.~\eqref{eq:Zm-match}.  Here we assign an error equal to the
magnitude of the final, order $\alpha_s^3$ error in this perturbative
expression, which is $\pm 0.103$.  While this may be a conservative estimate of the omitted
terms of order $\alpha_s^4$ and higher, it also is intended to include
the errors introduced by the order $\alpha_s^3$ estimate of the perturbative
running determine the intermediate SI step used to remove the $(a\mu)^2$
errors.

Finally we address the errors arising from our failure to extrapolate
to the limit of vanishing strange quark mass.  Recall, we have evaluated
the chiral limit in which both the valence quark mass which enters our
off-shell propagators and the dynamical light quark mass are extrapolated
to zero.  However, all of the gauge ensembles used in this calculation
were computed with a non-zero strange quark mass $m_s = 0.04$.  Since we are
matching our Green functions to those computed in perturbation theory in
the mass-independent, $m\rightarrow 0$, limit our non-zero value for $m_s$
implies an additional systematic error.  Because the dynamical quarks
enter only through loops, their effect is different from that of the
valence quarks discussed above.  They do not contribute chiral symmetry
breaking effects in our matrix elements.  However, because of low energy
chiral symmetry breaking, we do expect the dynamical quark masses to
appear linearly in a quantity such as $Z_m$ in the limit $ m\rightarrow 0$.
To estimate the size of this $O(m_s)$ effect, we begin with the size of
the observed linear dependence, $\partial Z_m/\partial m \approx 5.4$
which comes from both the calculated valence and light dynamical mass
dependence of $\Lambda_S$.  This is then multiplied by 1/2 because there
is only one flavor of strange quark and by $m_s=0.04$ giving an error
in $Z_m$ of $\pm 0.108$.  The total systematic error given in
Eq.~\ref{eq:Zm-result}, $\pm 0.150$, is then the sum of these three
errors in quadrature.

\subsection{Quark wavefunction renormalization and renormalization
group running}
\label{sub:Zq_running}

In Section~\ref{sec:ch_sym_breaking}, we calculated the ratio of
renormalization constants
\begin{equation}
\frac{Z_{q}}{Z_{A}} =
   \frac{1}{2}\left(\Lambda_{A}+\Lambda_{V}\right).
\label{eq:Zq_Za}
\end{equation}
To calculate $Z_{q}$, we multiply this quantity with $Z_{A}=0.7161(1)$
obtained in Ref.~\cite{lin:2007pt}.  Thus, we have evaluated the quantity
$Z_{q}$ in the RI/MOM scheme, which is shown in Table~\ref{tab:Zq-comb}.
To calculate $Z_{q}$ in the $\overline{\mathrm{MS}}$ scheme, we
follow a similar procedure as in the previous section, and start
by dividing $Z_q^{\textrm{RI/MOM}}$ by the perturbative running factor.
As shown in Appendix~\ref{sec:Zq run factor}, the functional form
of the running factor is quite similar to that of $Z_{m}$. The energy
scale $\mu_{0}$ where $Z_{q}^{\mathrm{SI}}$ is fixed to the
$Z_{q}^{\mathrm{RI/MOM}}$ value is again chosen such that
$\left(a\mu_{0}\right)^{2}=2$.

\begin{figure}[H]
\begin{centering}
\includegraphics[clip]{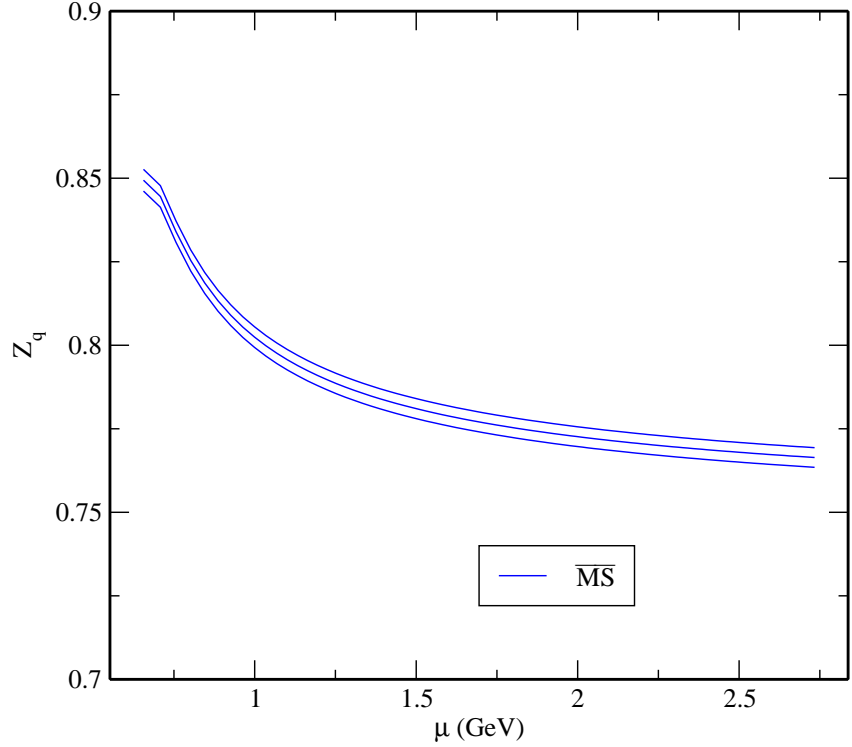}
\par\end{centering}

\caption{The wave function renormalization factor $Z_{q}$ expressed in the
$\overline{\mathrm{MS}}$ scheme.  These results are obtained by
applying the perturbative running factor to $Z_{q}^{\mathrm{SI}}$.
The value we are interested in is
$Z_{q}^{\overline{\mathrm{MS}}}\left(\mu=2\mbox{ GeV}\right)$.
The upper and lower curves show the statistical errors.
\label{fig:Zq-RGimp}}
\end{figure}

The calculated values of $Z_q^{\textrm{SI}}$ vary slightly with
momentum due to the presence of lattice artifacts. To remove these,
we again fit the dependence to the form $A+B\left(a\mu\right)^{2}$
and extrapolate to $a=0$.  The procedure is shown in
Figure~\ref{fig:Zq-RGfit}.  Finally, we take the scale-invariant
$Z_{q}^{\mathrm{SI}}$, run up to different scales in the RI/MOM scheme,
and then apply the perturbative matching factor
(Appendix~\ref{sec:Zq run factor}) to translate it to the
$\overline{\mathrm{MS}}$ scheme.  The $\overline{\mathrm{MS}}$
values are shown in Figure~\ref{fig:Zq-RGimp} and Table~\ref{tab:Zq-comb}.
Of particular interest, the value at $\mu=2\mbox{ GeV}$ is
\begin{equation}
Z_{q}^{\mathrm{\overline{MS}}}\left(2\mbox{ GeV}\right) =
      0.7726\pm0.0030\left(\mbox{stat}\right)\pm 0.0083
       \left(\mbox{sys}\right)
\label{eq:Zq result}
\end{equation}
The systematic errors are estimated using the same procedure
explained in Section~\ref{sub:Zm_running}.  They are the
sum in quadrature of the estimated errors arising from the
difference $\Lambda_A-\Lambda_V$ (0.0061), the use of a
perturbative matching factor accurate to order $\alpha^3$ (0.0045)
and our use of a non-zero sea quark mass (0.0035).

\subsection{Tensor Current Renormalization and Renormalization
Group Running}
\label{sub:Zt_running}

\begin{figure}[H]
\begin{centering}
\includegraphics[clip]{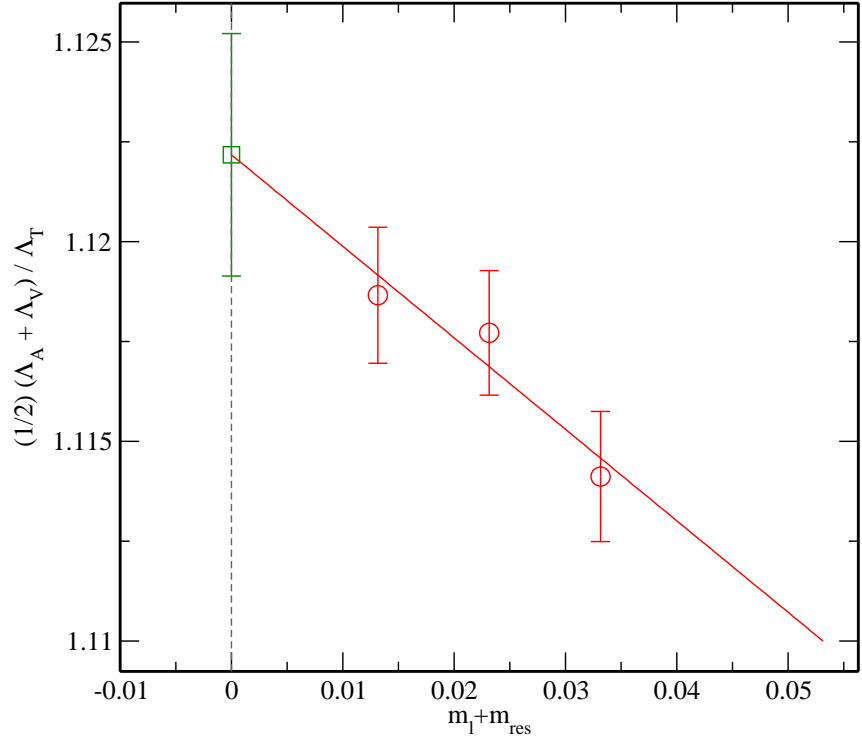}
\par\end{centering}

\caption{A plot of $\frac{1}{2}\left(\Lambda_{A}+\Lambda_{V}\right)/\Lambda_{T}$
as a function of quark mass as well as the linear extrapolation
to the chiral limit, at $(ap)^2=1.388$, or $\mu=2.04\mbox{ GeV}$} \label{fig:Z_T_chiral_limit}
\end{figure}

To calculate the tensor current renormalization constant $Z_T$,
we follow a procedure similar to those of the previous two
sections. For each dynamical quark mass, we combine the ratios
$Z_q/Z_T$ and $Z_q/Z_A$ in order to obtain the ratio of $Z_T$
to $Z_A$ in the RI/MOM scheme:
\[\frac{Z^{\rm RI/MOM}_T}{Z_A}(p)
   = \left[\frac{Z_T}{Z_q}(p)\right]\left[\frac{Z_q}{Z_A}(p)\right].\]
Ultimately, we use the independent hadronic matrix element
calculation of $Z_A$, which gives $Z_A = 0.7161(1)$, to obtain $Z_T$.
Table~\ref{tab:ZT-comb} shows the values obtained for the
$Z^{\rm RI/MOM}_T$ in the chiral limit for a range of lattice momenta.
As discussed above we have performed the chiral extrapolation using a
linear functional form, and Figure~\ref{fig:Z_T_chiral_limit} shows
this linear extrapolation at the lattice momentum $(ap)^2 = 1.388$.

\begin{figure}[H]
\begin{centering}
\includegraphics[clip]{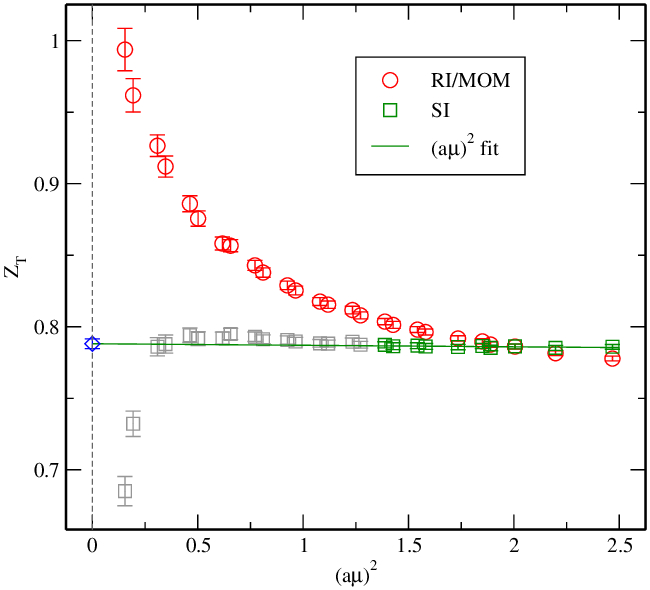}
\par\end{centering}

\caption{The quantities $Z_{T}^{\mathrm{RI/MOM}}\left(\mu\right)$ and
$Z_{T}^{\mathrm{SI}}\left(\mu\right)$ plotted versus the square of the
scale $a \mu$.  Here $Z_{T}^{\mathrm{SI}}\left(\mu\right)$ is obtained
by dividing $Z_{T}^{\mathrm{RI/MOM}}\left(\mu\right)$ by the predicted
perturbative running factor.  Shown also is the linear extrapolation of
$Z_{T}^{\mathrm{SI}}\left(\mu\right) = Z_{T}^{\mathrm{SI}}
+ c \left(a\mu\right)^{2}$ using the momentum region
$1.3<\left(a\mu\right)^{2}<2.5$ to remove lattice artifacts.}
\label{fig:ZT-RGfit}
\end{figure}

\begin{figure}[H]
\begin{centering}
\includegraphics[clip]{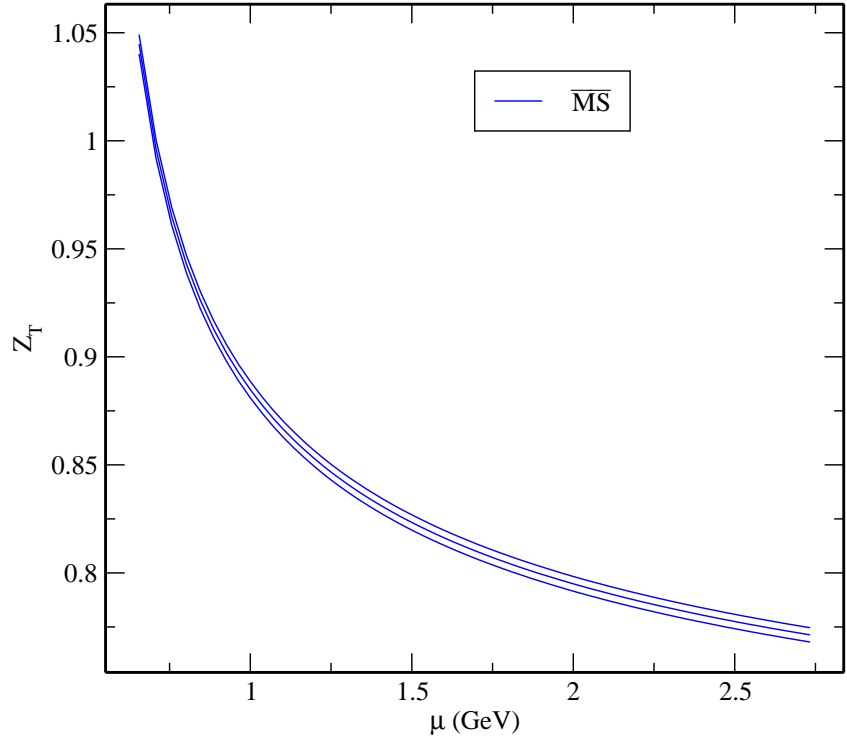}
\par\end{centering}

\caption{The wave function renormalization factor $Z_T$ expressed in the
$\overline{\mathrm{MS}}$ scheme.  These results are obtained by
applying the perturbative running factor to $Z_T^{\mathrm{SI}}$.
The value we are interested in is
$Z_{T}^{\overline{\mathrm{MS}}}\left(\mu=2\mbox{ GeV}\right)$.
The upper and lower curves show the statistical errors.}
\label{fig:ZT-RGimp}
\end{figure}

We obtain SI values for $Z_T$ in the chiral limit by dividing out
the tensor current perturbative running factor, the evaluation of which
is described in Appendix~\ref{sec:ZT run factor}.  Again, the SI
values obtained in this way exhibit a dependence on the lattice
momenta, and again we fit the momentum-dependent $Z^{\rm SI}_T$ to
the form $A + B(a\mu)^2$ and extrapolate to $(a\mu)^2\rightarrow 0$ to
remove the lattice artifacts, as shown in Figure~\ref{fig:ZT-RGfit}.
Finally, we run the scale-invariant $Z_T/Z_A$ back to different scales
in the RI/MOM scheme and use the perturbative matching factor
(Appendix~\ref{sec:ZT run factor}) to match to the $\overline{\rm MS}$
scheme.  The $\overline{\rm MS}$ values are shown in
Figure~\ref{fig:ZT-RGimp} and Table~\ref{tab:ZT-comb}. At $\mu = 2$GeV, we obtain:
\[
Z_T^{\overline{\rm MS}}(\rm 2 GeV) = 0.7950 \pm 0.0034(\rm stat)\pm 0.0150(\rm sys).
\]
The systematic errors are determined in the same fashion as in the
previous two sections.  Specifically the errors arising from the
difference $\Lambda_A-\Lambda_V$ (0.0054), the use of a perturbative
matching factor accurate to order $\alpha$ (0.014) and our use of
a non-zero sea quark mass (0.0003) are added in quadrature.

\section{Renormalization Coefficients for $B_{K}$
\label{sec:Renormalization-Coefficients-for-Bk}}

\subsection{General procedure for computing the mixing coefficients
\label{sub:bk mixing proc}}

By the renormalization of $B_K$ we mean the calculation
of the renormalization coefficient for the operator
\begin{equation}
\mathcal{O}_{VV+AA} = \left(\overline{s}\gamma^{\mu}\left(1-\gamma_{5}\right)d\right)
  \left(\overline{s}\gamma_{\mu}\left(1-\gamma_{5}\right)d\right)
\label{eq:O_VV+AA}
\end{equation}
which is the operator responsible for the mixing between $K^{0}$
and $\overline{K^{0}}$.  Since for finite $L_s$ our theory does not posses
exact chiral symmetry we must consider the possibility that this operator
can mix with the four other $\Delta S=2$ operators with a different
chiral structure:
\begin{align}
\mathcal{O}_{VV-AA} & =\left(\overline{s}\gamma^{\mu}\left(1-\gamma_{5}\right)d\right)
                       \left(\overline{s}\gamma_{\mu}\left(1+\gamma_{5}\right)d\right)
                           \label{eq:O_VV-AA}\\
\mathcal{O}_{SS+PP} & =\left(\overline{s}\left(1-\gamma_{5}\right)d\right)
                       \left(\overline{s}\left(1-\gamma_{5}\right)d\right)
                           \label{eq:O_SS+PP}\\
\mathcal{O}_{SS-PP} & =\left(\overline{s}\left(1-\gamma_{5}\right)d\right)
                       \left(\overline{s}\left(1+\gamma_{5}\right)d\right)
                           \label{eq:O_SS-PP}\\
\mathcal{O}_{TT} & =\left(\overline{s}\sigma^{\mu\nu}d\right)
                    \left(\overline{s}\sigma_{\mu\nu}d\right)
                           \label{eq:O_TT}
\end{align}
where they are labeled by the chirality structure of the even-parity
components.   The odd-parity components of these operators are not
important here since they don't contribute to
$K^{0}\leftrightarrow\overline{K^{0}}$ mixing.

For domain-wall fermions, the mixing of $\mathcal{O}_{VV+AA}$ with these
four operators with wrong chirality should be strongly suppressed
by $\mathcal{O}\left(m_{\mathrm{res}}^{2}\right)$. However, chiral
perturbation theory predicts that the $B$ parameters of the operators
with the wrong chirality diverge in the chiral limit~\cite{Aoki:2005ga,
Becirevic:2004fw}.  To address this issue, we will describe a
theoretical argument to estimate the size of the mixing terms and
an actual calculation of these chirality-violating mixing coefficients
on the 2+1 flavor dynamical lattices.

Following the Rome-Southampton prescription \cite{Martinelli:1994ty,Aoki:2005ga},
we first calculate the $5\times5$ matrix,
\begin{equation}
M_{ij} = \hat{P}_{j}\left[\Gamma_{i}^{\mathrm{latt}}\right]
       = \left(\Gamma_{i}^{\mathrm{latt}}\right)_{\alpha\beta\gamma\delta}^{ABCD}
         \left(\hat{P}_{j}\right)_{\beta\alpha\delta\gamma}^{BADC}
\label{eq:Mixing-matrix}
\end{equation}
where $\Gamma_{i}^{\mathrm{latt}}$ is the amputated, four-point Green
function. The Green functions are first averaged over all sources
and configurations, and then amputated using the averaged propagator,
in a procedure similar to the calculation of two-point amputated Green
functions $\Pi_{\Gamma,0}\left(p\right)$ in Section~\ref{sub:cal-lambda}.
$\hat{P}_{j}$ is a suitable projector, which projects out the component
with the expected chirality (for example, the projector corresponding
to $\mathcal{O}_{VV+AA}$ is $\gamma^{\mu}\otimes\gamma_{\mu}+
\gamma^{\mu}\gamma_{5}\otimes\gamma_{\mu}\gamma_{5}$).  The subscripts
$i,j\in\left\{ VV+AA,VV-AA,SS-PP,SS+PP,TT\right\} $.

It is straight-forward to calculate the mixing matrix at tree level
which we denote as:
\begin{equation}
F_{ij} = \hat{P}_{j}\left[\Gamma_{i}^{\mathrm{tree}}\right].
\label{eq:Mixing-F}
\end{equation}
The RI/MOM renormalization condition which we adopt is then:
\begin{equation}
\frac{1}{Z_{q}^{2}}Z_{ij}M_{jk}=F_{ik}
\label{eq:4q-renorm}
\end{equation}
or
\begin{equation}
\frac{1}{Z_{q}^{2}}Z=FM^{-1}.
\label{eq:Z_4q}
\end{equation}

\subsection{Theoretical argument for the suppression of mixing coefficients}

As can be seen from the structure of the four operators in
Eqs.~\eqref{eq:O_VV-AA}, \eqref{eq:O_SS+PP}, \eqref{eq:O_SS-PP} and
\eqref{eq:O_TT}, if they are to mix with
$\mathcal{O}_{VV+AA}$ defined in Eq.~\eqref{eq:O_VV+AA} then two quark
fields much change chirality from left- to right-handed.  For domain
wall fermions such a mixing can arise from the explicit breaking of
chiral symmetry coming from the finite separation between the left
and right walls.  The asymptotic behavior for large $L_s$ of the
resulting mixing coefficients can be estimated using the transfer
matrix $T$ for propagation in the $s$-direction introduced by Furman
and Shamir~\cite{Furman:1995ky}.  The large-$L_s$ limit is then
controlled by matrix elements of the operator $T^{L_s}$ and is
dominated by those four-dimensional fermion modes corresponding
to eigenvalues of the transfer matrix which lie near unity.

As described in detail in Ref.~\cite{Antonio:2007tr} and in the
original references cited therein, these fermion modes are believed
to fall into two classes: modes localized in space-time with
corresponding $T$ eigenvalues falling arbitrarily close to unity and
de-localized modes characterized by a mobility edge $\lambda_c > 0$
and with eigenvalues of $T$ lying below $e^{-\lambda_c}$
\cite{Golterman:2003qe, Golterman:2004cy,Golterman:2005fe,
Svetitsky:2005qa}.  Since two quarks must change chirality to
produce the required operator mixing, for the case of de-localized
modes such mixing will be suppressed by the two factors of
$e^{-\lambda_cL_s}$ needed for the propagation between the left
and right walls of these two fermions, consistent with our estimate
above that such effects should be of order $m_{\rm res}^2$.

However, the effects of the localized modes are more subtle.
We must address the possibility raised by Golterman and
Shamir~\cite{Golterman:2004mf} that the contribution of such a mode
to $m_{\rm res}$ is suppressed because such modes are relatively
rare and the necessary coincidence with the location of the operators
being mixed is unlikely.  However, if present, such a mode can mix
right- and left-handed fermions with little further suppression since
the corresponding $T$ eigenvalue may be very close to unity.
This raises the possibility that a single such mode, suppressed
by a single factor of $m_{\rm res}$ might be occupied by the
two different quark flavors to provide the double chirality
flip needed to mix the operators.  Fortunately, as argued in
Refs.~\cite{Christ:2005xh} and \cite{Aoki:2005ga}, this is not
possible because the mixing in question requires both a quark
and an anti-quark or two quarks of the same flavor to propagate
across the fifth dimension.  This requires two distinct modes
and hence incurs the double suppression which is well represented
by the $m_{\rm res}^2$ estimate above.  Note,
$m_{\rm res}^2 \approx 10^{-5}$ which will introduce $O(0.1\%)$
errors in current calculations of $B_K$~\cite{Antonio:2007pb}
and will be too small to be seen in non-perturbative studies
presented here.

\subsection{Lattice calculation of mixing coefficients}

\begin{figure}[H]
\begin{centering}
\includegraphics[clip]{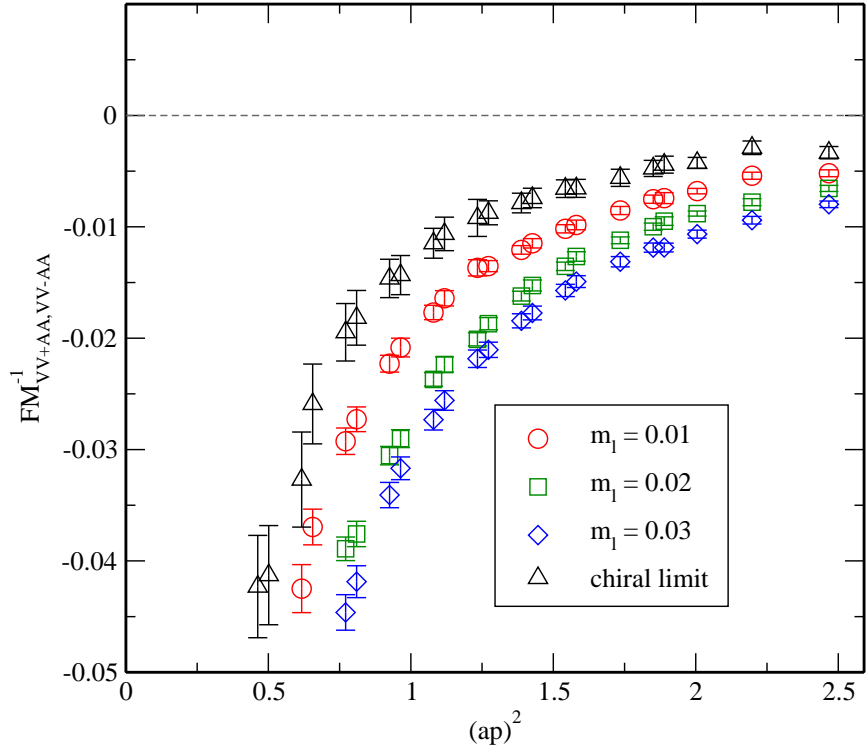}
\par\end{centering}
\caption{The mixing coefficient $FM_{VV+AA,VV-AA}^{-1}$ for our three
unitary mass values and linearly extrapolated to the chiral limit.
\label{fig:FMinv_VV+AA_VV-AA}}
\end{figure}

\begin{figure}[H]
\begin{centering}
\includegraphics[clip]{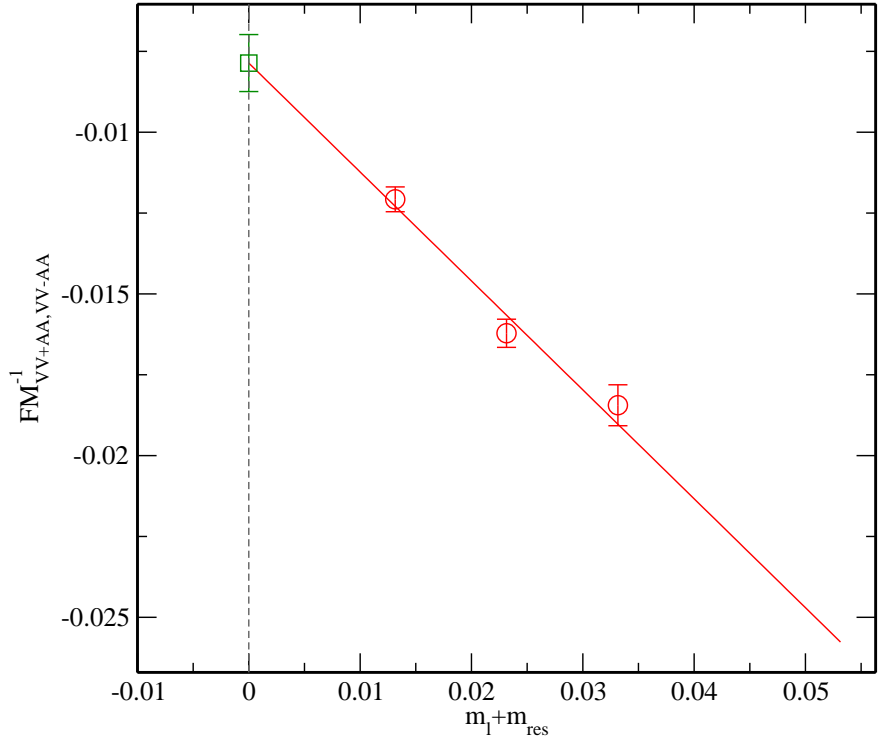}
\par\end{centering}
\caption{Linear extrapolation of the mixing coefficient $FM_{VV+AA,VV-AA}^{-1}$
to the chiral limit using the three unitary mass values, at the momentum scale
$\mu=2.04\mbox{ GeV}$.
\label{fig:FMinv_VV+AA_VV-AA_fit} }
\end{figure}

With the procedure described in Section~\ref{sub:bk mixing proc},
we can directly calculate the mixing coefficients.  In particular, we
have calculated the off-diagonal terms in the matrix $FM^{-1}$.
Figure~\ref{fig:FMinv_VV+AA_VV-AA} shows the mixing coefficient
$FM_{VV+AA,VV-AA}^{-1}$ at different unitary masses.  As in the
earlier discussion of the $\Lambda_{A}-\Lambda_{V}$ difference,
our use of exceptional external momenta permits both a linear and
quadratic mass dependence.  As was found in Ref.~\cite{Aoki:2005ga}
and suggested by the mass dependence seen in
Figure~\ref{fig:FMinv_VV+AA_VV-AA_fit} a linear dependence appears
reasonable and it is a linear form that we have used in determining
the chiral limit shown in Figure~\ref{fig:FMinv_VV+AA_VV-AA}.

As can be seen in Figure~\ref{fig:FMinv_VV+AA_VV-AA}, at the chosen
reference scale, $\mu\simeq2\mbox{ GeV}$ or $\left(ap\right)^{2}\simeq1.4$,
the mixing coefficient is about $0.7\%$ and decreases when the scale
is made larger.  Similar to the discussion in Section~\ref{sec:ch_sym_breaking},
we again propose that this non-zero mixing coefficient in the chiral
limit has its source in our use of exceptional momenta.  Again we
can determine the asymptotic behavior of the amplitude in question
by determining the least negative degree of divergence of a subgraph
$\Gamma^\prime$ through which all of the large external momenta
can be arranged to flow.  Here it is convenient to treat the operator
$O_{LL}$, which is evaluated at zero momentum, as an internal vertex
of dimension 6 rather than an unusual sort of external line.  This
alters the rules for computing the degree of divergence of a subgraph:
now any connected subgraph with $F$ external fermion lines and $B$
external boson lines in which this new $O_{LL}$ vertex appears, must
have degree of divergence $d=6-3F/2-B$ since $O_{LL}$ has a dimension
two higher than the usual renormalizable coupling.  (As before, the
degree of divergence of a disconnected graph is the sum of the
degrees of divergence of its connected components.)

\begin{figure}[H]
\begin{centering}
\includegraphics{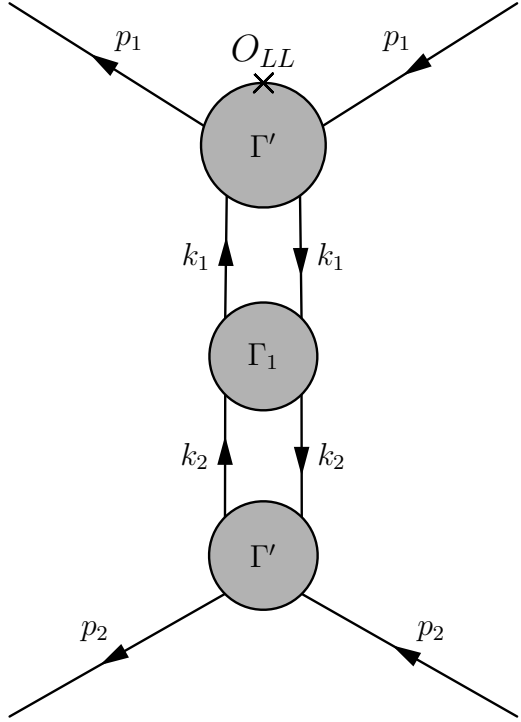}
\par\end{centering}
\caption{A possible identification of subgraphs appearing in the chirality violating
mixing between $O_{LL}$ and other four-quark operators.  The disconnected subdiagram
$\Gamma^\prime$ has degree of divergence $d=-2$ for the case of exceptional momenta
shown here.  This permits a complex pattern of low-energy, vacuum chiral symmetry
breaking coming from the low-energy, four-quark subgraph $\Gamma_1$ to enter such an
amplitude with only a mild $1/p^2$ suppression.\label{fig:mix_except}}
\end{figure}

As in the case of the vertex amplitude discussed in
Section~\ref{sec:ch_sym_breaking}, the appearance of exceptional momenta
does not change the asymptotic behavior in the large $\lambda$ limit
with external momenta $\lambda p_i$ for $1 \le i \le 4$.  Even for
our exceptional case $p_1=p_3=-p_2=-p_4$, $\lambda^d$ scaling with
$d = 6-4\cdot3/2=0$ is expected.  However, derivatives with respect to the
quark mass or the occurrence of factors of $\langle \overline{q} q\rangle$
will be strongly affected by this choice of external momenta.  As is
shown in Figure~\ref{fig:mix_except}, we can identify a disconnected
subgraph $\Gamma^\prime$ through which all the large external momenta
can be routed which has $d = (6-4\cdot3/2) + (4 - 4\cdot3/2) = -2$.
(Note, momentum conservation implies that if all of the large momenta can
be routed within a disconnected diagram then the choice of external momenta
must be exceptional.)  This $d=-2$ value implies a $1/p^2$ behavior with
only low momenta flowing through the omitted subgraph $\Gamma_1$.  Since
$\Gamma_1$ has four external lines it can translate standard QCD
vacuum symmetry breaking into the chiral symmetry breaking that is
required to produce the operator mixing shown in
Figure~\ref{fig:FMinv_VV+AA_VV-AA}.

\begin{figure}[H]
\begin{centering}
\includegraphics[clip]{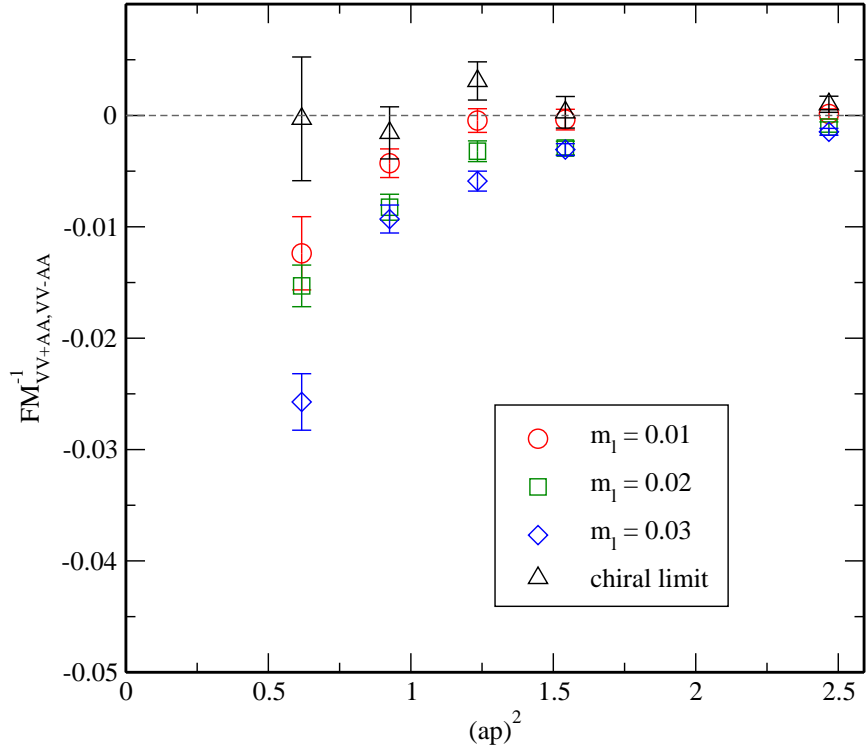}
\par\end{centering}
\caption{The mixing coefficient $FM_{VV+AA,VV-AA}^{-1}$ calculated at
non-exceptional momenta. When extrapolated to the chiral limit the
mixing coefficient vanishes, which shows that chiral symmetry breaking
as shown in Fig.~\ref{fig:FMinv_VV+AA_VV-AA} comes from the existence
of a low-energy sub-diagram that enters because of the special choice of
external momenta.  \label{fig:FMinv_VV+AA_VV-AA_nonexp}}
\end{figure}

Again, we confirm this conclusion, by recomputing the coefficient
$FM_{VV+AA,VV-AA}^{-1}$ at non-exceptional momenta, as shown in
Figure~\ref{fig:FMinv_VV+AA_VV-AA_nonexp}. With that choice of momenta the
mixing coefficient vanishes completely within our statistical accuracy.

\begin{figure}[H]
\begin{centering}
\includegraphics[clip]{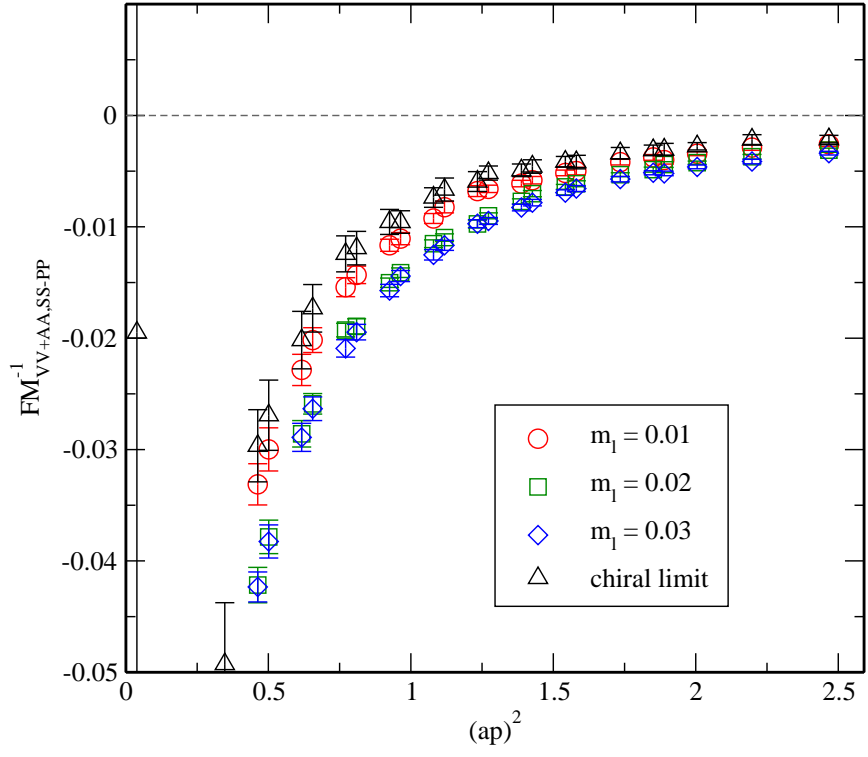}
\par\end{centering}
\caption{The mixing coefficient $FM_{VV+AA,SS-PP}^{-1}$ for unitary choices
of the mass.  \label{fig:FMinv_VV+AA_SS-PP} }
\end{figure}

\begin{figure}[H]
\begin{centering}
\includegraphics[clip]{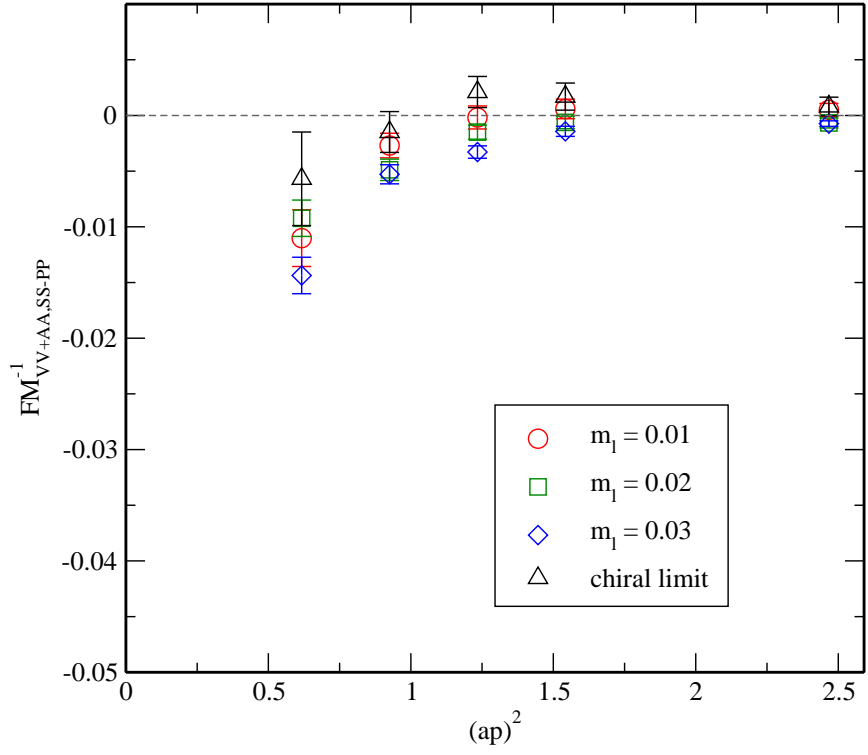}
\par\end{centering}
\caption{The mixing coefficient $FM_{VV+AA,SS-PP}^{-1}$ calculated at
non-exceptional momenta. When extrapolated to the chiral limit the
mixing coefficient vanishes, which shows that chiral symmetry breaking
as shown in Fig.~\ref{fig:FMinv_VV+AA_SS-PP} comes from the existence
of a low-energy sub-diagram that enters because of the special choice of
external momenta.  \label{fig:FMinv_VV+AA_SS-PP_nonexp}}
\end{figure}

\begin{figure}[H]
\begin{centering}
\includegraphics[clip]{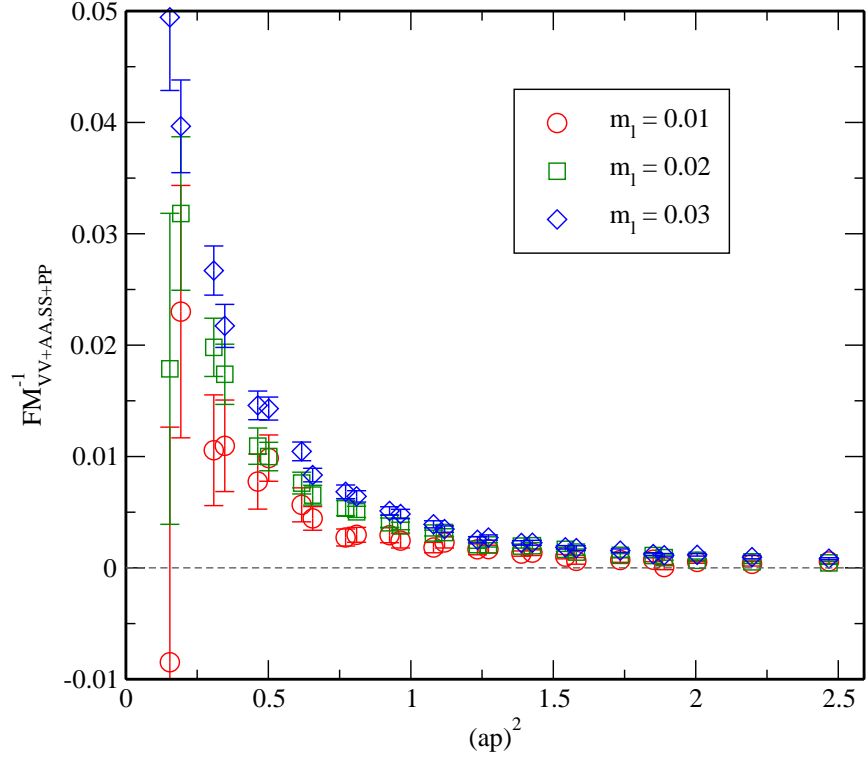}
\par\end{centering}
\caption{The mixing coefficient $FM_{VV+AA,SS+PP}^{-1}$ for unitary choices
of the mass.  The coefficients are very tiny over the region of medium to large
momenta.\label{fig:FMinv_VV+AA_SS+PP} }
\end{figure}

\begin{figure}[H]
\begin{centering}
\includegraphics[clip]{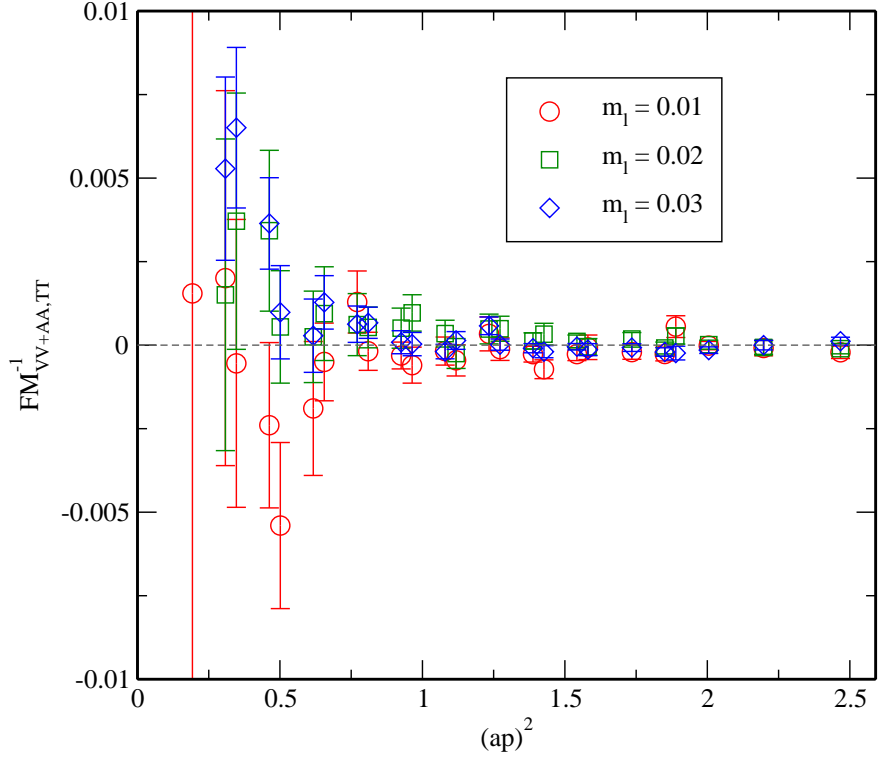}
\par\end{centering}
\caption{The mixing coefficient $FM_{VV+AA,TT}^{-1}$ for unitary choices
of the mass.  These coefficients agree well with zero. \label{fig:FMinv_VV+AA_TT} }
\end{figure}

The other chiral symmetry breaking mixing coefficients,
$FM_{VV+AA,SS\pm PP}^{-1}$ and $FM_{VV+AA,TT}^{-1}$ are very similar
to the case of $FM_{VV+AA,VV-AA}^{-1}$ just discussed.  These coefficients
are plotted in Figure~\ref{fig:FMinv_VV+AA_SS-PP} to
Figure~\ref{fig:FMinv_VV+AA_TT}.  Since our theoretical argument implies
that the mixing coefficients are very small, \emph{i.e.}
$\mathcal{O}\left(m_{\mathrm{res}}^{2}\right)$ and our numerical results
are consistent with this implication, it is safe to neglect them and
calculate the renormalization coefficient for $B_{K}$:
\begin{equation}
Z_{B_{K}}^{\mathrm{RI/MOM}}
 = \frac{Z_{VV+AA,VV+AA}}{Z_{A}^{2}}.
\label{eq:Z_B_K}
\end{equation}

\subsection{Calculation of $Z_{B_{K}}$ and renormalization group running
\label{sub:Zbk_running}}

\begin{figure}[H]
\begin{centering}
\includegraphics[clip]{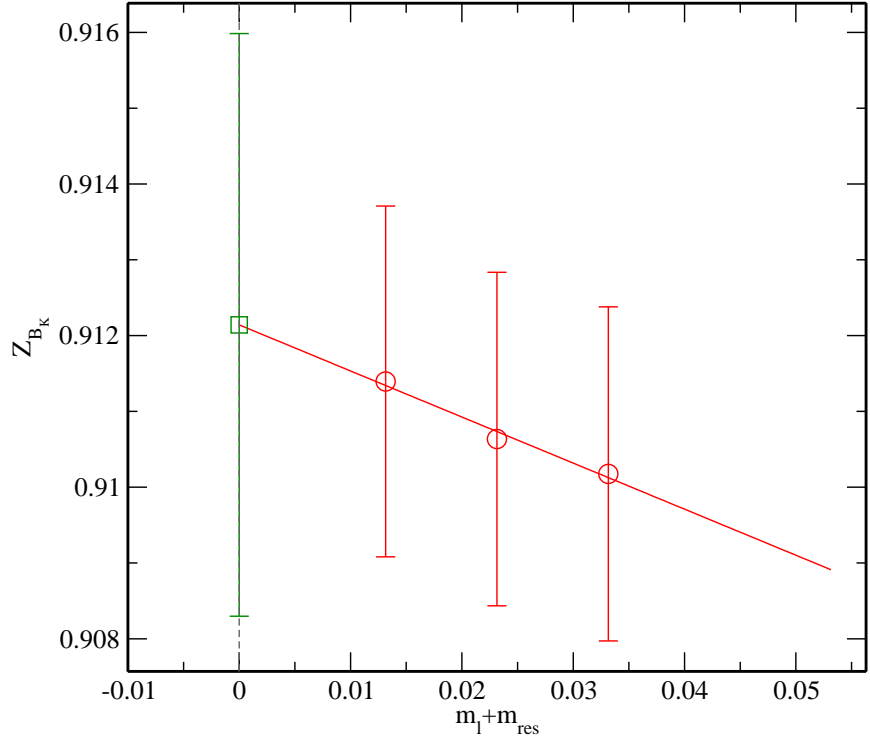}
\par\end{centering}
\caption{Linear extrapolation of $Z_{B_{K}}$ to the chiral limit using
unitary mass values and the scale $\mu=2.04\mbox{ GeV}$
.\label{fig:Z_bk_fit}}
\end{figure}

\begin{figure}[H]
\begin{centering}
\includegraphics[clip]{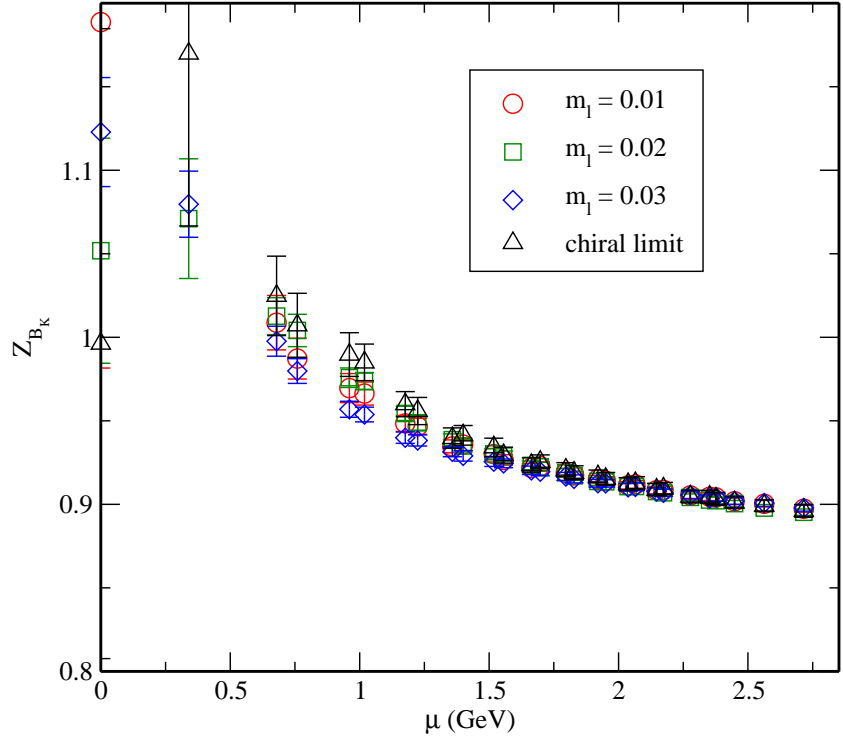}
\par\end{centering}
\caption{The renormalization factor $Z_{B_{K}}^{\mathrm{RI/MOM}}$ evaluated
for unitary mass values and extrapolated to the chiral limit.
\label{fig:Z_bk_chlim}}
\end{figure}

Using Eq.~\eqref{eq:Z_B_K}, the value of $Z_{q}/Z_{A}$ from
Section~\ref{sec:ch_sym_breaking} and the value of
$Z_{q}^{-2}Z_{VV+AA,VV+AA}=FM_{VV+AA,VV+AA}^{-1}$, we can calculate
the lattice values of $Z_{B_{K}}$ at different masses and
momenta, as shown in Table~\ref{tab:Zbk_mass}. To extrapolate
to the chiral limit, we again use a linear function, for the
same reasons as described in Section~\ref{sec:ch_sym_breaking}.
The linear mass fit at the scale $\mu=2\mbox{ GeV}$ is illustrated
in Figure~\ref{fig:Z_bk_fit}, and the value of $Z_{B_{K}}^{\mathrm{RI/MOM}}$
in the chiral limit is shown in Figure~\ref{fig:Z_bk_chlim} and
Table~\ref{tab:Zbk-comb}.

Similar to the procedure described in Section~\ref{sub:Zm_running},
in order to determine $Z_{B_{K}}^{\overline{\mathrm{MS}}}$ from
$Z_{B_{K}}^{\mathrm{RI/MOM}}$, we first divide the
$Z_{B_{K}}^{\mathrm{RI/MOM}}(\mu)$ by the predicted running factor at
one-loop order and obtain the quantity $Z_{B_{K}}^{\mathrm{SI}}(\mu)$.
Then we fit a quadratic function $A+B\left(a\mu\right)^{2}$ over
the region $1.3<\left(a\mu\right)^{2}<2.5$ to remove the
$\mathcal{O}\left(\left(a\mu\right)^{2}\right)$ dependence from
$Z_{B_{K}}^{\mathrm{SI}}$ induced by the lattice artifacts. Finally,
we restore its perturbative running in the $\overline{\mathrm{MS}}$
scheme to the scale $\mu=2\mbox{ GeV}$. The perturbative running and
matching factors are presented in Appendix~\ref{sec:Zbk run factor}.

\begin{figure}[H]
\begin{centering}
\includegraphics[clip]{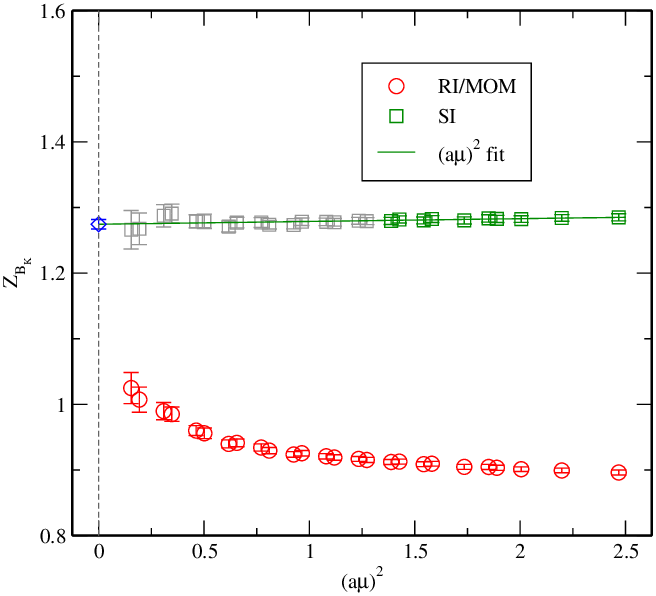}
\par\end{centering}
\caption{The quantities $Z_{B_{K}}^{\mathrm{RI/MOM}}$ and $Z_{B_{K}}^{\mathrm{SI}}$
plotted versus the square of the scale $a \mu$.  Here $Z_{B_{K}}^{\mathrm{SI}}$
is obtained by dividing $Z_{B_{K}}^{\mathrm{RI/MOM}}$ by the predicted
perturbative running factor.  Shown also is the linear extrapolation of
$Z_{B_{K}}^{\mathrm{SI}}(\mu) = Z_{B_{K}}^{\mathrm{SI}}
+ c \left(a\mu\right)^{2}$ using the momentum region
$1.3<\left(a\mu\right)^{2}<2.5$ to remove lattice artifacts.
\label{fig:ZbkRGfit} }
\end{figure}

\begin{figure}[H]
\begin{centering}
\includegraphics[clip]{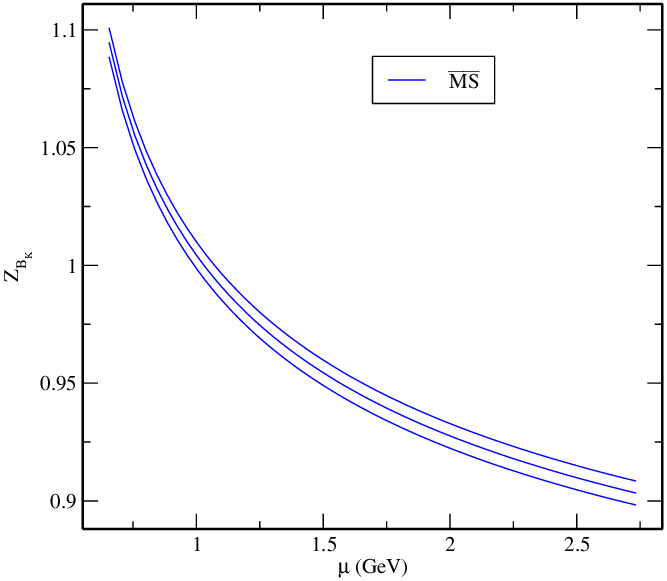}
\par\end{centering}
\caption{The renormalization factor $Z_{B_{K}}$ expressed in the
$\overline{\mathrm{MS}}$ scheme.  These results are obtained by
applying the perturbative running factor to $Z_{B_{K}}^{\mathrm{SI}}$.
The value we are interested in is
$Z_{B_{K}}^{\overline{\mathrm{MS}}}\left(\mu=2\mbox{ GeV}\right)$.
The upper and lower curves show the statistical errors.
\label{fig:ZbkRGimp} }
\end{figure}

The procedure of dividing by the running and removing the
$\left(a\mu\right)^{2}$ dependence is shown in Figure~\ref{fig:ZbkRGfit},
and the result of restoring the running in the $\overline{\mathrm{MS}}$
scheme is shown in Figure~\ref{fig:ZbkRGimp}.  Table~\ref{tab:Zbk-comb}
lists $Z_{B_{K}}^{\overline{\mathrm{MS}}}$ at different momentum scales.
The final $Z_{B_{K}}$ we need in the $\overline{\mathrm{MS}}$ scheme
and $\mu=2\mbox{ GeV}$ is
\begin{equation}
Z_{B_{K}}^{\overline{\mathrm{MS}}}\left(2\mbox{ GeV}\right)
   = 0.9276\pm0.0052(\mbox{stat})\pm0.0220(\mbox{sys}).
\label{eq:Zbk-result}
\end{equation}
The systematic error is calculated, following the same procedure as
has been used for the previous quantities, as a sum in quadrature of
the amount the result changes when
$\frac{1}{2}\left(\Lambda_{A}+\Lambda_{V}\right)$ is replaced by
$\Lambda_{A}$ (0.0131), the size of the highest order perturbative
correction being made (here of $O(\alpha_s)$) (0.0177) and the
effect of our non-zero value for $m_s$ in the calculation of
$Z_{B_K}$ (0.0007).

\section{Conclusions\label{sec:Conclusions}}

We have presented a study of the renormalization coefficients $Z_q$, $Z_m$
$Z_T$ and $B_{K}$ on the $16^3 \times 32$, 2+1 flavor dynamical domain-wall
fermion lattices with Iwasaki gauge action of $\beta=2.13$ and
$a^{-1}=1.729(28)\mbox{ GeV}$ generated by the RBC and UKQCD collaborations.
These coefficients are important components in calculations of a number
of important physical quantities reported elsewhere~\cite{Boyle:2007fn, Antonio:2007pb}
The procedure closely follows that used in an earlier study with quenched
lattice configurations\cite{Blum:2001sr,Aoki:2005ga}.  In addition to
providing the Z-factors necessary to support a variety of calculations
on these lattice configurations, this paper also presents a number of new
results which go beyond earlier work.

First, the troublesome double pole which appears in a quenched calculation
of the quantity $\Lambda_{S}$ because of topological near zero modes is
now highly suppressed by the 2+1 flavor determinant.  This allows us to
use $\Lambda_{S}$ for an accurate calculation of $Z_m$.  Second we have
identified the $O(5\%)$ large chiral symmetry breaking effects seen in
the off-shell Green functions $\Lambda_V$ and $\Lambda_A$ as caused by
our use of exceptional momenta.  We have advanced both a theoretical
discussion explaining the pattern of symmetry breaking which we have observed
and a calculation with non-exceptional momenta in which these effects
are dramatically reduced.

Third, for $Z_{B_K}$ we have presented both a theoretical argument and
numerical calculations showing the mixing coefficients with the operators
with the wrong chirality are very small so that the calculation of
$Z_{B_K}$ can be simplified by neglecting these mixing coefficients.
Finally we have exploited the earlier perturbative work of others and
evaluated the factors relating the normalization of operators defined in the
$\mathrm{\overline{MS}}$ and RI/MOM schemes determining $Z_m^{\mathrm{\overline{MS}}}$,
$Z_q^{\mathrm{\overline{MS}}}$, $Z_{B_K}^{\mathrm{\overline{MS}}}$ and
$Z_T^{\mathrm{\overline{MS}}}$ from their non-perturbative RI/MOM counterparts
to three, three, one and two loops respectively.

\begin{acknowledgments}
We thank our collaborators in the RBC and UKQCD collaborations for
assistance and useful discussions.  This work was supported by DOE grant
DE-FG02-92ER40699 and PPARC grants PPA/G/O/2002/00465 and PP/D000238/1.
We thank the University of Edinburgh, PPARC, RIKEN, BNL and the
U.S.\ DOE for providing the facilities on which this work was performed.
A.S. was supported by the U.S.\ Dept.\ of Energy under contracts
DE-AC02-98CH10886, and DE-FG02-92ER40716.
\end{acknowledgments}

\appendix

\section{The QCD $\beta$ Functions and the Running of $\alpha_{s}$\label{sec:alpha_s}}

The four-loop QCD beta functions is calculated in \cite{vanRitbergen:1997va}
and the conventions we use are the same as in \cite{Chetyrkin:1999pq}:
\begin{align}
\beta_{0}= & \frac{1}{4}\left(11-\frac{2}{3}n_{f}\right),\nonumber \\
\beta_{1}= & \frac{1}{16}\left(102-\frac{38}{3}n_{f}\right),\nonumber \\
\beta_{2}= & \frac{1}{64}\left(\frac{2857}{2}-\frac{5033}{18}n_{f}+\frac{325}{54}n_{f}^{2}\right),\nonumber \\
\beta_{3}= & \frac{1}{256}\left[\frac{149753}{6}+3564\zeta_{3}-\left(\frac{1078361}{162}+\frac{6508}{27}\zeta_{3}\right)n_{f}\right.\nonumber \\
 & \left.+\left(\frac{50065}{162}+\frac{6472}{81}\zeta_{3}\right)n_{f}^{2}+\frac{1093}{729}n_{f}^{3}\right].\label{eq:beta-function}\end{align}
To calculate the coupling constant $\alpha_{s}\left(\mu\right)$ at
any scales, we have used the four-loop (NNNLO) running formula for
$\alpha_{s}$~\cite{vanRitbergen:1997va}:\begin{align}
\frac{\partial a_{s}}{\partial\ln\mu^{2}} & =\beta\left(a_{s}\right)\nonumber \\
 & =-\beta_{0}a_{s}^{2}-\beta_{1}a_{s}^{3}-\beta_{2}a_{s}^{4}-\beta_{3}a_{s}^{5}+\mathcal{O}\left(a_{s}^{6}\right)\label{eq:a_s run}\end{align}
where $a_{s}=\alpha_{s}/\pi$. (We have changed the normalization
of $a_{s}$ to match the definition of the $\beta$-functions coefficients.) For a
numerical implementation, we start from the world-average value at
$\mu=M_{Z}$~\cite{Yao:2006px}, \begin{equation}
\alpha_{s}^{\left(5\right)}\left(M_{Z}\right)=0.1176\pm0.002,\label{eq:alpha_s-Mz}\end{equation}
where the superscript indicates that it is in the 5-flavor region,
and run $\alpha_{s}$ across the $m_{b}$ and $m_{c}$ threshold with
the matching conditions:
\begin{equation}
\alpha_{s}^{\left(5\right)}\left(m_{b}\right) =\alpha_{s}^{\left(4\right)}\left(m_{b}\right)\qquad\textrm{and}\qquad
\alpha_{s}^{\left(4\right)}\left(m_{c}\right) =\alpha_{s}^{\left(3\right)}\left(m_{c}\right)\,.\label{eq:alpha_s run match}\end{equation}
Having computed $\alpha_{s}^{\left(3\right)}\left(m_{c}\right)$,
we can calculate the coupling constant at any scale in the 3-flavor
theory. For example, \begin{equation}
\alpha_{s}^{\left(3\right)}\left(\mu=2\mbox{ GeV}\right)=0.2904.\label{eq:alpha_s 2GeV}\end{equation}

\section{Perturbative Running and Matching for $Z_{m}$\label{sec:Zm run factor}}

In \cite{Chetyrkin:1999pq}, the renormalization group equation for
$m_{\mathrm{ren}}\left(\mu\right)$ is solved to four-loop order (NNNLO)
. Using the solution with our definition of
the renormalization coefficients $Z_{m}$, we obtain:
\begin{equation}
Z_{m}^{\mathrm{SI}}=\frac{c\left(\alpha_{s}\left(\mu_{0}\right)/\pi\right)}{c\left(\alpha_{s}\left(\mu\right)/\pi\right)} Z_{m}^{\mathrm{RI/MOM}}\left(\mu\right)\label{eq:Zm-SI-again}\end{equation}
where the function $c\left(x\right)$ is given by:\begin{align}
c\left(x\right)= & \left(x\right)^{\bar{\gamma_{0}}}\left\{ 1+\left(\bar{\gamma_{1}}-\bar{\beta_{1}}\bar{\gamma_{0}}\right)x\right.\nonumber \\
 & +\frac{1}{2}\left[\left(\bar{\gamma_{1}}-\bar{\beta_{1}}\bar{\gamma_{0}}\right)^{2}+\bar{\gamma_{2}}+ \bar{\beta_{1}}^{2}\bar{\gamma_{0}}-\bar{\beta_{1}}\bar{\gamma_{1}}-\bar{\beta_{2}}\bar{\gamma_{0}}\right]x^{2}\nonumber \\
 & +\left[\frac{1}{6}\left(\bar{\gamma_{1}}-\bar{\beta_{1}}\bar{\gamma_{0}}\right)^{3}+\frac{1}{2} \left(\bar{\gamma_{1}}-\bar{\beta_{1}}\bar{\gamma_{0}}\right)\left(\bar{\gamma_{2}}+\bar{\beta_{1}}^{2} \bar{\gamma_{0}}-\bar{\beta_{1}}\bar{\gamma_{1}}-\bar{\beta_{2}}\bar{\gamma_{0}}\right)\right.\nonumber \\
 & \left.\qquad+\frac{1}{3}\left(\bar{\gamma_{3}}-\bar{\beta_{1}}^{3}\bar{\gamma_{0}}+2\bar{\beta_{1}} \bar{\beta_{2}}\bar{\gamma_{0}}-\bar{\beta_{3}}\bar{\gamma_{0}}+\bar{\beta_{1}}^{2} \bar{\gamma_{1}}-\bar{\beta_{2}}\bar{\gamma_{1}}-\bar{\beta_{1}}\bar{\gamma_{2}}\right)\right]x^{3}\nonumber \\
 & \left.+\mathcal{O}\left(x^{4}\right)\right\} ,\label{eq:ZmRI-ZmSI-c_x}\end{align}
with $\bar{\beta_{i}} =\frac{\beta_{i}}{\beta_{0}}$ and \begin{equation}
\bar{\gamma_{i}} =\frac{\gamma_{m}^{\mathrm{RI/MOM}\left(i\right)}}{\beta_{0}}\label{eq:gamma_bar}\end{equation}
The evaluation of the coefficients of the QCD $\beta$ function and the running of $\alpha_{s}$
are explained in Appendix~\ref{sec:alpha_s} and the anomalous dimensions
are \begin{align}
\gamma_{m}^{\mathrm{RI/MOM}\left(0\right)}= & 1\nonumber \\
\gamma_{m}^{\mathrm{RI/MOM}\left(1\right)}= & \frac{1}{16}\left(126-\frac{52}{9}n_{f}\right)\nonumber \\
\gamma_{m}^{\mathrm{RI/MOM}\left(2\right)}= & \frac{1}{64}\left[\left(\frac{20911}{3}-\frac{3344}{3}\zeta_{3}\right)+\left(-\frac{18386}{27}+ \frac{128}{9}\zeta_{3}\right)n_{f}+\frac{928}{81}n_{f}^{2}\right]\nonumber \\
\gamma_{m}^{\mathrm{RI/MOM}\left(3\right)}= & \frac{1}{256}\left[\left(\frac{300665987}{648}-\frac{15000871}{108}\zeta_{3}+\frac{6160}{3}\zeta_{5}\right)\right.\nonumber \\
 & \qquad+\left(-\frac{7535473}{108}+\frac{627127}{54}\zeta_{3}+\frac{4160}{3}\zeta_{5}\right)n_{f}\nonumber \\
 & \qquad\left.+\left(\frac{670948}{243}-\frac{6416}{27}\zeta_{3}\right)n_{f}^{2}-\frac{18832}{729}n_{f}^{3}\right]\,, \label{eq:anomalous-dim}\end{align}
where $n_{f}=3$.

When applying Eq. \eqref{eq:Zm-SI-again}, we need to choose a value
of $\mu_{0}$, where the SI value is calculated. The exact value
of $\mu_{0}$ is immaterial and for convenience we choose its value
such that \begin{equation}
\left(a\mu_{0}\right)^{2}=2\,.\label{eq:mu_0 choice}\end{equation}

To match the renormalization coefficients $Z_{m}$ from RI/MOM scheme
to $\overline{\mathrm{MS}}$ scheme, we have applied the three-loop
matching factor \cite{Chetyrkin:1999pq} obtaining:\begin{eqnarray}
\frac{Z_{m}^{\overline{\mathrm{MS}}}}{Z_{m}^{\mathrm{RI/MOM}}} & = & 1+\frac{\alpha_{s}}{4\pi}\left[-\frac{16}{3}\right]+\left(\frac{\alpha_{s}}{4\pi}\right)^{2} \left[-\frac{1990}{9}+\frac{152}{3}\zeta_{3}+\frac{89}{9}n_{f}\right]\nonumber \\
 &  & +\left(\frac{\alpha_{s}}{4\pi}\right)^{3} \left[-\frac{6663911}{648}+\frac{408007}{108}\zeta_{3}-\frac{2960}{9}\zeta_{5}+\frac{236650}{243}n_{f}\right.\nonumber \\
 &  & \left.-\frac{4936}{27}\zeta_{3}n_{f}+\frac{80}{3}\zeta_{4}n_{f}-\frac{8918}{729}n_{f}^{2}-\frac{32}{27} \zeta_{3}n_{f}^{2}\right].\label{eq:Zm-match}\end{eqnarray}

\section{Perturbative Running and Scheme Matching for $Z_{q}$\label{sec:Zq run factor}}

The renormalization group equation for $Z_{q}$ is very similar to
that for $Z_{m}$ \cite{Chetyrkin:1999pq} and we can reuse the solution
of the equation from Appendix \eqref{sec:Zm run factor} to write:\begin{equation}
Z_{q}^{\mathrm{SI}}=\frac{c^{\left[\gamma_{2}\right]}\left(\alpha_{s}\left(\mu_{0}\right)/\pi\right)}{c^{\left[\gamma_{2}\right]}\left(\alpha_{s}\left(\mu\right)/\pi\right)}Z_{q}^{\mathrm{RI/MOM}}\left(\mu\right)\label{eq:Zq-SI-again}\end{equation}
where the function $c^{\left[\gamma_{2}\right]}\left(x\right)$ has
exactly the same functional form as $c\left(x\right)$ defined
in Eq. \eqref{eq:ZmRI-ZmSI-c_x}, but with the coefficients $\bar{\gamma}_{i}$ of the anomalous dimension 
$\gamma_m$ replaced by those of $\gamma_{2}$:\begin{align}
\bar{\gamma_{i}} & =\frac{\gamma_{2}^{\mathrm{RI/MOM}\left(i\right)}}{\beta_{0}}\label{eq:gamma_bar_q}\end{align}
The coefficients of the anomalous dimension $\gamma_{2}$ are \cite{Chetyrkin:1999pq}:\begin{align}
\gamma_{2}^{\mathrm{RI/MOM}\left(0\right)} & =0\nonumber \\
\gamma_{2}^{\mathrm{RI/MOM}\left(1\right)} & =\frac{N^{2}-1}{16N^{2}}\left\{ \left[\frac{3}{8}+\frac{11}{4}N^{2}\right]+n_{f}\left[-\frac{1}{2}N\right]\right\} \nonumber \\
\gamma_{2}^{\mathrm{RI/MOM}\left(2\right)} & =\frac{N^{2}-1}{64N^{3}}\left\{ \left[\frac{3}{16}+\frac{25}{3}N^{2}+\frac{14225}{288}N^{4}-3N^{2}\zeta_{3}-\frac{197}{16}N^{4}\zeta_{3}\right]\right.\nonumber \\
 & \qquad\qquad+n_{f}\left[-\frac{1}{3}N-\frac{611}{36}N^{3}+2N^{3}\zeta_{3}\right]\nonumber \\
 & \qquad\qquad\left.+n_{f}^{2}\left[\frac{10}{9}N^{2}\right]\right\} \nonumber \\
\gamma_{2}^{\mathrm{RI/MOM}\left(3\right)} & =\frac{N^{2}-1}{256N^{4}}\left\{ \left[\frac{1027}{128}+\frac{7673}{384}N^{2}+\frac{174565}{1152}N^{4}+\frac{3993865}{3456}N^{6}\right.\right.\nonumber \\
 & \qquad\qquad\qquad+25\zeta_{3}+31N^{2}\zeta_{3}-\frac{10975}{64}N^{4}\zeta_{3}-\frac{111719}{192}N^{6}\zeta_{3}\nonumber \\
 & \qquad\qquad\qquad\left.-40\zeta_{5}-60N^{2}\zeta_{5}+\frac{5465}{64}N^{4}\zeta_{5}+\frac{20625}{128}N^{6}\zeta_{5}\right]\nonumber \\
 & \qquad\qquad+n_{f}\left[\frac{1307}{48}N+\frac{557}{144}N^{3}-\frac{172793}{288}N^{5}\right.\nonumber \\
 & \qquad\qquad\qquad-4N\zeta_{3}+2N^{3}\zeta_{3}+\frac{7861}{48}N^{5}\zeta_{3}\nonumber \\
 & \qquad\qquad\qquad\left.-30N^{3}\zeta_{5}-\frac{125}{4}N^{5}\zeta_{5}\right]\nonumber \\
 & \qquad\qquad+n_{f}^{2}\left[-\frac{521}{72}N^{2}+\frac{259}{3}N^{4}+6N^{2}\zeta_{3}-\frac{26}{3}N^{4}\zeta_{3}\right]\nonumber \\
 & \qquad\qquad\left.+n_{f}^{3}\left[-\frac{86}{27}N^{3}\right]\right\} \label{eq:gamma_2 anom-dim}\end{align}
where $N=3$, which represents the number of colors, and $n_{f}=3$.

When we match $Z_{q}$ from RI/MOM scheme to $\overline{\mathrm{MS}}$
scheme, the three-loop matching factor is \cite{Chetyrkin:1999pq}\begin{align}
\frac{Z_{q}^{\overline{\mathrm{MS}}}}{Z_{q}^{\mathrm{RI/MOM}}} & =1+\left(\frac{\alpha_{s}}{4\pi}\right)^{2}\left[-\frac{517}{18}+12\zeta_{3}+\frac{5}{3}n_{f}\right]\nonumber \\
 & \quad+\left(\frac{\alpha_{s}}{4\pi}\right)^{3}\left[-\frac{1287283}{648}+\frac{14197}{12}\zeta_{3}+\frac{79}{4}\zeta_{4} -\frac{1165}{3}\zeta_{5}\right.\nonumber \\
 & \qquad\qquad\qquad\left.+\frac{18014}{81}n_{f}-\frac{368}{9}\zeta_{3}n_{f}-\frac{1102}{243}n_{f}^{2}\right].\label{eq:Zq match}\end{align}

\section{Perturbative Running and Scheme Matching for $Z_{B_{K}}$\label{sec:Zbk run factor}}

To remove (restore) the perturbative renormalization group running
of $Z_{B_{K}}$, we use the one-loop renormalization group running
formula~\cite{Aoki:2005ga}: \begin{equation}
Z_{B_{K}}^{\mathrm{SI}}\left(n_{f}\right)=w_{\mathrm{scheme}}^{-1}\left(\mu,n_{f}\right)Z_{B_{K}}^{\mathrm{scheme}} \left(\mu,n_{f}\right)\label{eq:Z_bk_SI}\end{equation}
where\begin{equation}
w_{\mathrm{scheme}}^{-1}\left(\mu,n_{f}\right)=\alpha_{s}\left(\mu\right)^{-\gamma_{0}/2\beta_{0}}\left[1+ \frac{\alpha_{s}\left(\mu\right)}{4\pi}J_{\mathrm{scheme}}^{\left(n_{f}\right)}\right]\label{eq:w-RI-MOM}\end{equation}
and \begin{align}
J_{\mathrm{RI/MOM}}^{\left(n_{f}\right)} & =-\frac{17397-2070n_{f}+104n_{f}^{2}}{6\left(33-2n_{f}\right)^{2}}+8\ln2\label{eq:J-RI-MOM}\\
J_{\mathrm{\overline{MS}}}^{\left(n_{f}\right)} & =\frac{13095-1626n_{f}+8n_{f}^{2}}{6\left(33-2n_{f}\right)^{2}}\label{eq:J-MSbar}\end{align}
with $n_{f}=3$ in our analysis.

\section{Perturbative Running and Scheme Matching for $Z_T$\label{sec:ZT run factor}}

The anomalous dimension of the tensor current in the $\overline{\rm MS}$ scheme is given at three-loops in \cite{Gracey:2003yr},
  \begin{eqnarray}
    \gamma_T^{\overline{\rm MS}(0)} &=& \frac{1}{3}\,,\nonumber\\
    \gamma_T^{\overline{\rm MS}(1)} &=& \frac{1}{16}\frac{2}{27}(543-26n_f)\,,\nonumber\\
    \gamma_T^{\overline{\rm MS}(2)} &=& \frac{1}{64}\frac{2}{243}(\frac{1}{2}157665-4176\zeta_3-(2160\zeta_3+7860)n_f-54n_f^2)\,.
  \end{eqnarray}
  For consistency we have adjusted the normalization from that used in \cite{Gracey:2003yr} such that $\gamma_T^{\overline{\rm MS}}$ satisfies the generic
RG-equation for the renormalization constant $Z_\Gamma$ of
the quark bilinear $\bar{\psi}\Gamma\psi$,
  \begin{eqnarray}
    \frac{\partial \ln Z_\Gamma}
	{\partial \ln \mu^2}&=& \gamma_\Gamma(a_s)\nonumber\\
    &=&  -\gamma_\Gamma^{(0)}a_s - \gamma_\Gamma^{(1)}a_s^2 -
	\gamma_\Gamma^{(2)}a_s^3 + {\cal O}(a_s^4)\,,\label{eqn:gammaTdef}
  \end{eqnarray}
  with $a_s = \alpha_s/\pi$.

The perturbative running for the tensor current has also been computed at three loops in the 
$\rm RI/MOM^\prime$ scheme\cite{Gracey:2003yr}, and we use it to obtain the $\rm RI/MOM$ scheme anomalous dimension as follows. We consider the conversion function $C_\Gamma^{\rm RI/MOM^{(\prime)}}$ used to match the $\rm RI/MOM$ or $\rm RI/MOM^{\prime}$ scheme to the $\overline{\rm MS}$ scheme:
  \begin{eqnarray}
    Z_\Gamma^{\rm \overline{\rm MS}}&=&C_\Gamma^{\rm RI/MOM^{(\prime)}}
    Z_\Gamma^{\rm RI/MOM^{(\prime)}}\,.\label{eqn:conv_function}
  \end{eqnarray}
  Applying the above renormalization group equation (\ref{eqn:gammaTdef})
  we obtain
  \begin{eqnarray}
    \gamma_\Gamma^{\rm RI/MOM^{(\prime)}}&=&
    \gamma_\Gamma^{\overline{\rm MS}} - \frac{\partial \ln C_\Gamma^{\rm RI/MOM^{(\prime)}}}
	  {\partial \ln \mu^2}\,.
  \end{eqnarray}
  Since the only difference between
  the ${\rm RI/MOM}$ and $\rm RI/MOM^\prime$ schemes lies in the definition of the
  quark field renormalization constants $Z_2^{\rm RI/MOM^\prime}$ and $Z_2^{\rm RI/MOM}$,
  we write
  $C_\Gamma^{\rm RI/MOM^{(\prime)}}=C_\Gamma C_2^{\rm RI/MOM^{(\prime)}}$. The
  vertex part $C_\Gamma$ of the conversion function is common to both the
  $\rm RI/MOM$ and ${\rm RI/MOM^{\prime}}$ schemes.
  It follows that
  \begin{eqnarray}\label{eqn:master}
    \gamma_{\Gamma}^{\rm RI/MOM^{\prime}}-\gamma_{\Gamma}^{\rm RI/MOM} &=&
    \frac{\partial \ln C_2^{\rm RI/MOM}}
         {\partial \ln \mu^2}
	 -\frac{\partial \ln C_2^{\rm RI/MOM^{\prime}}}
         {\partial \ln \mu^2}\nonumber\\
	 &=& \gamma_2^{\rm RI/MOM} - \gamma_2^{\rm RI/MOM^{\prime}}\,.
  \end{eqnarray}
  Since both functions $\gamma_2^{\rm RI/MOM}$ and $\gamma_2^{\rm RI/MOM^{\prime}}$ are known
  \cite{Chetyrkin:1999pq}, we can now compute the anomalous dimension of the
  tensor current in the $\rm RI/MOM$ scheme from the known one in the
  $\rm RI/MOM^{\prime}$ scheme. We note that since the r.h.s. of (\ref{eqn:master}) is valid for any
  choice of $\Gamma$ on the l.h.s., one may use the identity
  \begin{eqnarray}
    \gamma_\Gamma^{\rm RI/MOM}& =& \gamma_\Gamma^{\rm RI/MOM^{\prime}} - (\gamma_{\Gamma^\prime}^{\rm RI/MOM^{\prime}}-\gamma_{\Gamma^\prime}^{\rm RI/MOM})\,.
  \end{eqnarray}

  In order to compute $\gamma_T^{\rm RI/MOM}$ here we have used $\gamma_2^{\rm RI/MOM}$ as in (\ref{eq:gamma_bar_q}) and $\gamma_2^{\rm RI/MOM^\prime}$ from \cite{Chetyrkin:1999pq}:
  \begin{eqnarray}
    \gamma_2^{\rm RI/MOM^\prime(0)} &=& 0\,,\nonumber\\
    \gamma_2^{\rm RI/MOM^\prime(1)} &=&  \frac{N^2-1}{16 N^2}\Bigg\{
    \left[
      \frac{3}{8}
      +\frac{11}{4}  N^{2}
      \right]
    \, + \,  \,n_f
    \left[
      -\frac{1}{2}  N
      \right]
    \,\Bigg\}{}\,,\nonumber\\
    \gamma_2^{\rm RI/MOM^\prime(2)} &=&  \frac{N^2 - 1}{64 N^3}\Bigg\{
    \left[
      \frac{3}{16}
      +\frac{233}{24}  N^{2}
      +\frac{17129}{288}  N^{4}
      -3  N^{2} \,\zeta_{3}
      -\frac{197}{16}  N^{4} \,\zeta_{3}
      \right]
    \nonumber\\
    &{+}&  \,n_f
    \left[
      -\frac{7}{12}  N
      -\frac{743}{36}  N^{3}
      +2  N^{3} \,\zeta_{3}
      \right]
    \nonumber\\
    &{+}&  \, n_f^{2}
    \left[
      \frac{13}{9}  N^{2}
      \right]
    \,\Bigg\}{}\,, \nonumber\\
    \gamma_2^{\rm RI/MOM^\prime(3)} &=&  \frac{N^2 - 1}{256 N^4}\Bigg\{
    \left[
      \frac{1027}{128}
+\frac{8069}{384}  N^{2}
+\frac{240973}{1152}  N^{4}
+\frac{5232091}{3456}  N^{6}
\right. \nonumber \\ &{}& \left.
\phantom{+}
+25  \,\zeta_{3}
+31  N^{2} \,\zeta_{3}
-\frac{12031}{64}  N^{4} \,\zeta_{3}
-\frac{124721}{192}  N^{6} \,\zeta_{3}
\right. \nonumber \\ &{}& \left.
  \phantom{+}
  -40  \,\zeta_{5}
  -60  N^{2} \,\zeta_{5}
+\frac{5465}{64}  N^{4} \,\zeta_{5}
+\frac{20625}{128}  N^{6} \,\zeta_{5}
\right]
\nonumber\\
&{+}&  \,n_f
\left[
  \frac{329}{12}  N
-\frac{1141}{144}  N^{3}
-\frac{113839}{144}  N^{5}
\right. \nonumber \\ &{}& \left.
  \phantom{+ \,n_f }
-4  N  \,\zeta_{3}
+5  N^{3} \,\zeta_{3}
+\frac{2245}{12}  N^{5} \,\zeta_{3}
\right. \nonumber \\ &{}& \left.
  \phantom{+ \,n_f }
-30  N^{3} \,\zeta_{5}
-\frac{125}{4}  N^{5} \,\zeta_{5}
\right]
\nonumber\\
&{+}&  \, n_f^{2}
\left[
  -\frac{515}{72}  N^{2}
+\frac{1405}{12}  N^{4}
+6  N^{2} \,\zeta_{3}
-\frac{32}{3}  N^{4} \,\zeta_{3}
\right]
\nonumber\\
&{+}&  \, n_f^{3}
\left[
  -\frac{125}{27}  N^{3}
\right]
\,\Bigg\}{}.
  \end{eqnarray}

  In this way we obtain the anomalous dimension:
  \begin{eqnarray}
    \gamma_T^{\rm RI/MOM(0)} &=& \frac{1}{3}\,,\nonumber\\
    \gamma_T^{\rm RI/MOM(1)} &=& \frac{1}{16}\frac{2}{27}(543-26n_f)\,,\nonumber\\
    \gamma_T^{\rm RI/MOM(2)} &=& \frac{1}{64}\frac{1}{243}(478821-117648\zeta(3)+
    6(384\zeta(3)-8713) n_f +928n_f^2)\,,
  \end{eqnarray}
  from which we compute the running of $Z_T$ using (\ref{eq:ZmRI-ZmSI-c_x}).

  Combining (\ref{eq:a_s run}) and (\ref{eqn:gammaTdef}) we compute the expression for the matching factor $C_T^{\rm RI/MOM}$. After expanding in $a_s$ we obtain:
  \begin{eqnarray}
    \frac{Z_T^{\overline{\rm MS}}}{Z_T^{\rm RI/MOM}}  = 1+\frac{1}{81}(-4866+1656\zeta(3)+259n_f)
    \left(\frac{\alpha_s}{4\pi}\right)^2\,.
  \end{eqnarray}

\bibliographystyle{apsrev}
\bibliography{references}

\begin{thebibliography}{29}
\expandafter\ifx\csname natexlab\endcsname\relax\def\natexlab#1{#1}\fi
\expandafter\ifx\csname bibnamefont\endcsname\relax
  \def\bibnamefont#1{#1}\fi
\expandafter\ifx\csname bibfnamefont\endcsname\relax
  \def\bibfnamefont#1{#1}\fi
\expandafter\ifx\csname citenamefont\endcsname\relax
  \def\citenamefont#1{#1}\fi
\expandafter\ifx\csname url\endcsname\relax
  \def\url#1{\texttt{#1}}\fi
\expandafter\ifx\csname urlprefix\endcsname\relax\def\urlprefix{URL }\fi
\providecommand{\bibinfo}[2]{#2}
\providecommand{\eprint}[2][]{\url{#2}}

\bibitem[{\citenamefont{Antonio et~al.}(2007{\natexlab{a}})}]{Antonio:2006px}
\bibinfo{author}{\bibfnamefont{D.~J.} \bibnamefont{Antonio}}
  \bibnamefont{et~al.} (\bibinfo{collaboration}{RBC and UKQCD}),
  \bibinfo{journal}{Phys. Rev.} \textbf{\bibinfo{volume}{D75}},
  \bibinfo{pages}{114501} (\bibinfo{year}{2007}{\natexlab{a}}),
  \eprint{hep-lat/0612005}.

\bibitem[{\citenamefont{Allton et~al.}(2007)}]{Allton:2007hx}
\bibinfo{author}{\bibfnamefont{C.}~\bibnamefont{Allton}} \bibnamefont{et~al.}
  (\bibinfo{collaboration}{RBC and UKQCD}), \bibinfo{journal}{Phys. Rev.}
  \textbf{\bibinfo{volume}{D76}}, \bibinfo{pages}{014504}
  (\bibinfo{year}{2007}), \eprint{hep-lat/0701013}.

\bibitem[{\citenamefont{Boyle}(2007)}]{Boyle:2007fn}
\bibinfo{author}{\bibfnamefont{P.}~\bibnamefont{Boyle}}
  (\bibinfo{collaboration}{RBC}) (\bibinfo{year}{2007}),
  \eprint{arXiv:0710.5880 [hep-lat]}.

\bibitem[{\citenamefont{Martinelli et~al.}(1995)\citenamefont{Martinelli,
  Pittori, Sachrajda, Testa, and Vladikas}}]{Martinelli:1994ty}
\bibinfo{author}{\bibfnamefont{G.}~\bibnamefont{Martinelli}},
  \bibinfo{author}{\bibfnamefont{C.}~\bibnamefont{Pittori}},
  \bibinfo{author}{\bibfnamefont{C.~T.} \bibnamefont{Sachrajda}},
  \bibinfo{author}{\bibfnamefont{M.}~\bibnamefont{Testa}}, \bibnamefont{and}
  \bibinfo{author}{\bibfnamefont{A.}~\bibnamefont{Vladikas}},
  \bibinfo{journal}{Nucl. Phys.} \textbf{\bibinfo{volume}{B445}},
  \bibinfo{pages}{81} (\bibinfo{year}{1995}), \eprint{hep-lat/9411010}.

\bibitem[{\citenamefont{Aoki et~al.}(2003)\citenamefont{Aoki, Izubuchi,
  Kuramashi, and Taniguchi}}]{Aoki:2002iq}
\bibinfo{author}{\bibfnamefont{S.}~\bibnamefont{Aoki}},
  \bibinfo{author}{\bibfnamefont{T.}~\bibnamefont{Izubuchi}},
  \bibinfo{author}{\bibfnamefont{Y.}~\bibnamefont{Kuramashi}},
  \bibnamefont{and}
  \bibinfo{author}{\bibfnamefont{Y.}~\bibnamefont{Taniguchi}},
  \bibinfo{journal}{Phys. Rev.} \textbf{\bibinfo{volume}{D67}},
  \bibinfo{pages}{094502} (\bibinfo{year}{2003}), \eprint{hep-lat/0206013}.

\bibitem[{\citenamefont{Aoki and Kuramashi}(2003)}]{Aoki:2003uf}
\bibinfo{author}{\bibfnamefont{S.}~\bibnamefont{Aoki}} \bibnamefont{and}
  \bibinfo{author}{\bibfnamefont{Y.}~\bibnamefont{Kuramashi}},
  \bibinfo{journal}{Phys. Rev.} \textbf{\bibinfo{volume}{D68}},
  \bibinfo{pages}{034507} (\bibinfo{year}{2003}), \eprint{hep-lat/0306008}.

\bibitem[{\citenamefont{Lin and Scholz}(2007)}]{lin:2007pt}
\bibinfo{author}{\bibfnamefont{M.}~\bibnamefont{Lin}} \bibnamefont{and}
  \bibinfo{author}{\bibfnamefont{E.~E.} \bibnamefont{Scholz}}
  (\bibinfo{collaboration}{RBC and UKQCD}) (\bibinfo{year}{2007}),
  \eprint{arXiv:0710.0536 [hep-lat]}.

\bibitem[{\citenamefont{Blum et~al.}(2002)}]{Blum:2001sr}
\bibinfo{author}{\bibfnamefont{T.}~\bibnamefont{Blum}} \bibnamefont{et~al.},
  \bibinfo{journal}{Phys. Rev.} \textbf{\bibinfo{volume}{D66}},
  \bibinfo{pages}{014504} (\bibinfo{year}{2002}), \eprint{hep-lat/0102005}.

\bibitem[{\citenamefont{Christ}(2006)}]{Christ:2005xh}
\bibinfo{author}{\bibfnamefont{N.}~\bibnamefont{Christ}}
  (\bibinfo{collaboration}{RBC and UKQCD}), \bibinfo{journal}{PoS}
  \textbf{\bibinfo{volume}{LAT2005}}, \bibinfo{pages}{345}
  (\bibinfo{year}{2006}).

\bibitem[{\citenamefont{Weinberg}(1960)}]{Weinberg:1959nj}
\bibinfo{author}{\bibfnamefont{S.}~\bibnamefont{Weinberg}},
  \bibinfo{journal}{Phys. Rev.} \textbf{\bibinfo{volume}{118}},
  \bibinfo{pages}{838} (\bibinfo{year}{1960}).

\bibitem[{\citenamefont{Itzykson and Zuber}(1980)}]{Itzykson:1980rh}
\bibinfo{author}{\bibfnamefont{C.}~\bibnamefont{Itzykson}} \bibnamefont{and}
  \bibinfo{author}{\bibfnamefont{J.~B.} \bibnamefont{Zuber}}
  (\bibinfo{year}{1980}), \bibinfo{note}{new York, Usa: Mcgraw-hill (1980) 705
  P.(International Series In Pure and Applied Physics)}.

\bibitem[{\citenamefont{Politzer}(1976)}]{Politzer:1976tv}
\bibinfo{author}{\bibfnamefont{H.~D.} \bibnamefont{Politzer}},
  \bibinfo{journal}{Nucl. Phys.} \textbf{\bibinfo{volume}{B117}},
  \bibinfo{pages}{397} (\bibinfo{year}{1976}).

\bibitem[{\citenamefont{Becirevic et~al.}(2000)\citenamefont{Becirevic,
  Gimenez, Lubicz, and Martinelli}}]{Becirevic:1999kb}
\bibinfo{author}{\bibfnamefont{D.}~\bibnamefont{Becirevic}},
  \bibinfo{author}{\bibfnamefont{V.}~\bibnamefont{Gimenez}},
  \bibinfo{author}{\bibfnamefont{V.}~\bibnamefont{Lubicz}}, \bibnamefont{and}
  \bibinfo{author}{\bibfnamefont{G.}~\bibnamefont{Martinelli}},
  \bibinfo{journal}{Phys. Rev.} \textbf{\bibinfo{volume}{D61}},
  \bibinfo{pages}{114507} (\bibinfo{year}{2000}), \eprint{hep-lat/9909082}.

\bibitem[{\citenamefont{Chetyrkin and Retey}(2000)}]{Chetyrkin:1999pq}
\bibinfo{author}{\bibfnamefont{K.~G.} \bibnamefont{Chetyrkin}}
  \bibnamefont{and} \bibinfo{author}{\bibfnamefont{A.}~\bibnamefont{Retey}},
  \bibinfo{journal}{Nucl. Phys.} \textbf{\bibinfo{volume}{B583}},
  \bibinfo{pages}{3} (\bibinfo{year}{2000}), \eprint{hep-ph/9910332}.

\bibitem[{\citenamefont{Aoki et~al.}(2006)}]{Aoki:2005ga}
\bibinfo{author}{\bibfnamefont{Y.}~\bibnamefont{Aoki}} \bibnamefont{et~al.},
  \bibinfo{journal}{Phys. Rev.} \textbf{\bibinfo{volume}{D73}},
  \bibinfo{pages}{094507} (\bibinfo{year}{2006}), \eprint{hep-lat/0508011}.

\bibitem[{\citenamefont{Becirevic}(2004)}]{Becirevic:2004fw}
\bibinfo{author}{\bibfnamefont{D.}~\bibnamefont{Becirevic}},
  \bibinfo{journal}{Nucl. Phys. Proc. Suppl.} \textbf{\bibinfo{volume}{129}},
  \bibinfo{pages}{34} (\bibinfo{year}{2004}).

\bibitem[{\citenamefont{Furman and Shamir}(1995)}]{Furman:1995ky}
\bibinfo{author}{\bibfnamefont{V.}~\bibnamefont{Furman}} \bibnamefont{and}
  \bibinfo{author}{\bibfnamefont{Y.}~\bibnamefont{Shamir}},
  \bibinfo{journal}{Nucl. Phys.} \textbf{\bibinfo{volume}{B439}},
  \bibinfo{pages}{54} (\bibinfo{year}{1995}), \eprint{hep-lat/9405004}.

\bibitem[{\citenamefont{Antonio et~al.}(2007{\natexlab{b}})}]{Antonio:2007tr}
\bibinfo{author}{\bibfnamefont{D.~J.} \bibnamefont{Antonio}}
  \bibnamefont{et~al.} (\bibinfo{year}{2007}{\natexlab{b}}),
  \eprint{arXiv:0705.2340 [hep-lat]}.

\bibitem[{\citenamefont{Golterman and Shamir}(2003)}]{Golterman:2003qe}
\bibinfo{author}{\bibfnamefont{M.}~\bibnamefont{Golterman}} \bibnamefont{and}
  \bibinfo{author}{\bibfnamefont{Y.}~\bibnamefont{Shamir}},
  \bibinfo{journal}{Phys. Rev.} \textbf{\bibinfo{volume}{D68}},
  \bibinfo{pages}{074501} (\bibinfo{year}{2003}), \eprint{hep-lat/0306002}.

\bibitem[{\citenamefont{Golterman
  et~al.}(2005{\natexlab{a}})\citenamefont{Golterman, Shamir, and
  Svetitsky}}]{Golterman:2004cy}
\bibinfo{author}{\bibfnamefont{M.}~\bibnamefont{Golterman}},
  \bibinfo{author}{\bibfnamefont{Y.}~\bibnamefont{Shamir}}, \bibnamefont{and}
  \bibinfo{author}{\bibfnamefont{B.}~\bibnamefont{Svetitsky}},
  \bibinfo{journal}{Phys. Rev.} \textbf{\bibinfo{volume}{D71}},
  \bibinfo{pages}{071502} (\bibinfo{year}{2005}{\natexlab{a}}),
  \eprint{hep-lat/0407021}.

\bibitem[{\citenamefont{Golterman
  et~al.}(2005{\natexlab{b}})\citenamefont{Golterman, Shamir, and
  Svetitsky}}]{Golterman:2005fe}
\bibinfo{author}{\bibfnamefont{M.}~\bibnamefont{Golterman}},
  \bibinfo{author}{\bibfnamefont{Y.}~\bibnamefont{Shamir}}, \bibnamefont{and}
  \bibinfo{author}{\bibfnamefont{B.}~\bibnamefont{Svetitsky}}
  (\bibinfo{year}{2005}{\natexlab{b}}), \eprint{hep-lat/0503037}.

\bibitem[{\citenamefont{Svetitsky et~al.}(2006)\citenamefont{Svetitsky, Shamir,
  and Golterman}}]{Svetitsky:2005qa}
\bibinfo{author}{\bibfnamefont{B.}~\bibnamefont{Svetitsky}},
  \bibinfo{author}{\bibfnamefont{Y.}~\bibnamefont{Shamir}}, \bibnamefont{and}
  \bibinfo{author}{\bibfnamefont{M.}~\bibnamefont{Golterman}},
  \bibinfo{journal}{PoS} \textbf{\bibinfo{volume}{LAT2005}},
  \bibinfo{pages}{129} (\bibinfo{year}{2006}), \eprint{hep-lat/0508015}.

\bibitem[{\citenamefont{Golterman and Shamir}(2005)}]{Golterman:2004mf}
\bibinfo{author}{\bibfnamefont{M.}~\bibnamefont{Golterman}} \bibnamefont{and}
  \bibinfo{author}{\bibfnamefont{Y.}~\bibnamefont{Shamir}},
  \bibinfo{journal}{Phys. Rev.} \textbf{\bibinfo{volume}{D71}},
  \bibinfo{pages}{034502} (\bibinfo{year}{2005}), \eprint{hep-lat/0411007}.

\bibitem[{\citenamefont{Antonio et~al.}(2007{\natexlab{c}})}]{Antonio:2007pb}
\bibinfo{author}{\bibfnamefont{D.~J.} \bibnamefont{Antonio}}
  \bibnamefont{et~al.} (\bibinfo{year}{2007}{\natexlab{c}}),
  \eprint{hep-ph/0702042}.

\bibitem[{\citenamefont{van Ritbergen et~al.}(1997)\citenamefont{van Ritbergen,
  Vermaseren, and Larin}}]{vanRitbergen:1997va}
\bibinfo{author}{\bibfnamefont{T.}~\bibnamefont{van Ritbergen}},
  \bibinfo{author}{\bibfnamefont{J.~A.~M.} \bibnamefont{Vermaseren}},
  \bibnamefont{and} \bibinfo{author}{\bibfnamefont{S.~A.} \bibnamefont{Larin}},
  \bibinfo{journal}{Phys. Lett.} \textbf{\bibinfo{volume}{B400}},
  \bibinfo{pages}{379} (\bibinfo{year}{1997}), \eprint{hep-ph/9701390}.

\bibitem[{\citenamefont{Yao et~al.}(2006)}]{Yao:2006px}
\bibinfo{author}{\bibfnamefont{W.~M.} \bibnamefont{Yao}} \bibnamefont{et~al.}
  (\bibinfo{collaboration}{Particle Data Group}), \bibinfo{journal}{J. Phys.}
  \textbf{\bibinfo{volume}{G33}}, \bibinfo{pages}{1} (\bibinfo{year}{2006}).

\bibitem[{\citenamefont{Gracey}(2003)}]{Gracey:2003yr}
\bibinfo{author}{\bibfnamefont{J.~A.} \bibnamefont{Gracey}},
  \bibinfo{journal}{Nucl. Phys.} \textbf{\bibinfo{volume}{B662}},
  \bibinfo{pages}{247} (\bibinfo{year}{2003}), \eprint{hep-ph/0304113}.

\bibitem[{\citenamefont{Dawson}(2003)}]{Dawson:2002nr}
\bibinfo{author}{\bibfnamefont{C.}~\bibnamefont{Dawson}}
  (\bibinfo{collaboration}{RBC}), \bibinfo{journal}{Nucl. Phys. Proc. Suppl.}
  \textbf{\bibinfo{volume}{119}}, \bibinfo{pages}{314} (\bibinfo{year}{2003}),
  \eprint{hep-lat/0210005}.

\bibitem[{\citenamefont{Aoki et~al.}(2007)\citenamefont{Aoki, Dawson, Noaki,
  and Soni}}]{Aoki:2006ib}
\bibinfo{author}{\bibfnamefont{Y.}~\bibnamefont{Aoki}},
  \bibinfo{author}{\bibfnamefont{C.}~\bibnamefont{Dawson}},
  \bibinfo{author}{\bibfnamefont{J.}~\bibnamefont{Noaki}}, \bibnamefont{and}
  \bibinfo{author}{\bibfnamefont{A.}~\bibnamefont{Soni}},
  \bibinfo{journal}{Phys. Rev.} \textbf{\bibinfo{volume}{D75}},
  \bibinfo{pages}{014507} (\bibinfo{year}{2007}), \eprint{hep-lat/0607002}.

\end{thebibliography}

\newpage{}

\begin{table}[H]

\caption{The four factors $Z_{S,P}$, $Z_{V,A}$, $Z_{T}$ and $Z_{B_K}$
by which the matrix elements of the bare lattice bilinear operators
and the ratio of matrix elements $B_K$ should be multiplied in order
to obtain the corresponding quantities renormalized in the RI/MOM or
$\overline{\textrm{MS}}$(NDR) schemes.  The RI/MOM quantities are
defined at a scale $\mu=2.037$ GeV, an available lattice momentum.
The $\overline{\textrm{MS}}$(NDR) quantities are provided at the
scale $\mu=2$\,GeV.  The first error given is statistical and the
second systematic.  This table summarizes the main results of this
paper. \label{tab:z_npr_summary}}

\begin{centering}
\begin{tabular}{ccccccc}
\hline
\hline Scheme&
Scale &
$Z_q$            &   $Z_{S,P}$      &    $Z_{V,A}$    &    $Z_{T}$          &  $Z_{B_{K}}$\tabularnewline
\hline
RI/MOM&
2.037 GeV &
0.8086(28)(74)    &  0.466(14)(31)  &    0.7161(1)    &    0.8037(22)(55)  & 0.9121(38)(129) \tabularnewline
$\overline{\textrm{MS}}$(NDR)&
2.00 GeV &
0.7726(30)(83)    &  0.604(18)(55)  &    0.7161(1)    &    0.7950(34)(150) & 0.9276(52)(220) \tabularnewline
\hline
\hline
&
&
&
&
\tabularnewline
\end{tabular}
\par\end{centering}
\end{table}
\begin{table}[H]

\caption{The factors, computed in perturbation theory, by which
the matrix elements of the bare lattice operators should be
multiplied in order to obtain those in the
$\overline{\textrm{MS}}$(NDR) scheme at the renormalization
scale $\mu=1.729$\,GeV. This table shows that the difference in the choice of
the strong coupling constant leads to large uncertainty in the
renormalization constants.\label{tab:alpha diff Z diff}}

\vspace{0.4cm}\begin{centering}
\begin{tabular}{ccccc}
\hline \hline Coupling& $Z_{S,P}(1.729\,\textrm{GeV})$&
$Z_{V,A}(1.729\,\textrm{GeV})$& $Z_{T}(1.729\,\textrm{GeV})$&
$Z_{B_{K}}(1.729\,\textrm{GeV})$\tabularnewline \hline
$\alpha_{\textrm{MF}}(1.729\,\textrm{GeV})$& 0.788& 0.801& 0.827&
0.979\tabularnewline
$\alpha^{\overline{\textrm{MS}}}(1.729\,\textrm{GeV})$& 0.672&
0.693& 0.737& 0.963\tabularnewline \hline \hline & & & &
\tabularnewline
\end{tabular}
\par\end{centering}
\end{table}

\begin{table}[H]

\caption{The perturbative renormalization constants at the
conventional scale of $\mu=2\,\mbox{GeV}$ by renormalization
group running from $\mu=1.729$\,GeV. The entries in the first
column indicate which coupling was used in matching between the
bare lattice operators and the $\overline{\textrm{MS}}$(NDR)
scheme at $\mu=1.729$\,GeV.\label{tab:latt pert Z}}

\vspace{0.4cm}\begin{centering}
\begin{tabular}{ccccc}
\hline \hline Coupling& $Z_{S,P}(2\,\textrm{GeV})$&
$Z_{V,A}(2\,\textrm{GeV})$& $Z_{T}(2\,\textrm{GeV})$&
$Z_{B_{K}}(2\,\textrm{GeV})$\tabularnewline \hline
$\alpha_{\textrm{MF}}(1.729\,\textrm{GeV})$& 0.822& 0.801& 0.813&
0.993\tabularnewline
$\alpha^{\overline{\textrm{MS}}}(1.729\,\textrm{GeV})$& 0.701&
0.693& 0.725& 0.977\tabularnewline \hline \hline & & & &
\tabularnewline
\end{tabular}
\par\end{centering}
\end{table}

\begin{table}[H]

\caption{The quantity $\frac{1}{12}\mathrm{Tr}(S_{\mathrm{latt}}^{-1})$ evaluated
at the unitary mass points, $m_\mathrm{val}=m_l$ and linearly extrapolated to the
chiral limit $m_l=-m_{\mathrm{res}}$. \label{tab:TrSinv_dyn}}

\begin{centering}

\begin{tabular}{ccccc}
\hline
\hline
$ (ap)^2 $ & $ m_{l}=0.01 $ & $ m_{l}=0.02 $ & $ m_{l}=0.03 $ & $ \mathrm{chiral~limit} $\\
\hline
0.347 & 0.0839(16) & 0.1141(16) & 0.1327(23) & 0.0524(34)\\
0.617 & 0.0558(13) & 0.0810(15) & 0.0980(20) & 0.0283(28)\\
0.810 & 0.0450(12) & 0.0692(14) & 0.0849(18) & 0.0187(28)\\
1.079 & 0.03744(82) & 0.0583(11) & 0.0741(16) & 0.0130(20)\\
1.234 & 0.0342(12) & 0.0543(13) & 0.0704(16) & 0.0105(26)\\
1.388 & 0.03203(84) & 0.0512(11) & 0.0665(15) & 0.0092(20)\\
1.542 & 0.03051(75) & 0.04873(97) & 0.0634(14) & 0.0087(18)\\
1.851 & 0.02640(92) & 0.04472(99) & 0.0597(13) & 0.0047(20)\\
2.005 & 0.02615(73) & 0.04354(92) & 0.0575(13) & 0.0054(18)\\
2.467 & 0.0236(10) & 0.0404(10) & 0.0540(13) & 0.0040(20)\\
\hline
\hline
\end{tabular}

\par\end{centering}
\end{table}

\begin{table}[H]

\caption{The five bare vertex amplitudes $\Lambda_{i}$, $i\in\left\{ S,P,V,A,T\right\} $
averaged over four sources, with $m_l=m_\mathrm{val}=0.01$\label{tab:Lmd-all-0.01}.}

\begin{centering}

\begin{tabular}{cccccc}
\hline
\hline
$ (ap)^2 $ & $ \Lambda_S $ & $ \Lambda_P $ & $ \Lambda_V $ & $ \Lambda_A $ & $ \Lambda_T $\\
\hline
0.347 & 2.125(86) & 6.72(19) & 1.1702(58) & 1.0675(43) & 0.8904(43)\\
0.617 & 1.945(51) & 4.45(11) & 1.1419(37) & 1.0938(30) & 0.9404(26)\\
0.810 & 1.856(37) & 3.677(81) & 1.1348(31) & 1.1025(27) & 0.9618(19)\\
1.079 & 1.758(27) & 3.022(57) & 1.1335(29) & 1.1135(27) & 0.9882(16)\\
1.234 & 1.715(24) & 2.792(50) & 1.1291(29) & 1.1137(27) & 0.9935(17)\\
1.388 & 1.677(21) & 2.600(43) & 1.1328(26) & 1.1191(24) & 1.0065(13)\\
1.542 & 1.642(19) & 2.448(38) & 1.1355(27) & 1.1240(25) & 1.0167(14)\\
1.851 & 1.599(16) & 2.239(32) & 1.1387(29) & 1.1301(27) & 1.0310(16)\\
2.005 & 1.578(15) & 2.154(28) & 1.1420(27) & 1.1342(26) & 1.0392(16)\\
2.467 & 1.532(13) & 1.979(23) & 1.1495(29) & 1.1434(29) & 1.0577(19)\\
\hline
\hline
\end{tabular}

\par\end{centering}
\end{table}

\begin{table}[H]

\caption{The five bare vertex amplitudes $\Lambda_{i}$, $i\in\left\{ S,P,V,A,T\right\} $
averaged over four sources, with $m_l=m_\mathrm{val}=0.02$\label{tab:Lmd-all-0.02}.}

\begin{centering}

\begin{tabular}{cccccc}
\hline
\hline
$ (ap)^2 $ & $ \Lambda_S $ & $ \Lambda_P $ & $ \Lambda_V $ & $ \Lambda_A $ & $ \Lambda_T $\\
\hline
0.347 & 1.828(45) & 5.09(14) & 1.1745(46) & 1.0412(28) & 0.8930(31)\\
0.617 & 1.774(30) & 3.600(82) & 1.1465(30) & 1.0838(21) & 0.9414(18)\\
0.810 & 1.721(24) & 3.052(61) & 1.1360(24) & 1.0943(19) & 0.9614(15)\\
1.079 & 1.655(19) & 2.590(45) & 1.1331(22) & 1.1069(20) & 0.9870(12)\\
1.234 & 1.637(16) & 2.428(40) & 1.1307(21) & 1.1083(20) & 0.9930(12)\\
1.388 & 1.608(15) & 2.283(33) & 1.1323(21) & 1.1141(19) & 1.0049(11)\\
1.542 & 1.581(14) & 2.175(30) & 1.1351(21) & 1.1199(20) & 1.0159(12)\\
1.851 & 1.552(11) & 2.019(24) & 1.1389(22) & 1.1275(21) & 1.0309(12)\\
2.005 & 1.532(11) & 1.955(23) & 1.1416(23) & 1.1315(22) & 1.0390(14)\\
2.467 & 1.4984(91) & 1.819(18) & 1.1498(26) & 1.1422(25) & 1.0580(17)\\
\hline
\hline
\end{tabular}

\par\end{centering}
\end{table}

\begin{table}[H]

\caption{The five bare vertex amplitudes $\Lambda_{i}$, $i\in\left\{ S,P,V,A,T\right\} $
averaged over four sources, with $m_l=m_\mathrm{val}=0.03$\label{tab:Lmd-all-0.03}.}

\begin{centering}

\begin{tabular}{cccccc}
\hline
\hline
$ (ap)^2 $ & $ \Lambda_S $ & $ \Lambda_P $ & $ \Lambda_V $ & $ \Lambda_A $ & $ \Lambda_T $\\
\hline
0.347 & 1.723(56) & 4.10(14) & 1.1809(56) & 1.0357(23) & 0.9020(24)\\
0.617 & 1.702(37) & 3.049(87) & 1.1457(37) & 1.0769(19) & 0.9451(17)\\
0.810 & 1.663(28) & 2.658(66) & 1.1356(31) & 1.0886(17) & 0.9642(14)\\
1.079 & 1.610(21) & 2.307(49) & 1.1325(27) & 1.1015(18) & 0.9883(12)\\
1.234 & 1.591(18) & 2.182(42) & 1.1294(25) & 1.1050(20) & 0.9951(12)\\
1.388 & 1.569(15) & 2.076(37) & 1.1312(25) & 1.1105(20) & 1.0061(11)\\
1.542 & 1.548(13) & 1.991(33) & 1.1337(26) & 1.1157(21) & 1.0161(12)\\
1.851 & 1.520(10) & 1.869(27) & 1.1366(26) & 1.1228(22) & 1.0300(13)\\
2.005 & 1.5065(96) & 1.820(25) & 1.1395(27) & 1.1271(24) & 1.0382(15)\\
2.467 & 1.4764(78) & 1.717(20) & 1.1464(27) & 1.1371(26) & 1.0561(17)\\
\hline
\hline
\end{tabular}

\par\end{centering}
\end{table}

\begin{table}[H]

\caption{\label{tab:nonexp-mom1} Groups of non-exceptional momenta satisfying
$p_{1}^{2}=p_{2}^{2}=\left(p_{1}-p_{2}\right)^{2}$.  The individual integers
$(n_x, n_y, n_z, n_t)$ should be multiplied by $2\pi/L_{d}$,
with $L_{x}=L_{y}=L_{z}=16$ and $L_{t}=32$.}

\begin{centering}
\begin{tabular}{ccc}
\hline
\hline$\left(ap\right)^{2}$&
$p_{1}$&
$p_{2}$\tabularnewline
\hline
0.617&
(1,1,1,2)&
(1,-1,1,2)\tabularnewline
&
(1,1,1,2)&
(1,1,-1,2)\tabularnewline
&
(1,1,1,2)&
(-1,1,1,2)\tabularnewline
&
(1,1,1,2)&
(1,1,1,-2)\tabularnewline
&
(1,1,1,2)&
(0,0,0,4)\tabularnewline
&
(1,1,1,2)&
(0,0,2,0)\tabularnewline
&
(1,1,1,2)&
(0,2,0,0)\tabularnewline
&
(1,1,1,2)&
(2,0,0,0)\tabularnewline
\hline
0.925&
(-1,-1,-2,0)&
(-2,-1,0,-2)\tabularnewline
&
(-1,-1,-2,0)&
(-2,-1,0,2)\tabularnewline
&
(-1,-1,-2,0)&
(-2,1,-1,0)\tabularnewline
&
(-1,-1,-2,0)&
(-1,-2,0,-2)\tabularnewline
&
(-1,-1,-2,0)&
(-1,-2,0,2)\tabularnewline
&
(-1,-1,-2,0)&
(-1,0,-1,-4)\tabularnewline
&
(-1,-1,-2,0)&
(-1,0,-1,4)\tabularnewline
&
(-1,-1,-2,0)&
(0,-1,-1,-4)\tabularnewline
&
(-1,-1,-2,0)&
(0,-1,-1,4)\tabularnewline
&
(-1,-1,-2,0)&
(0,1,-2,-2)\tabularnewline
&
(-1,-1,-2,0)&
(0,1,-2,2)\tabularnewline
&
(-1,-1,-2,0)&
(1,-2,-1,0)\tabularnewline
&
(-1,-1,-2,0)&
(1,0,-2,-2)\tabularnewline
&
(-1,-1,-2,0)&
(1,0,-2,2)\tabularnewline
\hline
\hline
&
&
\tabularnewline
\end{tabular}
\par\end{centering}
\end{table}

\begin{table}[H]

\caption{\label{tab:nonexp-mom2} Groups of non-exceptional momenta
satisfying $p_{1}^{2}=p_{2}^{2}=\left(p_{1}-p_{2}\right)^{2}$,
continuing Table~\ref{tab:nonexp-mom1}.}

\begin{centering}
\begin{tabular}{ccc}
\hline
\hline$\left(ap\right)^{2}$&
$p_{1}$&
$p_{2}$\tabularnewline
\hline
1.234&
(0,2,2,0)&
(2,2,0,0)\tabularnewline
&
(0,2,2,0)&
(0,2,0,4)\tabularnewline
&
(0,2,2,0)&
(0,0,2,4)\tabularnewline
&
(0,2,2,0)&
(-2,2,0,0)\tabularnewline
&
(0,2,2,0)&
(0,2,0,-4)\tabularnewline
&
(0,2,2,0)&
(2,0,2,0)\tabularnewline
&
(0,2,2,0)&
(0,0,2,-4)\tabularnewline
&
(0,2,2,0)&
(-2,0,2,0)\tabularnewline
\hline
1.542&
(1,1,2,4)&
(2,1,2,-2)\tabularnewline
&
(1,1,2,4)&
(1,-2,2,2)\tabularnewline
&
(1,1,2,4)&
(-2,1,2,2)\tabularnewline
&
(1,1,2,4)&
(-2,1,1,4)\tabularnewline
&
(1,1,2,4)&
(1,2,2,-2)\tabularnewline
&
(1,1,2,4)&
(1,-2,1,4)\tabularnewline
&
(1,1,2,4)&
(2,1,-1,4)\tabularnewline
&
(1,1,2,4)&
(1,2,-1,4)\tabularnewline
\hline
2.467&
(2,2,2,4)&
(2,2,-2,4)\tabularnewline
&
(2,2,2,4)&
(2,-2,2,4)\tabularnewline
&
(2,2,2,4)&
(-2,2,2,4)\tabularnewline
&
(2,2,2,4)&
(2,2,2,-4)\tabularnewline
\hline
\hline
&
&
\tabularnewline
\end{tabular}
\par\end{centering}
\end{table}

\begin{table}[H]

\caption{Results from fitting the coefficient for mass term in
$(\Lambda_{A}-\Lambda_{V})/[(\Lambda_{A}+\Lambda_{V})/2]$.
The linear dependence is assumed to be $c_{1}\frac{m\Lambda_{QCD}}{p^{2}}$
and the quadratic dependence is assumed to be $c_{2}\frac{m^{2}}{p^{2}}$.
The respective $\chi^{2}/d.o.f$ is also listed.  Both the coefficient
$c_1$ more nearly agreeing with its expected value of 1 and the smaller
$\chi^2$ suggest that the linear description is to be preferred.  We
use the value $\Lambda_\mathrm{QCD}=  319.5$ MeV.
\label{tab:LA-LV_fits}}

\begin{centering}

\begin{tabular}{ccccc}
\hline
\hline
$ (ap)^2 $ & $ c_{1} $ & $ (\chi^2/dof)_{1} $ & $ c_{2} $ & $ (\chi^2/dof)_{2} $\\
\hline
0.347 & -3.84(75) & 3.0(3.4) & -14.7(3.1) & 6.6(5.1)\\
0.617 & -3.33(67) & 2.2(2.8) & -12.9(2.8) & 5.4(4.4)\\
0.810 & -3.06(56) & 1.2(2.1) & -12.1(2.5) & 4.1(3.7)\\
1.079 & -3.01(42) & 0.4(1.3) & -12.4(2.0) & 3.3(3.6)\\
1.234 & -2.96(47) & 6.2(5.0) & -11.4(2.1) & 12.8(7.2)\\
1.388 & -2.58(36) & 1.7(2.7) & -10.5(1.7) & 6.2(4.8)\\
1.542 & -2.49(34) & 0.4(1.4) & -10.2(1.6) & 3.3(3.6)\\
1.851 & -2.33(35) & 0.06(41) & -9.5(1.6) & 1.8(2.4)\\
2.005 & -2.21(28) & 0.02(23) & -9.2(1.3) & 1.3(2.2)\\
2.467 & -1.89(32) & 0.01(22) & -7.4(1.4) & 0.7(1.5)\\
\hline
\hline
\end{tabular}

\par\end{centering}
\end{table}

\begin{table}[H]

\caption{Values for $\frac{1}{2}(\Lambda_{A}+\Lambda_{V})$, $\Lambda_{S}$,
and $\Lambda_{T}$ extrapolated to the chiral limit using a linear mass
fit. \label{tab:Lambda_fit}}

\begin{centering}

\begin{tabular}{cccc}
\hline
\hline
$ (ap)^2 $ & $ \frac{1}{2} (\Lambda_A + \Lambda_V) $ & $ \Lambda_S $ & $ \Lambda_T $\\
\hline
0.347 & 1.1211(56) & 2.28(14) & 0.8800(66)\\
0.617 & 1.1226(49) & 2.060(88) & 0.9363(36)\\
0.810 & 1.1228(41) & 1.952(66) & 0.9593(25)\\
1.079 & 1.1275(43) & 1.836(47) & 0.9873(22)\\
1.234 & 1.1242(44) & 1.784(43) & 0.9915(22)\\
1.388 & 1.1292(40) & 1.736(37) & 1.0061(18)\\
1.542 & 1.1333(43) & 1.694(32) & 1.0169(20)\\
1.851 & 1.1381(47) & 1.644(27) & 1.0319(24)\\
2.005 & 1.1417(46) & 1.617(25) & 1.0400(27)\\
2.467 & 1.1504(51) & 1.564(22) & 1.0593(33)\\
\hline
\hline
\end{tabular}

\par\end{centering}
\end{table}

\begin{table}[H]

\caption{The non-perturbative factor $Z_{m}^{\mathrm{RI/MOM}}$ as a function
of the scale $\mu$ calculated from $\Lambda_{S}$ and the corresponding values
for $Z_{m}^{\mathrm{\overline{MS}}}$.  Note that the values for
$Z_{m}^{\mathrm{\overline{MS}}}$ given in column three are obtained from
those in column two by applying the $\mathrm{RI/MOM-\overline{MS}}$
perturbative matching factors after the $O(a\mu)^2$ lattice artifacts
have been removed using an intermediate conversion to a scale-invariant
scheme as described in the text. \label{tab:Zm-comb}}

\begin{centering}

\begin{tabular}{ccc}
\hline
\hline
$ \mu (\mathrm{GeV}) $ & $ Z_m^{\mathrm{RI/MOM}} $ & $ Z_m^{\overline{\mathrm{MS}}} $\\
\hline
1.018 & 2.85(18) & 1.625(47)\\
1.358 & 2.56(11) & 1.758(51)\\
1.556 & 2.428(80) & 1.731(51)\\
1.796 & 2.273(56) & 1.690(49)\\
1.920 & 2.216(49) & 1.669(49)\\
2.037 & 2.146(42) & 1.651(48)\\
2.147 & 2.087(37) & 1.634(48)\\
2.352 & 2.018(29) & 1.605(47)\\
2.448 & 1.978(27) & 1.593(46)\\
2.716 & 1.899(22) & 1.562(46)\\
\hline
\hline
\end{tabular}

\par\end{centering}
\end{table}

\begin{table}[H]

\caption{The non-perturbative factor $Z_{q}^{\mathrm{RI/MOM}}$ as a function
of the scale $\mu$ calculated from
$\frac{1}{2}\left(\Lambda_{A}+\Lambda_{V}\right)$ and the corresponding values
for $Z_{q}^{\mathrm{\overline{MS}}}$.  Note that the values for
$Z_{q}^{\mathrm{\overline{MS}}}$ given in column three are obtained from
those in column two by applying the $\mathrm{RI/MOM-\overline{MS}}$
perturbative matching factors after the $O(a\mu)^2$ lattice artifacts
have been removed using an intermediate conversion to a scale-invariant
scheme as described in the text.
\label{tab:Zq-comb}}

\begin{centering}

\begin{tabular}{ccc}
\hline
\hline
$ \mu (\mathrm{GeV}) $ & $ Z_q^{\mathrm{RI/MOM}} $ & $ Z_q^{\overline{\mathrm{MS}}} $\\
\hline
1.018 & 0.8028(40) & 0.8010(31)\\
1.358 & 0.8039(35) & 0.7849(30)\\
1.556 & 0.8041(30) & 0.7798(30)\\
1.796 & 0.8074(31) & 0.7754(30)\\
1.920 & 0.8050(32) & 0.7736(30)\\
2.037 & 0.8086(28) & 0.7722(30)\\
2.147 & 0.8115(31) & 0.7710(30)\\
2.352 & 0.8150(34) & 0.7691(29)\\
2.448 & 0.8176(33) & 0.7684(29)\\
2.716 & 0.8238(37) & 0.7665(29)\\
\hline
\hline
\end{tabular}

\par\end{centering}
\end{table}

\begin{table}[H]

\caption{The non-perturbative factor $Z_{T}^{\mathrm{RI/MOM}}$ as a function
of the scale $\mu$ calculated from $\Lambda_{T}$ and the corresponding values
for $Z_{T}^{\mathrm{\overline{MS}}}$.  Note that the values for
$Z_{T}^{\mathrm{\overline{MS}}}$ given in column three are obtained from those
in column two by applying the $\mathrm{RI/MOM-\overline{MS}}$ perturbative
matching factors after the $O(a\mu)^2$ lattice artifacts have been removed
using an intermediate conversion to a scale-invariant scheme as described in
the text.
\label{tab:ZT-comb}}
\begin{centering}

\begin{tabular}{ccc}
\hline
\hline
$ \mu (\mathrm{GeV}) $ & $ Z_T^{\mathrm{RI/MOM}} $ & $ Z_T^{\overline{\mathrm{MS}}} $\\
\hline
1.018 & 0.9121(74) & 0.8812(38)\\
1.358 & 0.8583(46) & 0.8355(36)\\
1.556 & 0.8380(32) & 0.8194(35)\\
1.796 & 0.8177(27) & 0.8048(34)\\
1.920 & 0.8118(27) & 0.7986(34)\\
2.037 & 0.8037(22) & 0.7935(34)\\
2.147 & 0.7981(21) & 0.7892(34)\\
2.352 & 0.7899(18) & 0.7821(33)\\
2.448 & 0.7862(17) & 0.7791(33)\\
2.716 & 0.7779(16) & 0.7719(33)\\
\hline
\hline
\end{tabular}

\par\end{centering}
\end{table}

\begin{table}[H]
\caption{The quantity $Z_{B_{K}}^{\mathrm{RI/MOM}}$ evaluated at the unitary
points where $m_\mathrm{val}=m_l=m$.
\label{tab:Zbk_mass}}
\begin{centering}

\begin{tabular}{cccc}
\hline
\hline
$ \mu (\mathrm{GeV}) $ & $ m = 0.01 $ & $ m = 0.02 $ & $ m = 0.03 $\\
\hline
0.954 & 0.9663(69) & 0.9737(52) & 0.9538(44)\\
1.272 & 0.9347(39) & 0.9387(35) & 0.9315(30)\\
1.458 & 0.9266(31) & 0.9289(30) & 0.9245(26)\\
1.683 & 0.9189(25) & 0.9189(25) & 0.9167(24)\\
1.799 & 0.9151(22) & 0.9137(23) & 0.9126(23)\\
1.909 & 0.9114(23) & 0.9106(22) & 0.9102(22)\\
2.012 & 0.9085(22) & 0.9077(20) & 0.9078(21)\\
2.204 & 0.9045(23) & 0.9026(20) & 0.9035(21)\\
2.294 & 0.9018(20) & 0.9004(19) & 0.9020(20)\\
2.545 & 0.8974(21) & 0.8953(19) & 0.8978(19)\\
\hline
\hline
\end{tabular}

\par\end{centering}
\end{table}

\begin{table}[H]
\caption{The non-perturbative factor $Z_{B_{K}}^{\mathrm{RI/MOM}}$
as a function of the scale $\mu$ and the corresponding values for
$Z_{B_{K}}^{\overline{\mathrm{MS}}}$.  Note that the values for
$Z_{B_{K}}^{\overline{\mathrm{MS}}}$ given in column three are obtained from
those in column two by applying the $\mathrm{RI/MOM-\overline{MS}}$
perturbative matching factors after the $O(a\mu)^2$ lattice artifacts have
been removed using an intermediate conversion to a scale-invariant scheme as
described in the text. \label{tab:Zbk-comb}}
\begin{centering}

\begin{tabular}{ccc}
\hline
\hline
$ \mu (\mathrm{GeV}) $ & $ Z_{B_K}^{\mathrm{RI/MOM}} $ & $ Z_{B_K}^{\overline{\mathrm{MS}}} $\\
\hline
1.018 & 0.985(11) & 1.0016(56)\\
1.358 & 0.9397(61) & 0.9651(54)\\
1.556 & 0.9295(48) & 0.9507(54)\\
1.796 & 0.9208(42) & 0.9370(53)\\
1.920 & 0.9168(38) & 0.9311(52)\\
2.037 & 0.9121(38) & 0.9261(52)\\
2.147 & 0.9088(37) & 0.9217(52)\\
2.352 & 0.9045(39) & 0.9145(52)\\
2.448 & 0.9011(35) & 0.9114(51)\\
2.716 & 0.8961(38) & 0.9038(51)\\
\hline
\hline
\end{tabular}

\par\end{centering}
\end{table}

\newpage{}


\end{document}